\documentclass[a4paper,11pt]{article}
\pdfoutput=1 

\input{packages}

\title{A hierarchical Bayesian framework for cosmology using Type~1 AGN variability}

\author[a,b]{Júlia Laguna-Miralles,}
\author[a]{Vasily Belokurov}
\author[a,b,c]{Miles Cranmer}

\affiliation[a]{Institute of Astronomy, University of Cambridge}
\affiliation[b]{Kavli Institute for Cosmology, University of Cambridge}
\affiliation[c]{Department of Applied Mathematics and Theoretical Physics, University of Cambridge}

\emailAdd{jl2268@cam.ac.uk}

\abstract{

Independent luminosity-distance probes at redshifts above two are needed to test the
cosmic expansion history beyond the redshift range densely populated by Type~Ia supernovae. Type~1 active galactic nuclei (AGN) are abundant at these
redshifts, but their use for cosmology requires standardizable observables with
controlled scatter, evolution, and measurement uncertainty. We present a hierarchical Bayesian framework for
cosmology using AGN variability, based on the empirical anti-correlation between
optical/UV variability amplitude and luminosity. The framework targets the moderate-baseline regime of current
wide-field time-domain surveys, where individual light curves cannot generally identify the full long-timescale stochastic process, but can constrain
finite-window mean brightness and short-lag variability amplitude. Each light curve is fitted
independently to obtain posterior samples of these summaries, which are then importance-reweighted under a population model relating variability to
luminosity, rest-frame wavelength, intrinsic scatter, and the assumed distance--redshift relation. This two-stage construction propagates object-level uncertainty while avoiding repeated light-curve likelihood evaluations during
cosmological inference, making analyses of hundreds of thousands to millions of light curves per band feasible. Using Gaia DR3-like G-band simulations matched to real Gaia cadences, noise properties, and quality cuts, we show that finite-baseline light
curves are more robustly summarized by window-averaged brightness and short-lag variability than by the separate long-timescale parameters of the stochastic model. End-to-end closure tests recover the injected variability--luminosity relation,
intrinsic scatter, and distance--redshift parameters up to the expected calibration degeneracies. The present Gaia G-band study is therefore a proof-of-concept validation of a scalable AGN-variability distance framework, with the main gains expected from Gaia DR4, ZTF, DESI-selected AGN samples,
and Rubin/LSST-era data.

}


\begin{document}
\maketitle

\section{Introduction}
\label{sec:introduction}

\subsection{AGN as high-redshift cosmological probes}
\label{sec:intro_agn_probes}

Independent distance–redshift measurements provide a direct test of the late-time expansion history, complementary to Cosmic Microwave Background (CMB) and large-scale-structure constraints. This is especially relevant in view of current cosmological tensions. In flat \(\Lambda\)CDM, Planck CMB data infer \(H_0=67.4\pm0.5\,{\rm km\,s^{-1}\,Mpc^{-1}}\), whereas the SH0ES distance ladder with Cepheid-Supernova Type Ia (SN Ia) gives \(H_0=73.04\pm1.04\,{\rm km\,s^{-1}\,Mpc^{-1}}\), corresponding to a \(\simeq4.9\sigma\) tension \citep{Planck_2020,Riess_2022}. Recent DESI DR2 Baryon Acoustic Oscillation (BAO) measurements are consistent with flat \(\Lambda\)CDM when considered alone, but their combination with CMB and SN~Ia data has led to reported preferences for time-varying dark-energy extensions \citep{Abdul_Karim_2025,Lodha_2025}. The interpretation of these hints remains unsettled, with independent analyses emphasizing sensitivity to dark-energy parametrization, reconstruction method, data-set choices, and inter-data-set tensions \citep{Cortes_2024a,Cortes_2024b,JiangPedrotti_2024,Efstathiou_2025, Efstathiou_2025_SN, Ong_2026}. Hence, these results motivate independent standardizable luminosity probes with different astrophysical systematics across the SN--BAO redshift range. 

SN~Ia remain the benchmark late-time luminosity-distance probe. Pantheon+
contains \(1701\) light curves of \(1550\) spectroscopically confirmed
SNe~Ia over \(0.001<z<2.26\) \citep{Brout_2022,Scolnic_2022}. However, at higher
redshift, where SN~Ia samples become sparse, Type~1 AGN provide a natural
complement: they are luminous, numerous, and observable across a much larger
cosmic volume. The SDSS DR16 quasar catalogue contains \(750{,}414\)
broad-line quasars over \(0.1\lesssim z\lesssim6\)
\citep{Lyke_2020,WuShen_2022}; of
which the Gaia--SDSS sample analysed here provides \(2.15\times10^5\) objects with Gaia DR3 \(G\)-band
epoch photometry \citep{Gaia_DR3,eyer_2023gaia}. Nonetheless, AGN are not standard candles in their intrinsic luminosities. Their cosmological use must therefore rely on standardizable empirical or physical relations connecting observables to intrinsic luminosity.

Throughout, we use “AGN” and “quasar” interchangeably for unobscured Type 1 AGN.

\subsection{AGN standardization methods for luminosity distances}
\label{sec:intro_existing_agn_distances}

AGN have been proposed as luminosity-distance indicators through several
complementary standardization approaches. The most extensively developed is the
X-ray--UV quasar Hubble diagram, which exploits the non-linear relation between
the rest-frame monochromatic luminosities at \(2500\,\text{\AA}\) and \(2\,{\rm keV}\),
\(L_{\rm UV}\) and \(L_{\rm X}\), to infer luminosity distances out to
\(z\simeq7.5\) \citep[see, e.g.,][]{Risaliti_2015,Risaliti_2019,Lusso_2020}.
In the redshift range overlapping current SN~Ia samples, these distances are broadly consistent with the supernova Hubble diagram, while at \(z\gtrsim1.5\)--\(2\) reported departures from flat \(\Lambda\)CDM have been discussed, within the adopted parametrizations, as possible evidence for evolving dark energy \citep{Risaliti_2019,Lusso_2020,Lusso_2025}. At present, the method is limited by sample statistics and per-object scatter: cleaned samples contain \(N\simeq2.4\times10^3\) quasars, with an observed \(\simeq0.21\)--\(0.24\) dex dispersion in the 
\(L_{\rm X}\)--\(L_{\rm UV}\) relation. This corresponds to a typical Hubble-diagram scatter
of \(\simeq1.4\) mag, compared with \(\simeq0.15\) mag for post-standardized SN~Ia samples \citep{Lusso_2017,Lusso_2020,Lusso_2025,Brout_2022,Scolnic_2022}. Because the method requires matched optical/UV and X-ray measurements, further growth is tied to suitable X-ray coverage, making X-ray depth and uniformity a
key limiting factor. Related AGN distance indicators, such
as radio-loud quasar multi-band luminosity correlations and the C~{\sc iv} Baldwin effect, provide complementary routes but currently rely on more
specialized samples \citep{Huang_2024RLQ,Huang_2023CIV}. This motivates time-domain approaches that can exploit the much larger samples readily available from wide-field photometric surveys.

Optical/UV variability provides one such route. At fixed rest-frame wavelength and timescale, more luminous Type~1 AGN are less variable \citep[see, e.g.,][]{Hook_1994,VandenBerk_2004,Kelly_2009,MacLeod_2010,MacLeod_2012,Caplar_2017,Suberlak_2021, Arevalo_2023}. Because the luminosity inferred from an observed magnitude at a given redshift is cosmology-dependent, the variability-luminosity anti-correlation turns continuum variability from repeated imaging into a statistical distance
indicator. Most closely related to the present work, ~\cite{Dutra_2025} used
\(\sim20\)-yr multi-band light curves for 6992 spectroscopically confirmed
Type~1 AGN, assembled from SDSS, Pan-STARRS1, and ZTF photometry
\citep{Yu_2025}, to calibrate a variability--UV-luminosity relation and
construct an AGN Hubble diagram to \(z\simeq3.5\). Combined with
SN~Ia distances, their analysis reported a preference for evolving dark-energy
parameterizations over constant-\(w\) and flat \(\Lambda\)CDM models.

\subsection{A scalable framework for survey-scale AGN-variability cosmology}
\label{sec:intro_hierarchical_moderate_baseline}

The present work targets the complementary wide-survey regime, where moderate-baseline photometry is available for more than \(10^5\) AGN sources. We develop
a hierarchical Bayesian framework tailored to these data sets, formulating AGN variability cosmology as a population-level problem: propagating the finite-window brightness and variability information robustly constrained for
each object into a cosmological model.

In the surveys considered for this work, moderate-baseline photometry corresponds to the
light-curve regime probed by Gaia DR3, and extendable to ZTF data, for AGNs at \(0.5\lesssim z\lesssim4\). A typical Gaia DR3 \(G\)-band per-object observing span is \(\simeq2.6\) yr, which corresponds to \(\simeq0.5\)--\(1.7\) yr in the rest frame, while representative ZTF \(g/r\)-band
baselines of \(\simeq6.4\) yr correspond to \(\simeq1.3\)--\(4.3\) yr. These baselines contain substantial short-lag variability
information, but are typically too short to robustly recover individual
long-timescale stochastic-process parameters
\citep{Kozlowski_2017,Suberlak_2021,Hu_2024}.

The inference is therefore built around the quantities supported by finite-window
survey light curves, using a two-stage procedure designed to scale to catalogue-size AGN samples. In Stage~1, each AGN is fitted
independently with a local Gaussian-process variability model, and in Stage~2, the resulting object-level posterior samples are combined in a hierarchical population model relating variability, luminosity, redshift, apparent brightness, rest-frame wavelength, and intrinsic scatter. 

This shifts the goal from constructing precise distance moduli for individual AGN, to extracting a collective distance signal from a large quasar sample. The Gaia \(G\)-band implementation presented here validates the method with Gaia-like simulations matched to real cadence, noise, and analysis-domain
properties, demonstrating that short-lag variability quantities are recoverable from current survey photometry and can be propagated, with their uncertainties, into a scalable variability--luminosity distance framework. To our knowledge, this is the first AGN variability--cosmology framework to use short-timescale variability as the primary distance-calibration observable. It should therefore be viewed as a proof of concept for applications to Gaia DR4, ZTF, DESI-selected quasars with time-domain photometry, and Rubin/LSST-era surveys.

\subsection{Paper outline}
\label{sec:intro_outline}

The remainder of this paper is organized as follows. Section~\ref{sec:agnvariability} motivates AGN variability as a luminosity indicator and defines the finite-window brightness and short-lag variability summaries propagated in the analysis. 
Section~\ref{sec:statitsical_model} presents the statistical model, including the conceptual full generative construction, the conditional population relation targeted here, and the two-stage importance-reweighting scheme. 
Section~\ref{sec:results} validates the method with Gaia-like light-curve simulations and end-to-end closure tests. Section~\ref{sec:Discussion} discusses calibration, single-band limitations, and extensions, with the main conclusions summarized in section~\ref{sec:conclusions}.

\section{AGN variability as a luminosity distance indicator}
\label{sec:agnvariability}

Optical/UV variability is a defining property of unobscured AGN and has long
been used for time-domain characterization and selection
\citep[e.g.][]{Hook_1994,VandenBerk_2004,deVries_2005,Schmidt_2010}.
On rest-frame timescales from days to years, Type~1 AGN light curves are
typically aperiodic and stochastic. Two complementary tools are commonly used to quantify this variability: empirical structure functions and stochastic-process
modeling. 

\subsection{Structure functions}
\label{subsec:sf_background}

The structure function (SF) is a second-order lag-space summary of stochastic
variability, widely used in AGN time-domain studies
\citep[e.g.][]{Hook_1994,VandenBerk_2004,deVries_2005,MacLeod_2012,Kozlowski_2016}.
For a rest-frame magnitude process \(m:\mathbb{R}\rightarrow\mathbb{R}\)
and time lag \(\Delta t\in\mathbb{R}_{\ge0}\), we define
\begin{equation}
{\rm SF}^2(\Delta t)
=
\mathbb E
\left\{
\left[
m(t+\Delta t)-m(t)
\right]^2
\right\},
\label{eq:sf2_definition_background}
\end{equation}
where the expectation is over realizations of the process, or over an ensemble
of objects with the same second-order variability law. Since SF conventions differ in the
literature \citep{Kozlowski_2016}, throughout this work \({\rm SF}^2\) denotes
the second-moment definition in eq.~\eqref{eq:sf2_definition_background}, and
\({\rm SF}\) denotes its square root.

For a weakly stationary process with autocovariance
\(C(\Delta t)\equiv\operatorname{Cov}[m(t+\Delta t),m(t)]\) and \(\sigma_m^2\equiv C(0)\), the structure function satisfies
\begin{equation}
{\rm SF}^2(\Delta t)
=
2\left[
\sigma_m^2-C(\Delta t)
\right].
\label{eq:sf_autocov_background}
\end{equation}
Thus, for stationary variability, the SF and autocovariance encode the same
second-order information. If the process decorrelates at long lags:
\(C(\Delta t)\rightarrow0\) and
\({\rm SF}^2(\Delta t)\rightarrow2\sigma_m^2\).

In practice, empirical or binned SFs are estimated by grouping observed epoch pairs into rest-frame lag bins and correcting the pair differences for photometric noise. Such
pair-based estimates are useful diagnostics and have been central to empirical studies of
AGN variability \citep[e.g.,][]{VandenBerk_2004,deVries_2005,Wilhite_2008, Morganson_2014, Caplar_2017, Li_2018}. They are not, however, independent likelihood-level data. The same epochs enter many pairs, correlating different lag bins; irregular cadence and seasonal gaps imprint the survey window; and finite bins average over a range of rest-frame lags rather than measuring the SF at a single \(\Delta t\). At short lags, the noise correction is especially important: when intrinsic variability is comparable to the photometric uncertainties, misestimated errors can mimic either a noise floor or excess variance. Consequently, single-power-law fits can bias SF amplitudes or slopes unless both the noise term and the long-lag turnover are modelled \citep{Kozlowski_2016,Kovacevic_2021, Stone_2023}. For this reason, empirical SF results enter this work only by motivating the predictors included in the population relation; the object-level data are modelled directly with a stochastic-process light-curve likelihood.

\subsection{Stochastic-process modelling}
\label{subsec:carma_drw_background}

Stochastic-process models provide a likelihood-based alternative to binned SFs by fitting the light curve directly. A continuous-time autoregressive moving-average model, CARMA\((p,q)\), with autoregressive order \(p\) and moving-average order \(q<p\), defines a stationary Gaussian process (GP) whose second-order structure can be represented either by a time-domain covariance function or by the corresponding power spectral density (PSD). CARMA likelihoods naturally handle irregular sampling and
heteroscedastic uncertainties, while higher-order models can represent multiple characteristic timescales, PSD bends, or oscillatory structure
\citep{Kelly_2014}.

The local light-curve model is defined in rest-frame magnitude space, with latent process \(f:\mathbb{R}\to\mathbb{R}\). This follows standard optical quasar-variability analyses and is appropriate for the retained signal-to-noise range: the flux--magnitude transformation is locally nearly linear, and the propagated magnitude uncertainties are well approximated as Gaussian.
In flux units, a Gaussian process in magnitude corresponds to a lognormal latent process
\citep{Kelly_2009,MacLeod_2010,MacLeod_2012,Zu_2013,Suberlak_2021,Stone_2023}.

Our baseline local model is the CARMA\((1,0)\) process, also known as the
damped random walk (DRW) or Ornstein--Uhlenbeck process. The DRW remains a
standard baseline for optical/UV AGN variability because it is low-dimensional,
computationally efficient, and directly interpretable
\citep{Kelly_2009,Kozlowski_2010,MacLeod_2010,Zu_2013}.  In this model,
\begin{equation}
f(t)
\sim
{\rm GP}
\left(
m_0,
k_{\rm DRW}(t,t')
\right), \qquad k_{\rm DRW}(t,t')
=
\sigma^2
\exp
\left[
-\frac{|t-t'|}{\tau}
\right].
\label{eq:drw_cov_background}
\end{equation}
Here \(m_0\in\mathbb{R}\) is the stationary mean magnitude,
\(\sigma^2\in\mathbb{R}_{>0}\) is the stationary variance, and
\(\tau\in\mathbb{R}_{>0}\) is the damping timescale. The kernel is a function \(k_{\rm DRW}:\mathbb{R}\times\mathbb{R}\rightarrow\mathbb{R}\). For \(n\in\mathbb{N}\) observed epochs, let
\(\mathbf{t}\in\mathbb{R}^n\) be the rest-frame epoch vector, \(\mathbf{m}^{\rm obs}\in\mathbb{R}^n\) the observed magnitude vector,
and \(\mathbf{s}\in\mathbb{R}_{>0}^n\) the reported uncertainty vector. With an additional white-noise term \(\sigma_{\rm add}\in\mathbb{R}_{\ge0}\) to account for underestimated reported uncertainties, the marginal likelihood is
\begin{equation}
\mathbf{m}^{\rm obs}
\sim
\mathcal N
\left[
m_0\mathbf{1},
\; \mathbf C
\right],
\label{eq:LC0likelihood}
\end{equation}
where
\begin{equation}
    \mathbf C=\mathbf K_{\rm DRW}+\bm{\Sigma}, \qquad (\mathbf K_{\rm DRW})_{jj'}=k_{\rm DRW}(t_j,t_{j'}), \qquad \bm{\Sigma}= \operatorname{diag}\left(s^2_1,\dots,s^2_n\right) + \sigma_{{\rm add}}^2\mathbf I . 
    \label{eq:drwkernel}
\end{equation}
Thus \(\mathbf K_{\rm DRW},\bm{\Sigma}\in\mathbb{R}^{n\times n}\) and
\(\mathbf C\in\mathcal S_{++}^{n}\) \footnote{Here \(\mathcal S^n_{++}\) denotes the set of \(n\times n\)
symmetric positive-definite matrices.}.

The DRW covariance directly determines the corresponding structure function.
Combining eq.~\eqref{eq:drw_cov_background} with
eq.~\eqref{eq:sf_autocov_background} gives
\begin{equation}
{\rm SF}_{\rm DRW}^2(\Delta t)
=
2\sigma^2
\left[
1-
\exp\left(-\frac{|\Delta t|}{\tau}\right)
\right].
\label{eq:drw_sf_background}
\end{equation}
For \(|\Delta t|\ll\tau\),
\begin{equation}
{\rm SF}_{\rm DRW}^2(\Delta t)
\simeq
2D|\Delta t|,
\qquad
D \equiv \frac{\sigma^2}{\tau} \in \mathbb{R}_{>0}.
\label{eq:drw_short_lag_background}
\end{equation}
Thus \(D\) is the rest-frame short-lag variability rate: it has units of \({\rm mag}^2\,{\rm time}^{-1}\), and \({\rm SF}_{\rm DRW}\propto | \Delta t|^{1/2}\) for \(|\Delta t|\ll\tau\). For \(|\Delta t|\gg\tau\), the same model approaches the long-lag plateau \({\rm SF}_{\infty}=\sqrt{2}\sigma\). In frequency space, the corresponding PSD has a single characteristic frequency scale of order \((2\pi\tau)^{-1}\), below which it is approximately white-noise-like and above which it approaches a red-noise tail with slope \(-2\).

Departures from a single-break DRW have been documented outside the regime targeted in this work. On well-sampled optical timescales from roughly a month to a few years, quasar light curves are generally consistent with the DRW description, sensitive to the accuracy of the photometric-error estimates \citep{Zu_2013}. However, at the high-cadence end, Kepler AGN light curves with minute-scale sampling probe hour-to-day variability and show high-frequency PSD slopes steeper than the DRW expectation \citep{Mushotzky_2011,Kasliwal_2015}. At
shorter rest-frame wavelengths, far-UV and extreme-UV variability can show
wavelength- and timescale-dependent behaviour that is not captured by a simple
optical DRW extrapolation \citep{Zhu_2016}. And at the long-baseline end, decade-scale quasar light curves probe year-to-decade variability and reveal
baseline-dependent \(\tau\) estimates, long-term trends, and ensemble PSD shapes that are not always described by a single DRW \citep{Kozlowski_2017,Stone_2023}. More flexible stochastic models, including higher-order CARMA processes and CARMA\((2,1)\) damped harmonic oscillators, can capture additional characteristic timescales and more general PSD shapes, but
require sufficiently long and well-sampled light curves to be constrained reliably \citep{Kelly_2014,Yu_2022,Yu_2025}.

In this work, the DRW is therefore
used as a compact local likelihood for extracting a robust low-dimensional
variability summary over the rest-frame lag and wavelength range probed by our
data, rather than as a claim that AGN variability is exactly DRW on all
timescales and wavelengths.

\subsection{Finite-baseline limitations and summary statistics}
\label{subsubsec:drw_finite_baseline_background}

Let \(t_{\min},t_{\max}\in\mathbb{R}\) with \(t_{\max}>t_{\min}\) be the rest-frame endpoints of a light curve,
and define
\[
T_{\rm rf}
\equiv
t_{\max}-t_{\min}
=
\frac{t^{\rm obs}_{\max}-t^{\rm obs}_{\min}}{1+z} \in\mathbb{R}_{>0}.
\]
Finite survey baselines limit which DRW parameters are reliably identified. The long-timescale parameters \((\sigma,\tau)\) are weakly constrained
unless the light curve samples the structure-function turnover. Simulations indicate that robust object-level recovery of damping timescales generally requires \(T_{\rm rf}\gtrsim10\tau\), with the precise requirement depending on the prior, estimator, and fitting method
\citep{Kozlowski_2017,Suberlak_2021,Hu_2024}. For representative optical-quasar damping timescales
\(\tau\simeq500\)--\(750\) d, this corresponds to
\(T_{\rm rf}\sim14\)--\(21\) yr
\citep{MacLeod_2010,Suberlak_2021,Stone_2023}, far longer than the median
rest-frame baseline of the Gaia DR3 sample used in this work, \(T_{\rm rf}=1.05\) yr.
We therefore do not propagate \(\tau\), \(\sigma\), or the long-lag DRW plateau
to the cosmological hierarchy. Instead, we propagate the short-lag variability rate \(D\), defined in eq.~\eqref{eq:drw_short_lag_background}.  Section~\ref{sec:stage1_recovery_results} shows with Gaia-like simulations that \(D\) is recovered much more robustly than the separate long-timescale DRW parameters.

The finite-baseline limitation also affects the interpretation of the DRW mean. In eq.~\eqref{eq:drw_cov_background}, \(m_0\) is the stationary long-time
ensemble mean of \(f(t)\), whereas a finite light curve directly constrains the
mean level over its sampled window. We therefore use the latent finite-window
magnitude
\begin{equation}
m_{\rm win}
\equiv
\frac{1}{T_{\rm rf}}
\int_{t_{\min}}^{t_{\max}} f(t)\,dt ,
\label{eq:mwin_background}
\end{equation}
as the apparent-magnitude quantity propagated into the Bayesian hierarchy. This is the brightness level directly constrained by the observed window and contemporaneous with the variability summary \(D\).

For the DRW covariance in eq.~\eqref{eq:drw_cov_background}, the finite-window
magnitude has ensemble variance
\begin{equation}
v_{\rm win}
\equiv
{\rm Var}(m_{\rm win}-m_0)
=
\frac{2\sigma^2}{T_{\rm rf}^2}
\left[
T_{\rm rf}\tau
-
\tau^2
\left(
1-e^{-T_{\rm rf}/\tau}
\right)
\right].
\label{eq:mwin_variance}
\end{equation}
The limiting behaviour is \(v_{\rm win}\simeq \sigma^2\) for \(T_{\rm rf}\ll\tau\), and \(v_{\rm win}\simeq 2\sigma^2\tau/T_{\rm rf}\) for \(T_{\rm rf}\gg\tau\). Thus a short light curve does not generally recover the stationary mean: when \(T_{\rm rf}\ll\tau\), the observed window remains highly correlated and \(m_{\rm win}\) can differ from \(m_0\) by an amount of order \(\sigma\). Only when \(T_{\rm rf}\gg\tau\) does averaging over many effectively independent fluctuations drive \(m_{\rm win}\) toward \(m_0\). This is another manifestation of the finite-baseline problem: short light curves need not sample the stationary process, and inferred long-timescale DRW parameters can depend on baseline length, priors, and unmodelled long-term trends
\citep{Kozlowski_2017,Kozlowski_2021,Stone_2023,Hu_2024}. Related AGN
variability--cosmology work in the long-baseline, multi-band regime can model
the continuum level more directly, using flexible GP light-curve models,
detrending terms, and spectroscopic flux calibration tied to the inferred
light-curve mean \citep{Dutra_2025}. In the moderate-baseline setting targeted here, we
instead use \(m_{\rm win}\), rather than \(m_0\), as the luminosity-relevant
apparent magnitude in the hierarchy, and its recovery is studied in section~\ref{sec:stage1_recovery_results}.

\subsection{From finite-window magnitude to luminosity under a trial cosmology}
\label{sec:mag_to_luminosity}

For source \(i\), let \(m^{(b)}_{{\rm win},i}\) denote the finite-window apparent magnitude in photometric band \(b\). We keep band-dependent quantities explicit where needed, but write \(m_i\equiv m^{(b)}_{{\rm win},i}\in\mathbb{R}\) to simplify the notation. Linking a luminosity coordinate \(\ell_i\in\mathbb{R}\) to \(m_i\) requires assuming a trial cosmology that determines the distance--redshift relation, and a spectral energy distribution (SED) and passband convention. For the baseline analysis presented in this work, we use a fixed SED template shape; this determines the same-band \(K\)-correction \(K_b:\mathbb{R}_{>0}\to\mathbb{R}\) and a band-dependent constant \(a_b\in\mathbb{R}\). Details are provided in appendix~\ref{app:bandpass_quantities}.

We use the standard apparent-absolute magnitude relation,
\begin{equation}
m_i
=
M_b(\ell_i)
+
\mu_{\rm DM}(z_i,\theta_{\rm cos})
+
K_b(z_i),
\label{eq:general_magnitude_relation}
\end{equation}
where \(M_b:\mathbb{R}\to\mathbb{R}\) maps the luminosity coordinate \(\ell_i\) to the band absolute magnitude, and
\begin{equation}
\mu_{\rm DM}(z,\theta_{\rm cos})
=
5\log_{10}
\left[
\frac{d_L(z,\theta_{\rm cos})}{{\rm Mpc}}
\right]
+
25
\label{eq:DM}
\end{equation}
is the distance modulus of the trial cosmology.

In the baseline implementation, the coordinate \(\ell\) is the template continuum-normalization proxy at 3000 \AA, defined before emission-line components are added; it should therefore be interpreted as a model luminosity coordinate, not as an independently measured monochromatic luminosity. For a fixed SED shape, varying \(\ell\) only rescales the template luminosity:
\begin{equation}
M_b(\ell)=a_b-2.5\,\ell ,
\end{equation}
where \(a_b\) absorbs the template normalization, passband, and photometric zero point. Hence
\begin{align}
m_i
&=
a_b
+
\mu_{\rm DM}(z_i,\theta_{\rm cos})
+
K_b(z_i)
-
2.5\,\ell_i ,
\label{eq:magnitude_luminosity}
\\
\ell_i(m_i,z_i,\theta_{\rm cos})
&=
\frac{
a_b
+
\mu_{\rm DM}(z_i,\theta_{\rm cos})
+
K_b(z_i)
-
m_i
}{2.5}.
\label{eq:luminosity_magnitude}
\end{align}
Thus \(\ell_i\) is the luminosity coordinate associated with the same finite observing window over which the short-lag variability coordinate is inferred. Cosmology enters this conversion only through \(\mu_{\rm DM}\), while SED and
passband assumptions enter through \(a_b\) and \(K_b(z)\). Alternative SED prescriptions can be incorporated by recomputing \(a_b\), \(K_b(z)\), and \(\lambda_{\rm rf}^{\rm eff}(z)\); the calibration consequences of template choice and source-specific spectra are discussed in section~\ref{sec:Discussion}.

\subsection{Empirical motivation for the variability--luminosity model}
\label{subsec:agn_variability_correlations}

Optical/UV AGN variability shows empirical trends with luminosity, rest-frame wavelength, timescale, and accretion state, although the reported coefficients vary with the adopted variability statistic, lag range, wavelength coverage, sample selection, and fitting method. We therefore use the literature to motivate the predictors in our population model, not to set informative priors on their coefficients. Disc-continuum and reprocessing models provide physical motivation for many of these dependencies and reproduce some observed trends, but do not yet yield complete first-principles predictions for optical/UV variability over the parameter range considered here \citep{ShakuraSunyaev_1973,Krolik_1991,Cackett_2007,Cackett_2021,KubotaDone_2018,Hagen_2024}. Therefore, our baseline variability--luminosity relation is phenomenological. Here \(\lambda_{\rm rf}\) denotes rest-frame wavelength, and all lags and
variability timescales in this subsection are rest-frame quantities unless stated otherwise.

Variability amplitude increases with rest-frame time lag. Over finite lag ranges, empirical SF studies often approximate this as \({\rm SF}(\Delta t)\propto \Delta t^{\gamma}\), but the effective slope is not universal: \citep{VandenBerk_2004} found
\(\gamma=0.246\pm0.008\) on day-to-year rest-frame timescales, whereas the DRW short-lag limit gives \({\rm SF}\propto\Delta t^{1/2}\) before
flattening near the damping timescale. The measured lag dependence therefore depends on cadence, noise treatment, wavelength, luminosity, and proximity to
the turnover
\citep{MacLeod_2010,MacLeod_2012,Kozlowski_2016,Suberlak_2021,Stone_2023}.
This motivates using a variability coordinate tied to a common rest-frame lag regime; which in our framework is the short-lag variability rate \(D\).

Variability also increases toward shorter rest-frame wavelengths
\citep[e.g.,][]{VandenBerk_2004,MacLeod_2010,MacLeod_2012,SanchezSaez_2018,Patel_2025}.
In DRW analyses this appears mainly through the asymptotic amplitude:
\citep{MacLeod_2010} found
\({\rm SF}_{\infty}\propto\lambda_{\rm rf}^{-0.48}\), with a weaker
\(\tau\propto\lambda_{\rm rf}^{0.17}\) dependence. Fixed-timescale variance measurements suggest that this chromatic dependence can steepen on shorter
timescales. Let \(\mathcal{V}_{\rm int}(T_{\rm var})\) denote the noise-corrected intrinsic variance associated with rest-frame variability timescale \(T_{\rm var}\). After correcting for dilution by spectral components, \citep{Patel_2025} found the slope of \(\log_{10}\mathcal{V}_{\rm int}\) with \(\log_{10}\lambda_{\rm rf}\) to steepen from about \(-1.3\) at \(T_{\rm var}=300\,{\rm d}\) to about \(-2.6\) at \(T_{\rm var}=30\,{\rm d}\). 
Broadband measurements must therefore account for both continuum chromaticity and dilution by weakly variable components; this issue is discussed further in section~\ref{sec:Discussion}.

This chromatic trend is qualitatively expected in disc and reprocessing pictures, where shorter wavelengths arise from hotter, more compact radii and longer wavelengths from larger radii and longer light-travel times
\citep{ShakuraSunyaev_1973,Krolik_1991,Cackett_2007,Cackett_2021}. For cosmological use, however, it is also a nuisance: a fixed observed band samples \(\lambda_{\rm rf}\simeq\lambda_{\rm obs}/(1+z)\), so apparent redshift trends in broadband variability can be induced by chromatic variability rather than by genuine cosmic evolution. Existing multi-parameter
analyses generally find little evidence for an independent redshift dependence once wavelength, luminosity, accretion state, or related physical properties are included \citep{MacLeod_2010,SanchezSaez_2018}. We therefore include an explicit effective-rest-frame-wavelength term in the population model, rather than absorbing this chromatic effect into an empirical redshift correction.

The central empirical trend for our framework is the optical/UV variability--luminosity anti-correlation: at fixed rest-frame lag and wavelength, more luminous quasars are less variable \citep{VandenBerk_2004,MacLeod_2010,MacLeod_2012,Suberlak_2021,Arevalo_2023,Goncalves_2025}. In DRW analyses, \citep{Suberlak_2021} found approximately \({\rm SF}_{\infty}\propto L^{-0.30}\),  while fixed-timescale variance measurements show a timescale-dependent slope. In the same \(\mathcal{V}_{\rm int}\) notation, \citep{Arevalo_2023}
found the slope of \(\log_{10}\mathcal{V}_{\rm int}\) with \(\log_{10}L_{\rm bol}\) to steepen from about \(-0.33\) at \(T_{\rm var}=300\,{\rm d}\) to about \(-1.01\) at \(T_{\rm var}=30\,{\rm d}\). This reinforces the need to compare variability at a common rest-frame timescale.

Black-hole mass (\(M_{\rm BH}\)) and accretion state introduce additional variability trends, but the reported dependencies are metric-dependent. Characteristic timescales generally increase with \(M_{\rm BH}\), while amplitude trends with \(M_{\rm BH}\) depend
on the timescale considered and on whether luminosity or Eddington ratio (\(\lambda_{\rm Edd}\)) is controlled for \citep{Kelly_2009,MacLeod_2010,Burke_2021,Suberlak_2021,Arevalo_2023}. A more robust empirical trend is that high-\(\lambda_{\rm Edd}\) systems are less variable: for instance, \citep{Suberlak_2021} found \({\rm SF}_{\infty}\propto\lambda_{\rm Edd}^{-0.21}\), and \citep{Goncalves_2025} found a persistent anti-correlation between fractional variability and Eddington ratio across redshift bins.

We do not include \(M_{\rm BH}\) or \(\lambda_{\rm Edd}\) as first-order predictors in the baseline model because the available single-epoch virial mass estimates and bolometric corrections carry substantial statistical and systematic uncertainties, particularly for high-redshift quasars \citep{Vestergaard_2006,ShenKelly_2010,Shen_2013,Runnoe_2012,Krawczyk_2013}. Residual object-to-object variation associated with black-hole mass, accretion state, and unmodelled variability physics is represented by an intrinsic-scatter term inferred jointly with the population relation. The baseline relation therefore connects the propagated short-lag variability coordinate to luminosity and effective rest-frame wavelength, the two predictors directly available in our photometric analysis.

\section{Statistical Model: Hierarchical Bayesian Framework}
\label{sec:statitsical_model}

The statistical model connects AGN light-curve variability to the intrinsic luminosity scale implied by a trial cosmology. We separate the discussion into three levels: the conceptual full generative construction, the conditional population model targeted in this work, and the two-stage approximation used for scalable inference. Unless stated otherwise, this section is written for a fixed photometric band. The band-dependent SED/passband quantities are those defined in section~\ref{sec:mag_to_luminosity} and appendix~\ref{app:bandpass_quantities}; the baseline Gaia implementation uses \(b=G\), and repeated band labels are suppressed.

\subsection{Notation}
\label{subsec:notation}

For each quasar \(i\in\{1,\ldots,N\}\), let \(n_i\in\mathbb{N}\) be the number of retained epochs. We collect the retained single-band light-curve tuple as
\begin{equation}
y_i =
\left(\mathbf{t}^{\rm obs}_i,\mathbf{m}^{\rm obs}_i,\mathbf{s}_i\right)
\in \mathcal{Y}_i
\equiv
\mathbb{R}^{n_i}\times\mathbb{R}^{n_i}\times\mathbb{R}^{n_i}_{>0}.
\end{equation}
Here \(\mathbf{t}^{\rm obs}_i\in\mathbb{R}^{n_i}\) contains the reported observer-frame epochs, \(\mathbf{m}^{\rm obs}_i\in\mathbb{R}^{n_i}\) the observed apparent magnitudes, and \(\mathbf{s}_i\in\mathbb{R}^{n_i}_{>0}\) the reported magnitude uncertainties.
After choosing an arbitrary observer-frame reference epoch \(t^{\rm obs}_{i0}\), typically the first retained observation, we define the rest-frame epochs
\begin{equation}
t_{ij} =
\frac{t^{\rm obs}_{ij}-t^{\rm obs}_{i0}}{1+z_i},
\qquad
\mathbf{t}_i=(t_{i1},\ldots,t_{in_i})^\top\in\mathbb{R}^{n_i}.
\end{equation}
The spectroscopic redshift \(z_i\in\mathbb{R}_{>0}\) is treated as known. Strictly, a GP light-curve likelihood is a density for \(\mathbf m_i^{\rm obs}\) conditional on the rest-frame epochs \(\mathbf t_i\), reported uncertainties \(\mathbf s_i\), and the redshift used to define the rest-frame times. In this work, whenever such a likelihood is written as
a function of \(y_i\), the epochs and reported uncertainties contained in \(y_i\) are understood as conditioning information.

The local light-curve summaries entering the global model are scalars:
\begin{equation}
m_i\equiv m_{{\rm win},i}\in\mathbb{R},
\qquad
\eta_i\equiv
\log_{10}\!\left[\frac{D_i}{\mathrm{mag}^2\,\mathrm{day}^{-1}}\right]
\in\mathbb{R},
\end{equation}
where \(m_{{\rm win},i}\) and \(D_i=\sigma_i^2/\tau_i\) are the chosen-band
finite-window magnitude and short-lag variability rate defined in Eqs.~\eqref{eq:mwin_background} and
\eqref{eq:drw_short_lag_background}. These are chosen-band quantities, with the band label suppressed after fixing \(b\).

The remaining local light-curve parameters are collected in the nuisance coordinate vector
\begin{equation}
\bm{\nu}_i =
\left(
\frac{m_{0,i}}{\rm mag},
\ln\frac{\tau_i}{\mathrm{day}},
\ln\frac{\sigma_{{\rm add},i}}{\mathrm{mag}}
\right)^\top
\in\mathbb{R}^3.
\label{eq:local_nuisance_definition}
\end{equation}
Here \(m_{0,i}\in\mathbb{R}\), \(\tau_i\in\mathbb{R}_{>0}\), and \(\sigma_{{\rm add},i}\in\mathbb{R}_{>0}\) are the stationary DRW mean magnitude, damping time, and additional white-noise term for possible underestimation of the reported \(s_{ij}\). The DRW variance is then
\(\sigma_i^2=D_i\tau_i\).

The main global parameters of our variability--luminosity model are
\begin{equation}
\bm{\theta} = (\theta_{\rm cos},\theta_{\rm VL},\sigma_{\rm int})
\in \Theta_{\rm cos}\times\Theta_{\rm VL}\times\mathbb{R}_{>0},
\qquad
\theta_{\rm VL}=(\eta_{D,0},d_D,\alpha_D)\in\mathbb{R}^3 .
\label{eq:theta_definition}
\end{equation}
Here \(\theta_{\rm cos}\) denotes the sampled cosmological parameters. The variability--luminosity parameters \(\theta_{\rm VL}\) are dimensionless: \(\eta_{D,0}\) is the normalization of the \(\eta\) coordinate, and \(d_D\) and \(\alpha_D\) are the luminosity and rest-frame-wavelength slopes. The intrinsic scatter \(\sigma_{\rm int}\) is quoted in dex. The choice of predictors is motivated in section~\ref{subsec:agn_variability_correlations}. The conceptual forward model also contains luminosity-function, nuisance-population, and selection-function parameters, denoted by \(\bm{\psi}_{\Phi}\), \(\bm{\psi}_{\nu}\), and
\(\bm{\psi}_{S}\), respectively. For reference, the main recurring symbols used in the hierarchical model are
summarized in table~\ref{tab:notation_summary} in appendix~\ref{app:notation_summary}.

\subsection{Population relation}
\label{subsec:population_relation}

A fully generative point-process model would specify the redshift distribution, survey volume, object counts, and redshift-dependent selection. Here we condition instead on the SDSS DR16Q \citep{Lyke_2020} systemic redshifts of \citep{WuShen_2022} and neglect their uncertainties; for the catalogue entries used here, these uncertainties are small, with the vast majority below \(500\,{\rm km\,s^{-1}}\), or
\(\Delta z/(1+z)\lesssim1.7\times10^{-3}\).

At fixed \(z_i\), the full-forward model draws the intrinsic luminosity proxy from a quasar luminosity function,
\begin{equation}
\ell_i\mid z_i,\bm{\psi}_{\Phi}
\sim
p_{\bm{\psi}_{\Phi}}(\ell_i\mid z_i).
\end{equation}
The corresponding finite-window apparent magnitude
\(m_i\) is then set by the cosmology and bandpass model through
eq.~\eqref{eq:magnitude_luminosity}. The variability coordinate is drawn from an intrinsic population relation at fixed luminosity proxy and effective rest-frame wavelength,
\begin{equation}
\eta_i
\mid
\ell_i,z_i,\theta_{\rm VL},\sigma_{\rm int}
\sim
p_{\theta_{\rm VL},\sigma_{\rm int}}^{\rm pop}
\left[
\eta_i\mid \ell_i,\lambda_{\rm rf}^{\rm eff}(z_i)
\right],
\label{eq:full_forward_pop_relation}
\end{equation}
where \(\lambda_{\rm rf}^{\rm eff}(z_i)\) denotes the chosen-band effective rest-frame wavelength; its band-dependent construction is given in eq.~\eqref{eq:app_lambda_rf_eff}. Since \(\ell_i\) is explicit in this full-forward factorization, the population factor is parametrized by \(\theta_{\rm VL}\) and \(\sigma_{\rm int}\); \(\theta_{\rm cos}\) enters
through the mapping \(m_i=m_b(\ell_i,z_i;\theta_{\rm cos})\) in eq.~\eqref{eq:magnitude_luminosity} and through any apparent-magnitude selection.

The remaining local light-curve nuisance parameters are drawn from a population distribution conditional on \((\ell_i,z_i)\) and parameterized by \(\bm{\psi}_{\nu}\),
\begin{equation}
\bm{\nu}_i
\mid
\ell_i,z_i,\bm{\psi}_{\nu}
\sim
\pi^{\rm full}_{\bm{\psi}_{\nu}}(\bm{\nu}_i\mid \ell_i,z_i).
\label{eq:full_forward_nuisance_model}
\end{equation}

Given \((m_i,\eta_i,\bm{\nu}_i,z_i)\), the local DRW model defines the light-curve likelihood conditional on the finite-window magnitude. Equivalently,
\begin{equation}
p_{\rm LC}^{\rm win}
\left(
y_i\mid m_i,\eta_i,\bm{\nu}_i,z_i
\right)
\equiv
\frac{
p_{\rm GP}
\left(
y_i,m_i\mid\eta_i,\bm{\nu}_i,z_i
\right)
}{
p
\left(
m_i\mid\eta_i,\bm{\nu}_i,z_i
\right)
}.
\label{eq:gp_conditional_window_likelihood}
\end{equation}
Here \(p_{\rm GP}\) denotes the joint Gaussian density of \((\mathbf m_i^{\rm obs},m_i)\) implied by the local DRW model over the same
rest-frame window used to define \(m_i\). The denominator is the corresponding marginal density of the latent window average. Since \(m_i\) is a linear
functional of the latent GP, this marginal is Gaussian:
\begin{equation}
p(m_i\mid\eta_i,\bm{\nu}_i,z_i)
=
\mathcal N
\left(
m_i\mid m_{0,i},v_{{\rm win},i}
\right),
\end{equation}
where \(v_{{\rm win},i}\) is the finite-window variance from eq.~\eqref{eq:mwin_variance}, evaluated for the object-specific rest-frame
window \(T_{{\rm rf},i}\).

The joint density is
\begin{equation}
\begin{pmatrix}
\mathbf m_i^{\rm obs} \\
m_i
\end{pmatrix}
\Bigg|
\eta_i,\bm{\nu}_i,z_i
\sim
\mathcal N
\left[
\begin{pmatrix}
m_{0,i}\mathbf 1 \\
m_{0,i}
\end{pmatrix},
\begin{pmatrix}
\mathbf C_i & \mathbf c_{{\rm win},i} \\
\mathbf c_{{\rm win},i}^{\top} & v_{{\rm win},i}
\end{pmatrix}
\right],
\label{eq:gp_joint_gaussian}
\end{equation}
where \(\mathbf C_i\) is the observed-magnitude covariance matrix defined in eq.~\eqref{eq:drwkernel}. The vector \(\mathbf c_{{\rm win},i}\) contains the covariances between
\(m_i\) and the latent process at the observed epochs,
\begin{equation}
\left(\mathbf c_{{\rm win},i}\right)_j
\equiv
c_{{\rm win},ij}
=
\operatorname{Cov}
\left[
m_i,
f_i(t_{ij})
\right]
=
\frac{1}{T_{{\rm rf},i}}
\int_{a_i}^{b_i}
k_i(t,t_{ij})\,{\rm d}t .
\label{eq:cwin_def_background}
\end{equation}
For the DRW kernel, if \(t_{ij}\in[a_i,b_i]\), with \(a_i\) and \(b_i\) the rest-frame window limits and \(T_{{\rm rf},i}\equiv b_i-a_i\), this becomes
\begin{equation}
c_{{\rm win},ij}
=
\frac{\sigma_i^2\tau_i}{T_{{\rm rf},i}}
\left[
2
-
\exp\!\left(
-\frac{t_{ij}-a_i}{\tau_i}
\right)
-
\exp\!\left(
-\frac{b_i-t_{ij}}{\tau_i}
\right)
\right].
\label{eq:cwin_drw_background}
\end{equation}
Thus \(p_{\rm LC}^{\rm win}\) is the GP likelihood of the observed light curve conditioned on the latent window average being \(m_i\). The superscript ``win'' emphasizes conditioning on the finite-window magnitude rather than on the stationary DRW mean \(m_{0,i}\).

Let \(F_i=1\) denote final inclusion in the fitted sample in the full-forward construction. The corresponding selection probability may depend on the realized light curve, apparent magnitude, redshift, cadence, variability, and quality information; schematically,
\begin{equation}
\mathcal{S}^{\rm full}_{\bm{\psi}_S}(y_i,m_i,\eta_i,\bm{\nu}_i,z_i)
\equiv
p(F_i=1\mid y_i,m_i,\eta_i,\bm{\nu}_i,z_i,\bm{\psi}_S).
\end{equation}

Conditioning on the observed redshifts, define the full-forward integrand
\(\mathcal J_i\), with \(z_i\) and global parameters implicit, as
\begin{align}
\mathcal J_i(y,m,\eta,\ell,\bm{\nu})
&\equiv
p_{\rm LC}^{\rm win}
\left(
y\mid m,\eta,\bm{\nu},z_i
\right)
\mathcal{S}^{\rm full}_{\bm{\psi}_S}(y,m,\eta,\bm{\nu},z_i)
\nonumber\\
&\quad\times
p_{\theta_{\rm VL},\sigma_{\rm int}}^{\rm pop}
\left[
\eta\mid \ell,\lambda_{\rm rf}^{\rm eff}(z_i)
\right]
p_{\bm{\psi}_{\Phi}}(\ell\mid z_i)
\nonumber\\
&\quad\times
\pi^{\rm full}_{\bm{\psi}_{\nu}}(\bm{\nu}\mid \ell,z_i)
\delta
\left[
m-m_b(\ell,z_i,\theta_{\rm cos})
\right].
\label{eq:full_forward_integrand}
\end{align}
The selected-sample likelihood can then be written schematically as
\begin{equation}
\mathcal L_{\rm full}
\propto
\prod_{i=1}^{N}
\frac{\mathcal N_i}
{\mathcal A_i(z_i;\bm{\theta},
\bm{\psi}_{\Phi},\bm{\psi}_{\nu},\bm{\psi}_S)},
\qquad
\mathcal N_i
=
\int
\mathcal J_i(y_i,m,\eta,\ell,\bm{\nu})
\,d m\,d\eta\,d\ell\,d\bm{\nu},
\label{eq:full_forward_likelihood}
\end{equation}
\begin{equation}
\mathcal A_i(z_i;\bm{\theta},
\bm{\psi}_{\Phi},\bm{\psi}_{\nu},\bm{\psi}_S)
=
\int
\mathcal J_i(\tilde y,m,\eta,\ell,\bm{\nu})
\,d\tilde y\,d m\,d\eta\,d\ell\,d\bm{\nu}.
\label{eq:full_forward_selection_norm}
\end{equation}
The full generative model serves as the conceptual basis, but a direct implementation would require a luminosity-function model \(p_{\bm{\psi}_{\Phi}}\), a full-forward selection function \(\mathcal S^{\rm full}_{\bm{\psi}_S}\) matched to the parent population and analysis cuts, and a nuisance-population model \(\pi^{\rm full}_{\bm{\psi}_{\nu}}\). This is non-trivial because quasar luminosity functions depend on selection wavelength, obscuration, SED and
bolometric corrections, and faint-end or high-redshift extrapolations \citep[e.g.][]{Aird_2015,Kulkarni_2019,Shen_2020}. We therefore do not impose a separate luminosity-function prior in the baseline analysis, and instead target the conditional relation \(p_{\bm{\theta}}(\eta_i\mid m_i,z_i)\).

\begin{figure}[t]
\centering
\resizebox{\textwidth}{!}{%
\begin{tikzpicture}[
    x=1cm, y=1cm,
    >=Latex,
    every node/.style={font=\normalsize},
    obs/.style={
        draw=black!70, circle, fill=obsfill,
        minimum size=0.78cm, align=center, inner sep=0pt
    },
    latent/.style={
        draw=black!70, circle, fill=white,
        minimum size=0.78cm, align=center, inner sep=0pt
    },
    det/.style={
        draw=black!70, circle, double, double distance=0.8pt,
        fill=white, minimum size=0.78cm, align=center, inner sep=0pt
    },
    hyper/.style={
        draw=black!70, circle, fill=white,
        minimum size=0.78cm, align=center, inner sep=0pt
    },
    selectnode/.style={
        draw=black!70, rectangle, rounded corners=2pt,
        fill=obsfill, minimum width=0.90cm, minimum height=0.55cm,
        align=center, inner sep=2pt
    },
    dimobs/.style={
        draw=black!15, circle, fill=black!4, text=black!22,
        minimum size=0.78cm, align=center, inner sep=0pt
    },
    dimlatent/.style={
        draw=black!15, circle, fill=white, text=black!22,
        minimum size=0.78cm, align=center, inner sep=0pt
    },
    dimdet/.style={
        draw=black!15, circle, double, double distance=0.8pt,
        fill=white, text=black!22,
        minimum size=0.78cm, align=center, inner sep=0pt
    },
    dimhyper/.style={
        draw=black!15, circle, fill=white, text=black!22,
        minimum size=0.78cm, align=center, inner sep=0pt
    },
    dimselect/.style={
        draw=black!15, rectangle, rounded corners=2pt,
        fill=black!4, text=black!22,
        minimum width=0.90cm, minimum height=0.55cm,
        align=center, inner sep=2pt
    },
    edge/.style={
        ->, draw=black!72, line width=0.72pt,
        shorten >=1.5pt, shorten <=1.5pt
    },
    fulledge/.style={
        ->, draw=black!34, line width=0.65pt,
        shorten >=1.5pt, shorten <=1.5pt
    },
    dimedge/.style={
        ->, draw=black!12, line width=0.65pt,
        shorten >=1.5pt, shorten <=1.5pt
    },
    infer1/.style={
        ->, draw=stageoneorange, densely dashed, line width=1.05pt,
        shorten >=1.5pt, shorten <=1.5pt
    },
    infer2/.style={
        ->, draw=stagetwoblue, densely dashed, line width=1.05pt,
        shorten >=1.5pt, shorten <=1.5pt
    },
    plate/.style={
        draw=black!25, dotted, rounded corners=4pt,
        line width=0.70pt
    },
    stageboxone/.style={
        draw=stageoneorange, densely dashed, rounded corners=4pt,
        line width=1.05pt
    },
    stageboxtwo/.style={
        draw=stagetwoblue, densely dashed, rounded corners=4pt,
        line width=1.05pt
    },
    annoboxone/.style={
        draw=stageoneorange!75, rounded corners=4pt,
        fill=stageoneorange!6, line width=0.70pt
    },
    annoboxtwo/.style={
        draw=stagetwoblue!75, rounded corners=4pt,
        fill=stagetwoblue!6, line width=0.70pt
    },
    paneltitle/.style={
        font=\small\bfseries, align=center
    },
    panelsub/.style={
        font=\scriptsize, align=center
    }
]

\begin{scope}[shift={(0,0)}]
\node[paneltitle] at (2.95,7.20) {(a) Conceptual model};

\node[hyper] at (0.55,5.95) (aPsiPhi) {$\psi_\Phi$};
\node[hyper] at (1.75,5.95) (aThetaC) {$\theta_{\rm cos}$};
\node[hyper] at (3.05,5.95) (aThetaVL) {$\theta_{\rm VL}$};
\node[hyper] at (4.25,5.95) (aSigma) {$\sigma_{\rm int}$};
\node[hyper] at (5.45,5.95) (aPsiS) {$\psi_S$};

\draw[plate] (0.05,0.25) rectangle (5.95,5.15);
\node[anchor=south east, font=\scriptsize] at (5.82,0.33) {$i=1,\ldots,N$};

\node[obs]    at (0.75,4.45) (az) {$z_i$};
\node[obs]    at (2.05,4.45) (ased) {$\mathcal B_i$};
\node[det]    at (3.35,4.45) (alam) {$\lambda_{\rm rf, i}^{\rm eff}$};

\node[latent] at (0.85,2.85) (aL) {$\ell_i$};
\node[latent] at (2.50,2.85) (am) {$m_i$};
\node[latent] at (4.20,2.85) (aD) {$\eta_i$};

\node[obs]    at (3.15,1.10) (ay) {$y_i$};
\node[selectnode] at (5.15,1.10) (aI) {$F_i$};

\draw[fulledge] (aPsiPhi) to[out=-105,in=120] (aL);
\draw[edge] (aThetaC.south) .. controls (1.20,5.10) and (1.70,3.65) .. (am.125);
\draw[edge] (aThetaVL.south) .. controls (4.10,5.35) and (4.55,4.15) .. (aD.90);
\draw[edge] (aSigma) to[out=-90,in=65] (aD);

\draw[edge] (az) -- (aL);
\draw[edge] (az) to[out=-55,in=165] (am);
\draw[edge] (az) to[out=62,in=123] (alam);
\draw[edge] (ased) -- (alam);
\draw[edge] (ased) to[out=-85,in=100] (am);

\draw[edge] (aL) -- (am);
\draw[edge] (aL) to[out=52,in=128] (aD);
\draw[edge] (alam) -- (aD);

\draw[edge] (am) to[out=-75,in=145] (ay);
\draw[edge] (aD) to[out=-110,in=35] (ay);

\draw[fulledge] (aPsiS) to[out=-90,in=70] (aI);
\draw[fulledge] (az.east) to[bend left=5] (aI.north west);
\draw[fulledge] (am) to[out=-15,in=155] (aI);
\draw[fulledge] (aD) to[out=-20,in=125] (aI);
\draw[fulledge] (ay) -- (aI);

\node[panelsub] at (2.95,-0.38)
{Full forward dependencies};
\end{scope}

\begin{scope}[shift={(6.65,0)}]
\node[paneltitle] at (2.95,7.20) {(b) Stage--1 light-curve fits};

\node[dimhyper] at (0.55,5.95) (bPsiPhi) {$\psi_\Phi$};
\node[dimhyper] at (1.75,5.95) (bThetaC) {$\theta_{\rm cos}$};
\node[dimhyper] at (3.05,5.95) (bThetaVL) {$\theta_{\rm VL}$};
\node[dimhyper] at (4.25,5.95) (bSigma) {$\sigma_{\rm int}$};
\node[dimhyper] at (5.45,5.95) (bPsiS) {$\psi_S$};

\draw[plate] (0.05,0.25) rectangle (5.95,5.15);
\node[anchor=south east, font=\scriptsize, text=black!35] at (5.82,0.33) {$i=1,\ldots,N$};

\node[dimobs]    at (0.75,4.45) (bz) {$z_i$};
\node[dimobs]    at (2.05,4.45) (bsed) {$\mathcal B_i$};
\node[dimdet]    at (3.35,4.45) (blam) {$\lambda_{\rm rf, i}^{\rm eff}$};

\node[dimlatent] at (0.85,2.85) (bL) {$\ell_i$};
\node[latent, draw=stageoneorange, line width=0.95pt] at (2.50,2.85) (bm) {$m_i$};
\node[latent, draw=stageoneorange, line width=0.95pt] at (4.20,2.85) (bD) {$\eta_i$};

\node[obs]       at (3.15,1.10) (by) {$y_i$};
\node[dimselect] at (5.15,1.10) (bI) {$F_i$};

\draw[dimedge] (bThetaC.south) .. controls (1.20,5.10) and (1.70,3.65) .. (bm.125);
\draw[dimedge] (bThetaVL.south) .. controls (4.10,5.35) and (4.55,4.15) .. (bD.90);
\draw[dimedge] (bSigma) to[out=-90,in=65] (bD);
\draw[dimedge] (bz) -- (bL);
\draw[dimedge] (bz) to[out=-55,in=165] (bm);
\draw[dimedge] (bz) to[out=62,in=123] (blam);
\draw[dimedge] (bsed) -- (blam);
\draw[dimedge] (bsed) to[out=-85,in=100] (bm);
\draw[dimedge] (bL) -- (bm);
\draw[dimedge] (bL) to[out=52,in=128] (bD);
\draw[dimedge] (blam) -- (bD);
\draw[dimedge] (bm) to[out=-75,in=145] (by);
\draw[dimedge] (bD) to[out=-110,in=35] (by);

\draw[stageboxone] (1.82,2.39) rectangle (4.88,3.31);
\node[anchor=south west, font=\scriptsize\bfseries, text=stageoneorange]
    at (1.90,3.38) {Stage--1 proposal};

\draw[infer1] (by) to[out=120,in=-75] (bm);
\draw[infer1] (by) to[out=60,in=-105] (bD);

\node[annoboxone, minimum width=5.35cm, minimum height=0.90cm] at (2.95,-0.42) (banno) {};
\begin{scope}[shift={(0.92,-0.42)}]
    \draw[black!35, line width=0.25pt] (-0.45,-0.20) -- (-0.45,0.20);
    \draw[black!35, line width=0.25pt] (-0.45,-0.20) -- (0.55,-0.20);
    \draw[stageoneorange, line width=0.65pt]
        plot[smooth] coordinates {
        (-0.35,-0.04) (-0.22,0.08) (-0.08,-0.02) (0.08,0.12)
        (0.23,-0.06) (0.40,0.03) (0.52,-0.08)
        };
    \foreach \x/\y in {-0.35/-0.04,-0.22/0.08,-0.08/-0.02,0.08/0.12,0.23/-0.06,0.40/0.03,0.52/-0.08}{
        \fill[stageoneorange] (\x,\y) circle (0.018);
    }
\end{scope}
\node[anchor=west, font=\scriptsize, align=left] at (1.65,-0.42)
{Fit each light curve, export\\paired draws \((m_i^{(r)},\eta_i^{(r)})\)};
\end{scope}

\begin{scope}[shift={(13.30,0)}]
\node[paneltitle] at (2.95,7.20) {(c) Stage--2 population reweighting};

\node[dimhyper] at (0.55,5.95) (cPsiPhi) {$\psi_\Phi$};
\node[hyper, draw=stagetwoblue, line width=0.95pt] at (1.75,5.95) (cThetaC) {$\theta_{\rm cos}$};
\node[hyper, draw=stagetwoblue, line width=0.95pt] at (3.05,5.95) (cThetaVL) {$\theta_{\rm VL}$};
\node[hyper, draw=stagetwoblue, line width=0.95pt] at (4.25,5.95) (cSigma) {$\sigma_{\rm int}$};
\node[dimhyper] at (5.45,5.95) (cPsiS) {$\psi_S$};

\draw[plate] (0.05,0.25) rectangle (5.95,5.15);
\node[anchor=south east, font=\scriptsize, text=black!35] at (5.82,0.33) {$i=1,\ldots,N$};

\node[obs]       at (0.75,4.45) (cz) {$z_i$};
\node[obs]       at (2.05,4.45) (csed) {$\mathcal B_i$};
\node[det]       at (3.35,4.45) (clam) {$\lambda_{\rm rf, i}^{\rm eff}$};

\node[latent]    at (0.85,2.85) (cL) {$\ell_i$};
\node[latent, draw=stagetwoblue, line width=0.95pt] at (2.50,2.85) (cm) {$m_i$};
\node[latent, draw=stagetwoblue, line width=0.95pt] at (4.20,2.85) (cD) {$\eta_i$};

\node[dimobs]    at (3.15,1.10) (cy) {$y_i$};
\node[dimselect] at (5.15,1.10) (cI) {$F_i$};

\draw[edge] (cThetaC.south) .. controls (1.20,5.10) and (1.70,3.65) .. (cm.125);
\draw[edge] (cThetaVL.south) .. controls (4.10,5.35) and (4.55,4.15) .. (cD.90);
\draw[edge] (cSigma) to[out=-90,in=65] (cD);
\draw[edge] (cz) -- (cL);
\draw[edge] (cz) to[out=-55,in=165] (cm);
\draw[edge] (cz) to[out=62,in=123] (clam);
\draw[edge] (csed) -- (clam);
\draw[edge] (csed) to[out=-85,in=100] (cm);
\draw[edge] (cL) -- (cm);
\draw[edge] (cL) to[out=52,in=128] (cD);
\draw[edge] (clam) -- (cD);
\draw[dimedge] (cm) to[out=-75,in=145] (cy);
\draw[dimedge] (cD) to[out=-110,in=35] (cy);

\draw[stageboxtwo] (1.05,5.48) rectangle (4.95,6.35);
\node[anchor=south west, font=\scriptsize\bfseries, text=stagetwoblue]
    at (1.15,6.42) {Stage--2 global inference};

\draw[infer2] (cm.60) .. controls +(0.55,1.00) and +(0.35,-0.70) .. (cThetaC.330);
\draw[infer2] (cD.120) .. controls +(0.55,1.05) and +(0.45,-0.70) .. (cThetaVL.315);
\draw[infer2] (cD) to[out=70,in=-95] (cSigma);
\draw[infer2] (cm.215) to[out=220,in=-40] (cL.320);

\node[annoboxtwo, minimum width=5.35cm, minimum height=0.90cm] at (2.95,-0.42) (canno) {};
\begin{scope}[shift={(1.00,-0.42)}]
    \draw[black!35, line width=0.25pt] (-0.45,-0.22) -- (-0.45,0.22);
    \draw[black!35, line width=0.25pt] (-0.45,-0.22) -- (0.55,-0.22);
    \draw[stagetwoblue, line width=0.55pt] (-0.35,0.13) -- (0.50,-0.12);
    \foreach \x/\y in {-0.35/0.10,-0.24/0.06,-0.13/0.04,0.00/-0.02,0.12/-0.01,0.25/-0.08,0.40/-0.10,0.50/-0.15}{
        \fill[stagetwoblue] (\x,\y) circle (0.017);
    }
\end{scope}
\node[anchor=west, font=\scriptsize, align=left] at (1.85,-0.42)
{Reweight posterior draws,\\infer (\(\theta_{\rm cos},\theta_{\rm VL},\sigma_{\rm int}\))};
\end{scope}

\node[
    draw=black!18,
    rounded corners=7pt,
    fill=black!4,
    minimum width=18.90cm,
    minimum height=1.25cm
] at (9.62,-1.82) {};

\node[obs, minimum size=0.30cm] at (1.15,-1.56) {};
\node[latent, minimum size=0.30cm] at (3.35,-1.56) {};
\node[det, minimum size=0.30cm] at (5.55,-1.56) {};
\node[hyper, minimum size=0.30cm] at (7.75,-1.56) {};

\draw[edge] (9.55,-1.56) -- (10.10,-1.56);
\draw[fulledge] (12.15,-1.56) -- (12.70,-1.56);
\draw[infer1] (14.95,-1.56) -- (15.50,-1.56);
\draw[infer2] (17.15,-1.56) -- (17.70,-1.56);

\node[font=\scriptsize, align=center] at (1.15,-2.05) {Observed};
\node[font=\scriptsize, align=center] at (3.35,-2.05) {Latent};
\node[font=\scriptsize, align=center] at (5.55,-2.05) {Deterministic};
\node[font=\scriptsize, align=center] at (7.75,-2.05) {Global\\Parameter};

\node[font=\scriptsize, align=center] at (9.82,-2.05) {Generative};
\node[font=\scriptsize, align=center] at (12.42,-2.05) {Full Forward\\Dependency};
\node[font=\scriptsize, align=center, text=stageoneorange] at (15.22,-2.05) {Stage--1};
\node[font=\scriptsize, align=center, text=stagetwoblue] at (17.42,-2.05) {Stage--2};

\end{tikzpicture}%
}
\caption{
Probabilistic graphical model for the AGN variability--luminosity hierarchy and two-stage inference scheme. Panel (a) shows the conceptual full-forward model, where \(\mathcal B_i\) denotes the SED/passband information. The local nuisance vector \(\bm{\nu}_i\) and its hyperparameters \(\bm{\psi}_\nu\) are suppressed for readability, but enter the full-forward density and Stage--1 proposal as described in the text. Grey arrows mark full-forward dependencies that are omitted from the conditional target; in section~\ref{subsec:method_conditional_model}, \(F_i=1\) is treated by conditioning on parent-sample membership \(E_i=1\) and modelling retention after analysis cuts as \(I_i=1\). Panel (b) shows independent Stage--1 light-curve fits producing paired draws of \((m_i,\eta_i)\), with \(\bm{\nu}_i\) retained in the proposal but not displayed. Panel (c) shows Stage--2 importance reweighting, which infers \(\bm{\theta}=(\theta_{\rm cos},\theta_{\rm VL},\sigma_{\rm int})\) under the conditional population model.
}
\label{fig:pgm}
\end{figure}
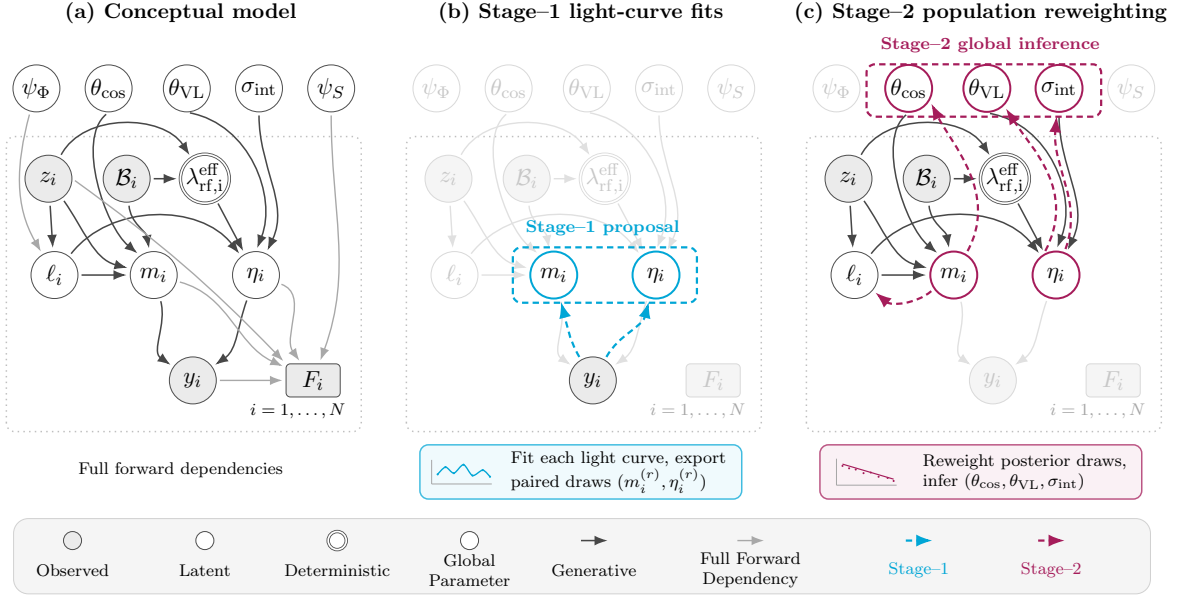

Figure~\ref{fig:pgm} summarizes the main full-forward dependencies and the two-stage approximation used for scalable inference described in section~\ref{subsec:inference}.

\subsection{Conditional model}
\label{subsec:method_conditional_model}

The baseline analysis replaces the full-forward population model with the
conditional variability distribution at fixed finite-window apparent magnitude
and redshift.  We define
\begin{equation}
p_{\bm{\theta}}\!\left(\eta_i\mid m_i,z_i\right)
\equiv
p_{\theta_{\rm VL}, \sigma_{\rm int}}^{\rm pop}
\left[
\eta_i
\mid
\ell_i(m_i,z_i,\theta_{\rm cos}),
\lambda_{\rm rf}^{\rm eff}(z_i)
\right],
\label{eq:conditional_induced_density}
\end{equation}
where \(\ell_i(m_i,z_i,\theta_{\rm cos})\) is the luminosity implied by the
observed finite-window magnitude under the trial cosmology, as given by eq.~\eqref{eq:luminosity_magnitude}.  Thus the conditional density depends on
\(\theta_{\rm VL}\) and \(\sigma_{\rm int}\) through the intrinsic variability--luminosity relation, and
on \(\theta_{\rm cos}\) through the magnitude--luminosity conversion.

Let \(\ell_{\rm piv}\) and \(\lambda_{\rm piv}\) be fixed pivots chosen to reduce posterior covariance. We model
\begin{equation}
\eta_i
\mid
m_i,z_i,\bm{\theta}
\sim
\mathcal N
\left(
\mu_{\eta,i}(\bm{\theta}),
\sigma_{\rm int}^2
\right),
\label{eq:conditional_D_model}
\end{equation}
\begin{equation}
\mu_{\eta,i}(\bm{\theta})
=
\eta_{D,0}
+
d_D
\left[
\ell_i(\theta_{\rm cos})-\ell_{\rm piv}
\right]
-
\alpha_D
\log_{10}
\left[
\frac{\lambda_{\rm rf}^{\rm eff}(z_i)}{\lambda_{\rm piv}}
\right],
\label{eq:D_luminosity_relation}
\end{equation}
where \(\ell_i(\theta_{\rm cos})\equiv
\ell_i(m_i,z_i,\theta_{\rm cos})\).

The conditional formulation also makes explicit how the full-forward fitted-sample event is treated in the Gaia DR3--SDSS implementation. We decompose \(F_i=1\) into two steps. First, \(E_i=1\) denotes membership in the outer epoch-photometry parent sample, consisting of SDSS DR16Q Type~1 AGN with Gaia \(G\)-band epoch photometry. The variability--luminosity relation calibrated below should therefore be read as conditional on this parent sample. Second, within this parent sample, \(I_i=1\) denotes retention after the additional analysis cuts. We let \(c_i\in\mathcal C_i\) collect observed sample-definition information, such as cadence, epoch count, photometric precision, sky coverage, and quality flags.

Within the \(E_i=1\) parent sample, we assume that the retained-analysis cuts do not depend directly on the latent variability coordinate once \((m_i,z_i,c_i)\) are fixed, and that \(c_i\) is not an additional astrophysical predictor of the intrinsic population relation:
\begin{align}
p(I_i=1\mid \eta_i,m_i,z_i,c_i,E_i=1)
&=
p(I_i=1\mid m_i,z_i,c_i,E_i=1),
\label{eq:selection_conditional_independence}
\\
p_{\bm{\theta}}(\eta_i\mid m_i,z_i,c_i,E_i=1)
&=
p_{\bm{\theta}}(\eta_i\mid m_i,z_i,E_i=1).
\label{eq:c_population_assumption}
\end{align}
These are sample-definition assumptions, not automatic consequences of conditioning. They are plausible when \(c_i\) describes data availability and light-curve quality, which affect usability and uncertainty, rather than astrophysical variability at fixed \((m_i,z_i)\). They would fail for variability-selected samples, cuts on recovered \(D_i\), variability signal-to-noise, or quality variables that trace an unmodelled physical
subpopulation. We therefore treat
Eqs.~\eqref{eq:selection_conditional_independence}--\eqref{eq:c_population_assumption} as working assumptions whose practical consequences are checked with the sample-definition and closure diagnostics in
sections ~\ref{subsec:real_data_quality_cuts} and
\ref{sec:full_closure_results}.

Under Eqs.~\eqref{eq:selection_conditional_independence}--\eqref{eq:c_population_assumption},
\begin{align}
p(\eta_i\mid m_i,z_i,c_i,I_i=1,E_i=1,\bm{\theta})
&\propto
p(I_i=1\mid \eta_i,m_i,z_i,c_i,E_i=1)
p_{\bm{\theta}}(\eta_i\mid m_i,z_i,c_i,E_i=1)
\nonumber\\
&=
p(I_i=1\mid m_i,z_i,c_i,E_i=1)
p_{\bm{\theta}}(\eta_i\mid m_i,z_i,E_i=1).
\end{align}
The first factor is independent of \(\eta_i\) and cancels in the normalization over \(\eta_i\), giving
\begin{equation}
p(\eta_i\mid m_i,z_i,c_i,I_i=1,E_i=1,\bm{\theta})
=
p_{\bm{\theta}}(\eta_i\mid m_i,z_i,E_i=1).
\end{equation}
In the remainder of the paper we write this parent-sample conditional density simply as \(p_{\bm{\theta}}(\eta_i\mid m_i,z_i)\).
Thus the additional retained-analysis selection factor for \(I_i=1\) cancels from the conditional likelihood. Apparent-magnitude, redshift, cadence, and quality cuts can affect which objects enter the analysis, but they do not require a Malmquist-type normalization in the conditional likelihood provided they do not select on \(\eta_i\) at fixed \((m_i,z_i,c_i,E_i=1)\) \citep{Kelly_2007,March_2018,Mantz_2019,Mandel_2019}. This is the relevant condition for the Gaia DR3--SDSS implementation: the outer epoch-photometry parent sample is conditioned on, while the retained-sample cuts are chosen to control data quality and posterior support. Section~\ref{subsec:real_data_quality_cuts} documents the retained-cut design, and Section~\ref{sec:full_closure_results} tests the two-stage recovery under Gaia-like cadence, noise, and the adopted cuts.

The exact target for the implemented conditional hierarchy is
\begin{equation}
\mathcal{L}^{\rm exact}_{\rm cond,\pi_{\nu}}(\bm{\theta})
=
\prod_{i=1}^{N}
\int
p_{\rm GP}
\left(
y_i,m_i\mid\eta_i,\bm{\nu}_i,z_i
\right)
p_{\bm{\theta}}
\left(
\eta_i\mid m_i,z_i
\right)
\pi_\nu(\bm{\nu}_i)
\,d m_i\,d\eta_i\,d\bm{\nu}_i .
\label{eq:exact_conditional_likelihood}
\end{equation}
The subscript \(\pi_\nu\) indicates that this target is defined with respect to a fixed nuisance reference measure, which is independent of cosmology. The only \(\bm{\theta}\)-dependent factor inside the integral is \(p_\theta(\eta_i\mid m_i,z_i)\), through the variability--luminosity and intrinsic scatter parameters and the cosmology-dependent mapping from \((m_i,z_i)\) to luminosity.  Thus eq.~\eqref{eq:exact_conditional_likelihood} uses the GP factor as the local light-curve measure and \(p_{\bm{\theta}}(\eta_i\mid m_i,z_i)\) as the conditional population relation, without a luminosity function or marginal apparent-magnitude model.

\subsection{Inference}
\label{subsec:inference}

We evaluate eq.~\eqref{eq:exact_conditional_likelihood} with a scalable two-stage importance-sampling approximation. For each object \(i\), let \(R_i\in\mathbb{N}\) be the number of retained Stage--1 posterior draws and define the local sample space
\begin{equation}
\Omega_i\equiv\mathbb{R}\times\mathbb{R}\times\mathbb{R}^3,
\qquad
\omega_i=(m_i,\eta_i,\bm{\nu}_i)\in\Omega_i .
\end{equation}
Stage~1 fits each light curve independently under a broad reference prior and exports the paired sequence
\begin{equation}
\left\{
\omega_i^{(r)}
=
\left(
m_i^{(r)},\eta_i^{(r)},\bm{\nu}_i^{(r)}
\right)
\in\Omega_i
\;\middle|\;
r=1,\ldots,R_i
\right\}.
\end{equation}
The pairing between \(m_i^{(r)}\), \(\eta_i^{(r)}\), and \(\bm{\nu}_i^{(r)}\) is retained, so that Stage~2 propagates the local posterior correlations rather than using marginal summaries or point
estimates. Stage~2 uses this sequence as an importance-sampling proposal and reweights it by the conditional population model
\(p_{\bm{\theta}}(\eta_i\mid m_i,z_i)\). This propagates the local DRW and latent finite-window-magnitude uncertainty into the global variability--luminosity and cosmological inference without repeated GP likelihood evaluations inside the global sampler.

\subsubsection{Stage 1: local light-curve fitting}

Stage-1 fits each quasar light curve independently under a broad, redshift-independent reference prior \(p_1(\eta_i,\bm{\nu}_i)\). The global variability--luminosity relation is not used; \(z_i\) enters only through the observer-frame to rest-frame time conversion.

We factor the Stage-1 prior as
\begin{equation}
p_{1,\nu}(\bm{\nu}_i)
\equiv
\int p_1(\eta_i,\bm{\nu}_i)\,d\eta_i
=
\pi_\nu(\bm{\nu}_i),
\qquad
p_1^\eta(\eta_i\mid\bm{\nu}_i)
=
\frac{p_1(\eta_i,\bm{\nu}_i)}{p_{1,\nu}(\bm{\nu}_i)} .
\label{eq:stage1_prior_factorization}
\end{equation}
Thus the Stage-1 nuisance marginal matches the nuisance reference measure in
eq.~\eqref{eq:exact_conditional_likelihood}, while the conditional interim prior \(p_1^\eta(\eta_i\mid\bm{\nu}_i)\) may be flat or non-flat and may depend on \(\bm{\nu}_i\).

The stationary-mean DRW likelihood for the observed magnitude vector is
\begin{equation}
p_{\rm LC}^{(0)}(y_i\mid \eta_i,\bm{\nu}_i,z_i)
=
\mathcal N
\left(
\mathbf m_i^{\rm obs}
\mid
m_{0,i}\mathbf 1,
\mathbf C_i
\right),
\label{eq:stage1_lc_likelihood}
\end{equation}
as defined in eq.~\eqref{eq:LC0likelihood}. This likelihood is the \(y_i\)-marginal of the joint Gaussian density
\(p_{\rm GP}(y_i,m_i\mid\eta_i,\bm{\nu}_i,z_i)\) used in the conditional target. Equivalently,
\begin{equation}
p_{\rm GP}
\left(
y_i,m_i\mid\eta_i,\bm{\nu}_i,z_i
\right)
=
p
\left(
m_i\mid y_i,\eta_i,\bm{\nu}_i,z_i
\right)
p_{\rm LC}^{(0)}
\left(
y_i\mid\eta_i,\bm{\nu}_i,z_i
\right).
\label{eq:gp_joint_density}
\end{equation}

From the joint Gaussian construction in eq.~\eqref{eq:gp_joint_gaussian}, the conditional distribution of \(m_i\) is analytic:
\begin{equation}
p
\left(
m_i\mid y_i,\eta_i,\bm{\nu}_i,z_i
\right) =
\mathcal N
\left(
m_i\mid
\mu_{{\rm win},i},
s_{{\rm win},i}^2
\right),
\label{eq:mwin_conditional_background}
\end{equation}
with
\begin{equation}
\mu_{{\rm win},i}
=
m_{0,i}
+
\mathbf c_{{\rm win},i}^{\top}
\mathbf C_i^{-1}
\left(
\mathbf m_i^{\rm obs}
-
m_{0,i}\mathbf 1
\right), \qquad
s_{{\rm win},i}^2
=
v_{{\rm win},i}
-
\mathbf c_{{\rm win},i}^{\top}
\mathbf C_i^{-1}
\mathbf c_{{\rm win},i}.
\label{eq:mwin_conditional}
\end{equation}

The resulting independent Stage-1 posterior is
\begin{equation}
p_{\rm S1}(\eta_i,\bm{\nu}_i\mid y_i,z_i)
=
\frac{
p_{\rm LC}^{(0)}(y_i\mid \eta_i,\bm{\nu}_i,z_i)
p_1(\eta_i,\bm{\nu}_i)
}{
Z_{1,i}
},
\label{eq:stage1_parameter_posterior}
\end{equation}
where
\begin{align}
Z_{1,i}
&=
\int
p_{\rm LC}^{(0)}
\left(
y_i\mid\eta_i,\bm{\nu}_i,z_i
\right)
p_1(\eta_i,\bm{\nu}_i)
\,d\eta_i\,d\bm{\nu}_i
\nonumber\\
&=
\int
p_{\rm GP}
\left(
y_i,m_i\mid\eta_i,\bm{\nu}_i,z_i
\right)
p_1(\eta_i,\bm{\nu}_i)
\,d m_i\,d\eta_i\,d\bm{\nu}_i .
\label{eq:stage1_evidence_joint}
\end{align}

For each retained Stage-1 draw \((\eta_i^{(r)},\bm{\nu}_i^{(r)})\), we sample
\begin{equation}
m_i^{(r)}
\sim
p(m_i\mid y_i,\eta_i^{(r)},\bm{\nu}_i^{(r)},z_i) = \mathcal N
\left(
\mu_{{\rm win},i}^{(r)},
s_{{\rm win},i}^{2\,(r)}
\right).
\end{equation}
Sampling \(m_i^{(r)}\), rather than replacing it by
\(\mu_{{\rm win},i}^{(r)}\), propagates the latent GP uncertainty in \(m_{i}\). Together these draws define the augmented Stage--1 proposal density on
\(\Omega_i\):
\begin{align}
q_i:\Omega_i\rightarrow\mathbb{R}_{\ge0},
\qquad
q_i(\omega_i)
&\equiv
p_{\rm S1}^{\rm aug}(m_i,\eta_i,\bm{\nu}_i\mid y_i,z_i)
\nonumber\\
&=
p
\left(
m_i\mid y_i,\eta_i,\bm{\nu}_i,z_i
\right)
p_{\rm S1}
\left(
\eta_i,\bm{\nu}_i\mid y_i,z_i
\right)
\label{eq:stage1_extended_proposal}
\\
&=
\frac{
p_{\rm GP}
\left(
y_i,m_i\mid \eta_i,\bm{\nu}_i,z_i
\right)
p_1(\eta_i,\bm{\nu}_i)
}{
Z_{1,i}
},
\label{eq:stage1_joint_proposal}
\end{align}
where the last equality uses Eqs.~\eqref{eq:gp_joint_density} and
\eqref{eq:stage1_parameter_posterior}. This augmented proposal is the distribution reweighted in Stage-2.

\subsubsection{Stage 2: global importance-sampling reweighting}
\label{subsec:stage2inference}

Stage-2 samples the global parameters \(\bm{\theta}\) by reweighting the
Stage-1 proposal: the interim variability prior \(p_1^\eta(\eta_i\mid\bm{\nu}_i)\)  is replaced by the conditional population model \(p_{\bm{\theta}}(\eta_i\mid m_i,z_i)\), while the fixed nuisance reference measure is retained.

Using the proposal density in eq.~\eqref{eq:stage1_joint_proposal} and the
Stage-1 prior factorization in eq.~\eqref{eq:stage1_prior_factorization}, the
object-level factor in eq.~\eqref{eq:exact_conditional_likelihood} becomes
\begin{align}
\mathcal{L}^{\rm cond}_{i,\pi_{\nu}}(\bm{\theta})
&=
\int
p_{\rm GP}(y_i,m_i\mid \eta_i,\bm{\nu}_i,z_i)\,
p_{\bm{\theta}}(\eta_i\mid m_i,z_i)\,
\pi_{\nu}(\bm{\nu}_i)
\,d m_i\,d\eta_i\,d\bm{\nu}_i
\nonumber\\[0.5em]
&=
Z_{1,i}
\,
\mathbb E_{\omega_i\sim q_i}
\left[
\frac{
p_{\bm{\theta}}(\eta_i\mid m_i,z_i)\,
\pi_{\nu}(\bm{\nu}_i)
}{
p_1(\eta_i,\bm{\nu}_i)
}
\right]
\nonumber\\[0.5em]
&=
Z_{1,i}
\,
\mathbb E_{\omega_i\sim q_i}
\left[
\frac{
p_{\bm{\theta}}(\eta_i\mid m_i,z_i)
}{
p_1^\eta(\eta_i\mid\bm{\nu}_i)
}
\frac{
\pi_{\nu}(\bm{\nu}_i)
}{
p_{1,\nu}(\bm{\nu}_i)
}
\right].
\label{eq:stage2_importance_identity}
\end{align} 

In the baseline implementation,  \(p_{1,\nu}=\pi_{\nu}\), so the nuisance factor cancels. Since \(Z_{1,i}\) is independent of \(\bm{\theta}\), it is dropped from the posterior evaluation. The
Stage--2 object likelihood is then estimated as
\begin{equation}
\widetilde{\mathcal L}_i(\bm{\theta})
=
\frac{1}{R_i}
\sum_{r=1}^{R_i}
r_{ir}(\bm{\theta}), \qquad r_{ir}(\bm{\theta})
=
\frac{
p_{\bm{\theta}}
\left(
\eta_i^{(r)}
\mid
m_i^{(r)},z_i
\right)
}{
p_1^\eta
\left(
\eta_i^{(r)}
\mid
\bm{\nu}_i^{(r)}
\right)
}.
\label{eq:stage2_object_likelihood_general}
\end{equation}

Using the Gaussian conditional model in eq.~\eqref{eq:conditional_D_model}, the
numerator in eq.~\eqref{eq:stage2_object_likelihood_general} is
\begin{equation}
p_{\bm{\theta}}
\left(
\eta_i^{(r)}
\mid
m_i^{(r)},z_i
\right)
=
\mathcal N
\left[
\eta_i^{(r)}
\mid
\mu_{\eta,i}^{(r)}(\bm{\theta}),
\sigma_{\rm int}^2
\right],
\end{equation}
where
\begin{align}
\mu_{\eta,i}^{(r)}(\bm{\theta})
&=
\eta_{D,0}
+
d_D
\left[
\ell_i^{(r)}(\theta_{\rm cos})-\ell_{\rm piv}
\right]
-
\alpha_D
\log_{10}
\left[
\frac{\lambda_{\rm rf}^{\rm eff}(z_i)}{\lambda_{\rm piv}}
\right],
\\
\ell_i^{(r)}(\theta_{\rm cos})
&\equiv
\ell_i(m_i^{(r)},z_i,\theta_{\rm cos})
=
\frac{
a_b
+
\mu_{\rm DM}(z_i,\theta_{\rm cos})
+
K_b(z_i)
-
m_i^{(r)}
}{2.5}.
\label{eq:stage2mean}
\end{align}

The full Stage--2 posterior is evaluated through the product likelihood
\begin{align}
p(\bm{\theta}\mid \{y_i,z_i\}_{i=1}^N)
&\propto
p(\bm{\theta})\,
\widetilde{\mathcal L}(\bm{\theta}),
\qquad
\widetilde{\mathcal L}(\bm{\theta})
\equiv
\prod_{i=1}^{N}
\widetilde{\mathcal L}_i(\bm{\theta}),
\label{eq:stage2_posterior}
\\
\log \widetilde{\mathcal L}(\bm{\theta})
&=
\sum_{i=1}^{N}
\left[
\operatorname{logsumexp}_{r=1}^{R_i}
\log r_{ir}(\bm{\theta})
-
\log R_i
\right].
\label{eq:stage2_loglike_general}
\end{align}
The second line is the same finite-sample product evaluated in log space to avoid numerical underflow.

In the implementation, Stage--1 samples \(D_i\) in natural-log coordinates, whereas Stage--2 evaluates
densities in \(\eta_i=\log_{10}[D_i/({\rm mag}^2\,{\rm day}^{-1})]\). Each retained draw is therefore converted as
\begin{equation}
\eta_i^{(r)}
=
\frac{1}{\ln 10}
\ln
\left[
\frac{D_i^{(r)}}{{\rm mag}^2\,{\rm day}^{-1}}
\right],
\end{equation}
before evaluating the Stage-2 likelihood. For the baseline priors,
\(p_1^\eta(\eta_i\mid\bm{\nu}_i)\) is constant over the retained posterior
support, so the denominator in
eq.~\eqref{eq:stage2_object_likelihood_general} is independent of
\(\bm{\theta}\) and is dropped. The object likelihood then reduces to
\begin{equation}
\widetilde{\mathcal L}_i(\bm{\theta})
\propto
\frac{1}{R_i}
\sum_{r=1}^{R_i}
\mathcal N
\left[
\eta_i^{(r)}
\mid
\mu_{\eta,i}^{(r)}(\bm{\theta}),
\sigma_{\rm int}^2
\right].
\label{eq:stage2_object_likelihood_flat}
\end{equation}
For non-flat interim priors, or when an interim prior density is written in the
natural-log coordinate rather than in \(\eta_i\), the denominator must be
retained with the appropriate Jacobian,
\begin{equation}
p_1^\eta(\eta_i\mid\bm{\nu}_i)
=
(\ln 10)\,
p_1^{\ln D}
\left(
\ln 10\,\eta_i
\mid
\bm{\nu}_i
\right),
\label{eq:prior_density_coordinate_change}
\end{equation}
where \(p_1^{\ln D}\) denotes the interim prior density with respect to
\(d\ln[D_i/({\rm mag}^2\,{\rm day}^{-1})]\).

The two-stage approximation requires four conditions: the retained Stage--1 draws must represent \(q_i\) adequately; each \(m_i^{(r)}\) must be a conditional GP draw paired with \((\eta_i^{(r)},\nu_i^{(r)})\); the Stage--2 target must remain the conditional relation \(p_\theta(\eta_i\mid m_i,z_i)\); and the Stage--1 proposal must overlap the relevant Stage--2 posterior support. The draw pairing and target definition hold by construction. Local recovery, finite-\(R_i\) proposal quality, overlap, and weight concentration are checked in sections ~\ref{sec:stage1_recovery_results} and ~\ref{sec:full_closure_stage2}.

\subsection{Calibration and single-band degeneracies}
\label{subsec:model_degeneracies}

At fixed cosmological shape, changing the Hubble constant \(H_0\) adds a redshift-independent offset to all distance moduli, and hence shifts all inferred luminosity proxies \(\ell_i\) by a constant. In the conditional mean of eq.~\eqref{eq:D_luminosity_relation}, this shift is exactly degenerate with the normalization \(\eta_{D,0}\): a global luminosity offset can be absorbed by a compensating shift in the variability--luminosity intercept. Thus an AGN-only variability--luminosity analysis can constrain the relative distance--redshift shape, but setting the absolute \(H_0\) scale requires an external calibration of the relation intercept. This is analogous to the zero-point degeneracy in standard-candle Hubble diagrams, where SN~Ia distances require an external absolute calibration, for example from Cepheid-calibrated SN Ia distances \citep{Riess_2022} or tip-of-the-red-giant-branch calibrations \citep{Freedman_2019}. We therefore fix \(H_0\) in the baseline analysis; calibration options are discussed in section~\ref{subsec:abslumcalibration}.

A second degeneracy is specific to the single-band implementation. Substituting the luminosity mapping into the Stage-2 mean from eq.~\eqref{eq:stage2mean} shows that the data must separate the cosmological redshift dependence from the chromatic term,
\[
\frac{d_D}{2.5}\,
\mu_{\rm DM}(z_i,\theta_{\rm cos})
\qquad \text{and} \qquad
-\alpha_D
\log_{10}
\left[
\frac{\lambda_{\rm rf}^{\rm eff}(z_i)}{\lambda_{\rm piv}}
\right].
\]
For a fixed observed band, \(\lambda_{\rm rf}^{\rm eff}(z)\) is largely a redshift coordinate. The chromatic coefficient \(\alpha_D\) can therefore mimic part of the cosmological redshift dependence. In the single-band cosmological tests presented here, we fix \(\alpha_D\) to a fiducial value when varying cosmology. Breaking this degeneracy requires wavelength leverage, for example from multi-band light curves, multi-epoch spectroscopy, or an external prior on the chromatic variability law. We return to this point in section~\ref{sec:Discussion}.

\subsection{Implementation}
\label{subsec:implementation}

Stage--1 consists of independent per-object GP fits, parallelized over light curves. The DRW likelihood is evaluated with \texttt{celerite2}, using a \texttt{RealTerm} kernel and an additional white-noise jitter term added in quadrature to the reported photometric uncertainties \citep{Foreman-Mackey_2017}. The per-object posteriors are sampled with the
affine-invariant ensemble sampler \texttt{emcee}
\citep{Goodman_2010,Foreman-Mackey_2013}.
We use 32 walkers and retain 2000 thinned posterior draws per source after 200-500 burn-in samples for each tested prior; burn-in sensitivity was checked on stress-biased Gaia \(G\)-band chains.

Stage-2 is implemented in \texttt{NumPyro}/\texttt{JAX}
\citep{Bradbury_2021,Phan_2019} and sampled on NVIDIA A100 GPUs with Hamiltonian Monte
Carlo using the No-U-Turn Sampler \citep{HoffmanGelman_2011}. The sampler
operates directly on the retained Stage-1 posterior draws and evaluates the
importance-reweighted likelihood in eq.~\eqref{eq:stage2_loglike_general}. The representative production runs
use four independent chains randomly initialized from the prior support, with 3000
warm-up steps and 1500 post-warm-up samples per chain. We monitor convergence with standard \(\hat R\) and effective-sample-size diagnostics
\citep{GelmanRubin_1992,Vehtari_2021}.

\section{Results}
\label{sec:results}

We validate the framework with Gaia-like \(G\)-band simulations matched to the Gaia cadence/noise templates, redshift range, and quality cuts of the Gaia DR3--SDSS
Type~1 AGN light-curve sample. The validation proceeds in two steps. First, object-level mocks test recovery of the propagated light-curve summaries \((m_i, \eta_i)\) from Gaia-like data under the DRW model. Second, an end-to-end proof-of-concept closure test injects a known cosmology, variability--luminosity relation, and intrinsic scatter, generates Gaia-like light curves over the selected analysis domain, processes the simulated light curves through the  two-stage pipeline, and tests recovery of the global cosmological, variability--luminosity, and intrinsic-scatter parameters. 

\subsection{Gaia--SDSS cadence and noise template sample}
\label{subsec:real_data_quality_cuts}

We start from a Gaia DR3--SDSS DR16Q positional cross-match within \(0.3^{\prime\prime}\), yielding \(486{,}994\) matches \citep{Gaia_DR3,Lyke_2020,WuShen_2022}. For sample characterization, we compute auxiliary photometric \(L_{3000}\) estimates from SDSS DR16 \textit{ugriz} photometry using code provided by M.~J. Temple, based on the Type~1 quasar SED framework of \citep{Temple_2021} and assuming a flat \(\Lambda\)CDM cosmology with \(H_0=71.0\,{\rm km\,s^{-1}\,Mpc^{-1}}\), \(\Omega_{\rm m}=0.27\), and \(\Omega_\Lambda=0.73\). Requiring finite redshifts, finite luminosity estimates, and \(\log_{10}(L_{\rm bol}/{\rm erg\,s^{-1}})\ge42\) leaves \(485{,}451\) characterized objects. These auxiliary \(L_{3000}\) estimates are used only for diagnostics; they are distinct from the Stage-2 luminosity coordinate \(\ell\), which is inferred from the \(m\) draws using the fixed SED/passband convention described in appendix~\ref{app:bandpass_quantities}.

\begin{figure}[t]
    \centering
    \includegraphics[width=1\linewidth]{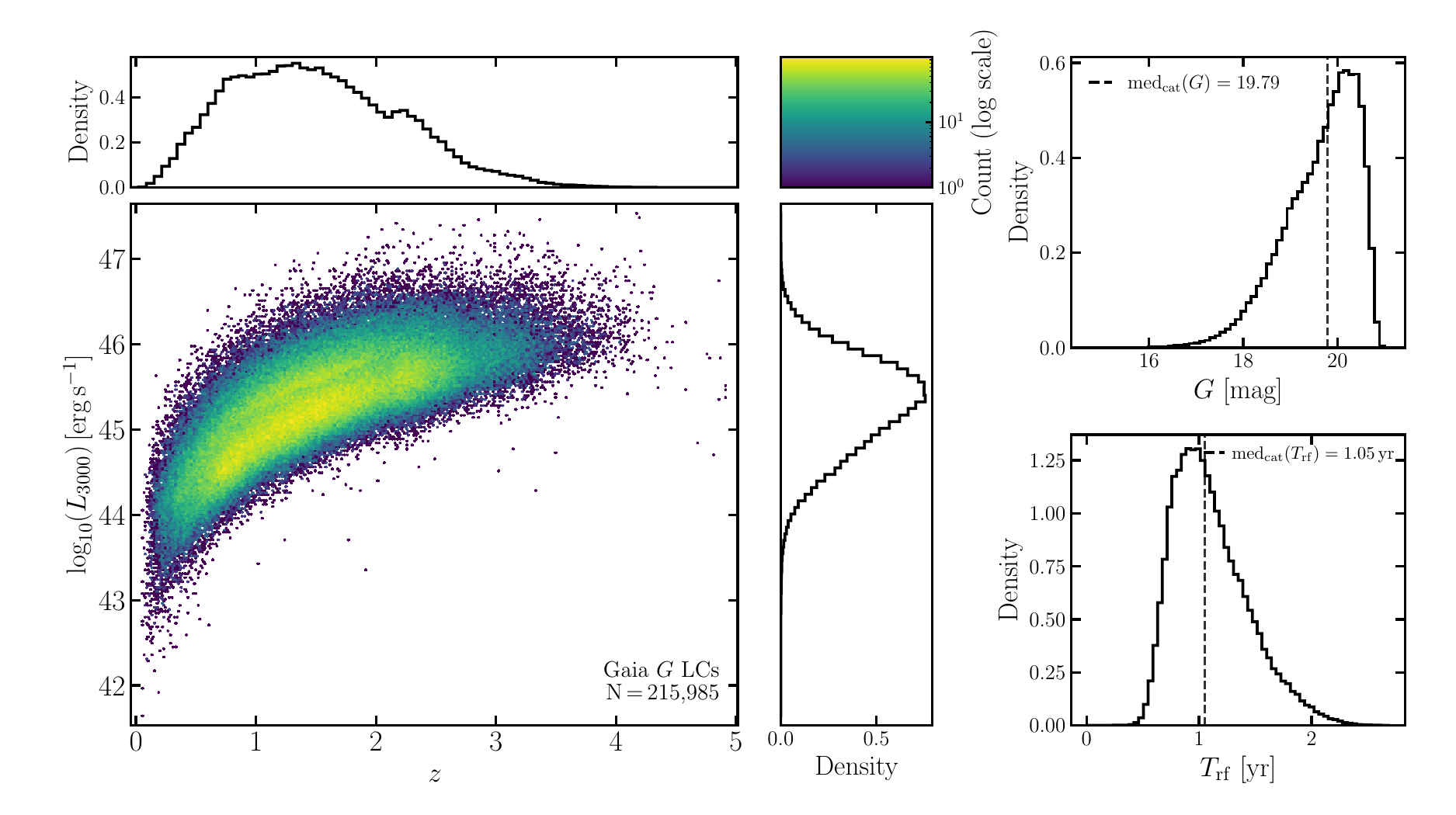}
    \caption{
    Gaia--SDSS DR16Q parent sample with Gaia \(G\)-band epoch photometry
    (\(N=215{,}985\)). The main panel shows redshift versus auxiliary photometric \(L_{3000}\);
    marginal panels show redshift, \(L_{3000}\), Gaia \(G\) magnitude, and
    rest-frame \(G\)-band baseline.
    }
    \label{fig:gaia_sdss_sample}
\end{figure}

We then restrict to characterized objects with Gaia DR3 \(G\)-band epoch photometry, a DR3 variability/time-domain product rather than a complete epoch-photometry release for all Gaia--SDSS AGN matches, yielding \(215{,}985\) DR16Q Type~1 AGN \citep{eyer_2023gaia}. This subset defines the empirical \(E_i=1\) parent sample used for both the
Stage--1 mock-recovery tests and the end-to-end closure tests; the implications
of this parent-sample conditioning for cosmological interpretation are discussed
in section~\ref{subsec:chromaticcalibrationandSED}. Figure~\ref{fig:gaia_sdss_sample} summarizes this parent sample in redshift, auxiliary \(L_{3000}\), Gaia \(G\) magnitude, and rest-frame \(G\)-band baseline. We use Gaia \(G\)-band light curves only in the baseline analysis: for the faint AGN considered here, typically \(G\gtrsim18\) as shown in figure~\ref{fig:gaia_sdss_sample}, the \(G_{\rm BP}\) and \(G_{\rm RP}\) epoch photometry have substantially lower signal-to-noise.

At the epoch level, we retain only transits with finite observing time,
positive finite Gaia \(G\)-band transit-averaged flux, and positive finite flux uncertainty, and assign the magnitude uncertainty by first-order
propagation, \(s_G=(2.5/\ln10)(\sigma_F/F)\). We also remove transits with
\(\texttt{variability\_flag\_g\_reject}=\mathrm{True}\), the Gaia DR3
DataLink flag marking \(G\)-band epochs rejected by the variability-processing cleaning chain
\citep{rimoldini_2023gaia,eyer_2023gaia}. Starting from \(9{,}223{,}348\) raw Gaia \(G\)-band transits, the numerical validity mask removes \(147{,}228\) epochs, all of which are also Gaia-flagged,
and the Gaia variability-processing flag removes a further \(272{,}241\) epochs. The epoch-level retained table therefore contains \(8{,}803{,}879\) \(G\)-band transits before source-level cuts. We then require \(N_G\ge10\) retained epochs as a minimal light-curve-quality cut. This acts only on the observed epoch count, a component of the sample-definition record \(c_i\) in section~\ref{subsec:method_conditional_model}, and removes only \(16\) sources,
leaving \(215{,}969\) objects.

We veto known changing-look AGN (CLAGN) using a literature compilation based on
\citep{Wang_2026}; after de-duplication, the veto list contains \(223\) unique entries. Matching by SDSS-style name, supplemented by a \(1^{\prime\prime}\) sky match and redshift validation, removes \(31\) known CLAGNs. After the \(N_G\ge10\) requirement and CLAGN veto, the Stage--1 catalogue contains \(215{,}938\) sources. This conservative veto removes previously reported large accretion-state
transitions, without attempting to identify new CLAGN candidates from the Gaia
light curves.

Together with the epoch-level validity and Gaia rejection flags, these retained-source cuts define the data-quality and epoch-availability part of the retained-analysis selection \(I_i=1\): they act on observed data quality, epoch availability, and externally identified state changes, not on recovered Gaia variability summaries. The additional analysis-domain cuts used in the closure tests are specified in section~\ref{sec:full_closure_construction}.

\subsection{Stage--1 recovery}
\label{sec:stage1_recovery_results}

We test recovery of the two local summaries propagated to Stage--2, the finite-window magnitude \(m_i\equiv m_{{\rm win},i}\) and the short-lag coordinate \(\eta_i\), from Gaia-like light curves generated under the same DRW GP likelihood used by Stage--1. The mocks use real Gaia \(G\)-band cadence and noise templates after the quality cuts described in 
section~\ref{subsec:real_data_quality_cuts}. Each template supplies observer-frame transit times, redshift, and per-epoch
magnitude uncertainties, which are converted to rest-frame quantities for both
simulation and fitting.

To separate DRW-parameter recovery from cadence/noise variation, we use a fixed
set of 800 Gaia templates, stratified by \(G_{\rm med}\), the median retained Gaia \(G\)-band magnitude, and by \(N_G\in\mathbb{N}\), the retained epoch count. We draw 50 templates from each of the \(4\times4\) cells listed in appendix~\ref{app:mockdatamatchrealdata}, and reuse the same templates at every injected DRW grid point. The selected templates retain their real redshifts and therefore their real rest-frame baselines, reproducing the usable Gaia template
pool in redshift, apparent magnitude, baseline, and epoch-count distributions.

For each selected template, we simulate stationary DRW light curves in magnitude
space over the grid
\[
(\tau_i^{\rm true},\sigma_i^{\rm true})
\in
\{5,20,80,300,1000,3000,10000\}\,{\rm d}
\times
\{0.03,0.06,0.10,0.14,0.20,0.30\}\,{\rm mag}.
\]
This gives \(42\) injected parameter cells and \(33{,}600\) mock light curves. The central grid spans typical optical quasar amplitudes and damping timescales \citep{Kelly_2009,MacLeod_2010,Suberlak_2021,Stone_2022}, while the largest amplitudes and longest timescales deliberately stress-test the prior-sensitive finite-baseline regime \citep{Kozlowski_2017,Hu_2024}.

For each grid point, the injected short-lag coordinate is
\(
\eta_i^{\rm true}
=
\log_{10}
\left[
\frac{(\sigma_i^{\rm true})^2/\tau_i^{\rm true}}
{{\rm mag}^2\,{\rm day}^{-1}}
\right].
\)
Mock generation follows the stationary-mean-anchored mode described in
appendix~\ref{app:DRWmockLCgeneration}: we set
\(m_{0,i}^{\rm true}=G_{{\rm med},i}\), inject no additional white-noise
jitter, and record the latent finite-window average
\(m_{{\rm win},i}^{\rm true}\).

Each mock light curve is refit under the Stage--1 reference priors in table~\ref{tab:s1_mock_priors}, using natural-log coordinates for positive local parameters. These priors define object-level proposals, not alternative astrophysical population models. They compare conventional \((\sigma,\tau)\) DRW parametrizations with priors that sample the short-lag coordinate \(D\) directly, and test sensitivity to the weakly constrained \(\sigma_i^2\)--\(\tau_i\) direction expected for finite baselines 
\citep{Kelly_2009,MacLeod_2010,Kozlowski_2017,Suberlak_2021,Yu_2022,Hu_2024, Yu_2026_EztaoX}. Priors~1--2 are
log-uniform in \((\sigma,\tau)\), with Prior~2 truncating the very long-\(\tau\)
volume not constrained by Gaia DR3 baselines. Priors~3--5 are log-uniform in \((D,\tau)\), with Prior~5 acting as a conservative stress test of boundary effects and proposal coverage.

\begin{table}[t]
\centering
\scriptsize
\setlength{\tabcolsep}{2pt}
\renewcommand{\arraystretch}{1.25}

\begin{tabular*}{\linewidth}{@{\extracolsep{\fill}} l l l l @{}}
\toprule
\multicolumn{1}{c}{\raisebox{0.99\normalbaselineskip}[0pt][0pt]{\textbf{Label}}} &
\multicolumn{1}{c}{\shortstack[c]{\textbf{Stage--1 sampled}\\\textbf{coordinates}}} &
\multicolumn{1}{c}{\raisebox{0.99\normalbaselineskip}[0pt][0pt]{\textbf{Stage--1 reference prior}}} &
\multicolumn{1}{c}{\shortstack[c]{\textbf{Stage--2 interim-prior}\\\textbf{factor}}} \\
\midrule

Prior 1 &
\(m_0,\ \ln\sigma,\ \ln\tau,\ \ln\sigma_{\rm add}\) &
\(\begin{array}[t]{@{}l@{}}
\sigma_{\rm DRW} \sim {\rm LogUniform}(0.01,2)\ {\rm mag} \\
\tau_{\rm DRW} \sim {\rm LogUniform}(10^{-4},10^{6})\ {\rm d}
\end{array}\)
&
\(\displaystyle
p_1^\eta(\eta\mid\bm{\nu})
=
\frac{\ln 10}{2\ln(200)}
\)
\\

\midrule

Prior 2 &
\(m_0,\ \ln\sigma,\ \ln\tau,\ \ln\sigma_{\rm add}\) &
\(\begin{array}[t]{@{}l@{}}
\sigma_{\rm DRW} \sim {\rm LogUniform}(0.01,2)\ {\rm mag} \\
\tau_{\rm DRW} \sim {\rm LogUniform}(10^{-4},10^{4})\ {\rm d}
\end{array}\)
&
\(\displaystyle
p_1^\eta(\eta\mid\bm{\nu})
=
\frac{\ln 10}{2\ln(200)}
\)
\\

\midrule

Prior 3 &
\(m_0,\ \ln D,\ \ln\tau,\ \ln\sigma_{\rm add}\) &
\(\begin{array}[t]{@{}l@{}}
D \sim {\rm LogUniform}(10^{-10},10^{4.6})\,
{\rm mag}^2\,{\rm d}^{-1} \\
\tau_{\rm DRW} \sim {\rm LogUniform}(10^{-4},10^{6})\ {\rm d}
\end{array}\)
&
\(\displaystyle
p_1^\eta(\eta\mid\bm{\nu})
=
\frac{1}{14.6}
\)
\\

\midrule

Prior 4 &
\(m_0,\ \ln D,\ \ln\tau,\ \ln\sigma_{\rm add}\) &
\(\begin{array}[t]{@{}l@{}}
D \sim {\rm LogUniform}(10^{-10},10^{4})\,
{\rm mag}^2\,{\rm d}^{-1} \\
\tau_{\rm DRW} \sim {\rm LogUniform}(1,10^{6})\ {\rm d}
\end{array}\)
&
\(\displaystyle
p_1^\eta(\eta\mid\bm{\nu})
=
\frac{1}{14}
\)
\\

\midrule

Prior 5 &
\(m_0,\ \ln D,\ \ln\tau,\ \ln\sigma_{\rm add}\) &
\(\begin{array}[t]{@{}l@{}}
D \sim {\rm LogUniform}(10^{-15.2},10^{11.3})\,
{\rm mag}^2\,{\rm d}^{-1} \\
\tau_{\rm DRW} \sim {\rm LogUniform}(10^{-2.6},10^{6.5})\ {\rm d}
\end{array}\)
&
\(\displaystyle
p_1^\eta(\eta\mid\bm{\nu})
=
\frac{1}{26.5}
\)
\\

\bottomrule
\end{tabular*}
\caption{
Stage-1 reference priors used for the mock-recovery tests. All priors use \(m_0\sim{\rm Uniform}(10,25)\) and \(\sigma_{\rm add}\sim{\rm LogUniform}(10^{-8},1)\,{\rm mag}\). For Priors~3--5,
\(\sigma_{\rm DRW}^2=D\tau_{\rm DRW}\) is derived rather than separately bounded. In the baseline hierarchy \(p_{1,\nu}=\pi_{\nu}\); for the
priors listed here, \(p_1^\eta\) is constant with respect to \(d\eta\) on the
retained support and therefore contributes only an overall
\(\bm{\theta}\)-independent factor.
}
\label{tab:s1_mock_priors}
\end{table}

For each quantity \(x\) represented in the retained Stage--1 draw set, we compute the draw median \(x_{i,50}
\) and report the median residual, root-mean-square error (RMSE), and nominal \(68\%\) coverage \(C_{68}\), defined as the fraction of injected values within the corresponding marginal \(16\)--\(84\) percentile interval of the same draws. These medians are calibration diagnostics only: Stage--2 propagates
posterior draws, not point estimates.

Table~\ref{tab:s1_mock_recovery_gaia} summarizes recovery in a Gaia-supported diagnostic interval of baseline ratio,
\(-0.5\le\log_{10}(T_{{\rm rf},i}/\tau_i^{\rm true})\le1.2\), containing
\(14{,}406\) mock light curves. For typical optical-quasar damping timescales of \(10^2\)--\(10^3\) d, this interval overlaps the baseline ratios most relevant to Gaia DR3 and extends toward the \(T_{\rm rf}/\tau\simeq10\) turnover-sampling regime used as a recovery diagnostic. Across the reference priors, the median residuals and nominal coverages are stable for the propagated summaries. The finite-window magnitude has negligible median bias, with \(|{\rm median}\,\Delta m_{\rm win}|\lesssim0.1\,{\rm mmag}\) and \(66\)--\(68\%\) coverage. The short-lag coordinate has \(|{\rm median}\,\Delta\eta|\lesssim0.05\,{\rm dex}\) and \(68\)--\(74\%\) coverage. The broadest interim prior mainly broadens the \(\eta\) residual distribution,
rather than shifting the median recovery. Diagnostics over the full 42-cell injected \((\tau,\sigma)\) grid lead to the same qualitative conclusion, with RMSEs more sensitive to the deliberately
included extreme baseline-ratio stress-test cases.

\begin{table}[t]
\centering
\begin{tabular}{lcccc}
\hline
Prior &
\(\Delta\eta\)/RMSE &
\(C_{68}(\eta)\) &
\(\Delta m_{\rm win}\)/RMSE &
\(C_{68}(m_{\rm win})\) \\
\hline
Prior 1 & -0.02 / 0.42 dex & 72.3\% & -0.06 / 14.1 mmag & 66.9\% \\
Prior 2 & +0.03 / 0.49 dex & 72.2\% & -0.04 / 14.1 mmag & 67.6\% \\
Prior 3 & -0.03 / 0.44 dex & 74.1\% & -0.06 / 14.1 mmag & 66.4\% \\
Prior 4 & -0.04 / 0.58 dex & 68.1\% & -0.06 / 19.6 mmag & 66.5\% \\
Prior 5 & -0.05 / 1.14 dex & 68.5\% & -0.03 / 14.1 mmag & 66.3\% \\
\hline
\end{tabular}
\caption{
Stage-1 mock recovery in the Gaia-relevant baseline-ratio interval. Entries give median residual, RMSE and \(C_{68}\), defined in the text.
}
\label{tab:s1_mock_recovery_gaia}
\end{table}

\begin{figure}[t]
    \centering
    \includegraphics[width=1\linewidth]{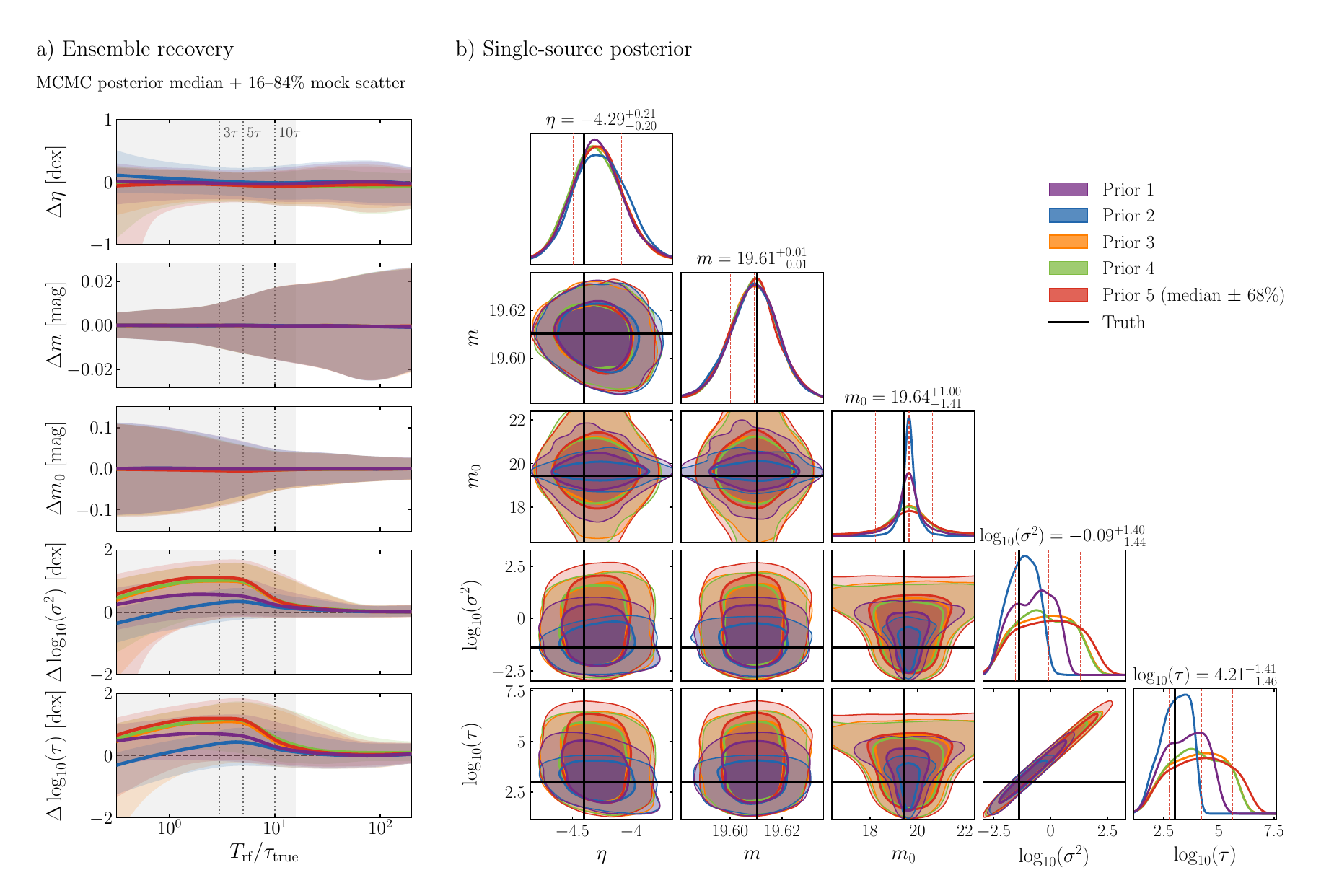}
    \caption{
    Panel (a) summarizes ensemble Stage--1 recovery over the mock grid as a function of rest-frame baseline coverage,
    \(T_{{\rm rf},i}/\tau_i^{\rm true}\). Curves show median residuals, defined as posterior median minus truth, and shaded bands show the 16th--84th percentile scatter across mocks. Curves are shape-preserving interpolations for display; table values are computed directly from posterior draws. Vertical dotted lines mark \(T_{{\rm rf},i}/\tau_i^{\rm true}=3,5,10\), and the grey band marks the Gaia-relevant baseline-ratio interval used in table~\ref{tab:s1_mock_recovery_gaia}. Panel (b) shows a single representative Gaia-like mock posterior under the Stage--1 reference priors, with injected truths marked in black.
    }
    \label{fig:whyDbestwithbaseline}
\end{figure}

Prior sensitivity appears mainly in \(m_0\), \(\tau\), and \(\sigma^2\), as expected from finite-baseline DRW identifiability and illustrated in figure~\ref{fig:whyDbestwithbaseline}. Before the DRW turnover is sampled, the likelihood mainly constrains \(D=\sigma^2/\tau\), so different \((\sigma,\tau)\) pairs can yield nearly indistinguishable short-lag variability. Truncating poorly sampled long-\(\tau\) volume (e.g Prior 2) can regularize apparent \(\sigma\) and \(\tau\) recovery, but this should not be interpreted as independent identification of both long-timescale parameters from Gaia-like light curves. This is not a limitation for Stage--2, where \(\sigma\) and \(\tau\) enter only through the proposal for \((m,\eta)\). This interpretation is consistent with the contemporaneous information-theoretic
analysis of \citep{Brewer_2026}, who find that quasar light curves constrain short-term CAR(1)/DRW volatility much more robustly than the characteristic timescale.

Figure~\ref{fig:whyDbestwithbaseline} also shows the expected contrast between the stationary DRW mean \(m_0\) and the finite-window magnitude \(m\). As \(T_{\rm rf}/\tau_{\rm true}\) increases, the light curve samples more independent DRW fluctuations and \(m_0\) becomes better identified. The finite-window magnitude behaves differently because it is the latent average over the Gaia observing window. At small \(\tau\), or equivalently large \(T_{\rm rf}/\tau_{\rm true}\),  the process can decorrelate between Gaia visits, so unobserved between-epoch fluctuations contribute to the conditional uncertainty of this window average. Nevertheless, over the Gaia-supported regime tested here, \(m_i\) remains much more tightly constrained than \(m_0\), as seen both in the ensemble recovery and in the representative single-source posterior. This behaviour is advantageous for Stage--2: the hierarchy propagates the full posterior uncertainty in \(m_i\), whereas using \(m_0\) would require baselines long enough to identify the stationary mean, precisely the regime where long-term trends and departures from a single stationary DRW become increasingly relevant.

\subsection{End-to-end Stage-1-Stage-2 closure tests}
\label{sec:full_closure_results}

We perform end-to-end closure tests within the Gaia DR3--SDSS epoch-photometry parent sample, \(E_i=1\), by generating Gaia-like mock light curves from a known conditional variability--luminosity model, processing them through the Stage--1--Stage--2 inference pipeline of section~\ref{subsec:inference}, and imposing the retained-analysis selection \(I_i=1\) required for real-data applications. These tests assume that the adopted conditional DRW population model is the data-generating model; they are therefore closure tests of the implemented likelihood, not tests of model adequacy for real quasars. Because they include real Gaia cadence templates, epoch-level photometric noise, finite-window stochasticity, Stage--1 posterior structure, and the same redshift, apparent-magnitude, and light-curve-quality requirements needed for real-data analyses, they test whether the pipeline recovers the injected global parameters after the practical \(I_i=1\) selection is applied.

\subsubsection{Closure construction}
\label{sec:full_closure_construction}

We first generate a candidate pool larger than the final closure catalogue, allowing the selected mocks to be matched without replacement to real Gaia cadence/noise templates. Candidate redshifts and luminosity coordinates are drawn from a rectangular closure domain,
\[
z_i\sim\mathcal U(z_{\min},z_{\max}),\qquad
\ell_i\sim\mathcal U(\ell_{\min},\ell_{\max}),
\]
with \(z_{\min}=0.5\), \(z_{\max}=3.5\), \(\ell_{\min}=44.0\), and \(\ell_{\max}=47.0\).
These uniform draws only populate the controlled \((z,\ell)\) region; they are not a luminosity-function model, and the Stage--2 likelihood remains conditional as described in section~\ref{subsec:method_conditional_model}.

Given the injected cosmology \(\theta_{\rm cos}^{\rm true}\), the true finite-window apparent magnitude is generated with the same
magnitude--luminosity mapping used in Stage--2,
\[
m_{{\rm win},i}^{\rm true}
=
a_b
+
\mu_{\rm DM}(z_i,\theta_{\rm cos}^{\rm true})
+
K_b(z_i)
-
2.5\,\ell_i .
\]
The same SED/passband convention from appendix \ref{app:bandpass_quantities} defines \(K_b(z_i)\) and \(\lambda_{\rm rf}^{\rm eff}(z_i)\). No additional scatter is added to the magnitude--luminosity mapping; population scatter enters only through the variability coordinate via
\(\epsilon_{{\rm int},i}\in\mathbb{R}\), with
\(\epsilon_{{\rm int},i}\sim
\mathcal{N}(0,\sigma_{\rm int,true}^2)\):
\begin{equation}
\begin{aligned}
\eta_i^{\rm true}
&=
\eta_{D,0}^{\rm true}
+
d_D^{\rm true}(\ell_i-\ell_{\rm piv})
-
\alpha_D^{\rm true}
\log_{10}\!\left(
\frac{\lambda_{\rm rf}^{\rm eff}(z_i)}{\lambda_{\rm piv}}
\right)
+
\epsilon_{{\rm int},i}.
\end{aligned}
\label{eq:closure_eta_true}
\end{equation}

For the closure runs reported here we use
\(\lambda_{\rm piv}=3000\,\text{\AA}\) and
\(\ell_{\rm piv}=45.5\). The wavelength pivot is the same \(3000\,\text{\AA}\) continuum-normalization anchor used in the SED/mock
construction. The luminosity pivot is a centering convention, chosen to be
representative of the retained Gaia-like closure samples and to reduce
covariance between \(\eta_{D,0}\) and \(d_D\). As a re-centering check, nearby
reasonable pivot choices produced the expected shift in \(\eta_{D,0}\), with
no material change to the recovered slopes or distance--redshift parameters.

For the distance--redshift relation, we use the Chevallier--Polarski--Linder (CPL) dark-energy parametrization \citep{Chevallier_2001,Linder_2003},
\begin{equation}
w(z) = w_0 + w_a \frac{z}{1+z},
\label{eq:cpl_parametrization}
\end{equation}
assuming spatial flatness. The fiducial injected cosmology is a Planck-like flat \(\Lambda\)CDM model within this parametrization:
\[
H_0^{\rm true}=67.4\,{\rm km\,s^{-1}\,Mpc^{-1}},
\qquad
\Omega_m^{\rm true}=0.315,
\qquad
w_0^{\rm true}=-1,
\qquad
w_a^{\rm true}=0 .
\]

The injected variability--luminosity parameters are
\[
\eta_{D,0}^{\rm true}=-4.49,
\qquad
d_D^{\rm true}=-0.85,
\qquad
\alpha_D^{\rm true}=2.51,
\qquad
\sigma_{\rm int,true}=0.12 .
\]
These injected values are chosen from a preliminary real-data Stage--2 fit using the same reference cosmology and fixed SED/passband convention described in appendix~\ref{app:bandpass_quantities}. They are used only to place the mock catalogue in the empirical Gaia-quasar regime, and are not imposed on the closure fit, which must recover them from the simulated light curves and Stage–1 posterior samples. The interpretation of these Gaia-only calibration choices is discussed in section~\ref{sec:Discussion}.

The intrinsic scatter is defined at the latent population level, whereas
scatter in Stage--1 posterior summaries also includes finite-cadence recovery
error and posterior geometry. We adopt
\(\sigma_{\rm int,true}=0.12\) dex as a mock-calibration choice, yielding a selected, refitted mock catalogue with signal-to-noise and apparent-scatter properties comparable to those seen in preliminary Gaia-quasar analyses.

To generate light curves, the injected short-lag variability rate
\(D_i=10^{\eta_i^{\rm true}}\,{\rm mag}^2\,{\rm day}^{-1}\)
must be split into a DRW damping time and amplitude. Since Stage-2 models only \(D_i\), this split is treated as a light-curve nuisance prescription. We draw
\begin{equation}
\log_{10}\left(\frac{\tau_i}{\rm day}\right)
\sim
\mathcal N\!\left[\log_{10}(750),\,0.35^2\right],
\qquad
30<\tau_i/{\rm day}<3000,
\label{eq:closure_tau_distribution}
\end{equation}
and set
\[
\sigma_i=(D_i\tau_i)^{1/2},
\qquad
{\rm SF}_{\infty,i}=\sqrt{2}\,\sigma_i .
\]
Objects are not rejected based on the implied \(\sigma_i\), \({\rm SF}_{\infty,i}\), or \(T_{{\rm rf},i}/\tau_i\); these quantities are used only as diagnostics, and alternative prescriptions are discussed in section \ref{sec:full_closure_stage2}. 

For the closure mocks, the injected \((z_i,\ell_i,\theta_{\rm cos}^{\rm true})\)
set \(m_{i}^{\rm true}\), which is used both in the Stage--2 luminosity mapping and as the finite-window constraint for DRW light-curve generation; appendix~\ref{app:DRWmockLCgeneration} gives the generation procedure.
For matching, each candidate mock carries its injected \((z_i,\ell_i,m_i^{\rm true})\), while each real Gaia template is represented by \((z_j^{\rm real},\ell_j^{\rm real},m_j^{\rm real})\), where \(m_j^{\rm real}\) is the Stage--1 posterior-median finite-window magnitude and \(\ell_j^{\rm real}\) is the median obtained from the magnitude draws using the reference cosmology and fixed SED/passband convention. We match without replacement in standardized \((z,\ell,m_i)\) space, weighting \(m_i\) most strongly because Gaia epoch uncertainties depend primarily on apparent magnitude; appendix~\ref{app:mockdatamatchrealdata} gives the details.

For the Stage--2 closure analysis, we apply the fiducial Gaia analysis-domain requirements:
\(
0.5\le z\le 3.5\) and \(m_{50}\equiv{\rm median}_r(m_i^{(r)})\le20.
\)
Combined with the epoch- and source-level quality filters described in section~\ref{subsec:real_data_quality_cuts}, these cuts define \(I_i=1\) for the closure catalogue. The closure tests therefore evaluate recovery after applying the same retained-analysis selection needed for the real-data fit.
The lower-redshift cut reduces sensitivity to the regime where unresolved host-galaxy light is expected to have the largest impact on the inferred nuclear brightness and variability amplitude; we test this choice by repeating the analysis up to \(z_{\min}=0.7\). The upper bound \(z_{\max}=3.5\) avoids pushing the single-band Gaia \(G\)-band mapping further into strongly IGM-attenuated far-UV and Lyman-continuum wavelengths, where the \(K\)-correction and effective-wavelength predictor depend increasingly on the assumed quasar SED extrapolation and IGM treatment 
\citep[e.g.][]{Madau_1995,Inoue_2014}.

Because the apparent-magnitude requirement uses the Stage--1 posterior median \(m_{50}\), the retained catalogue size can depend weakly on the Stage--1 reference prior. Across comparable prior runs, the variation after the redshift cut is only \(0.004\%\), indicating that this posterior-median magnitude cut is not a material source of prior-dependent sample selection.

\begin{figure}[t]
    \centering
    \includegraphics[width=0.9\linewidth]{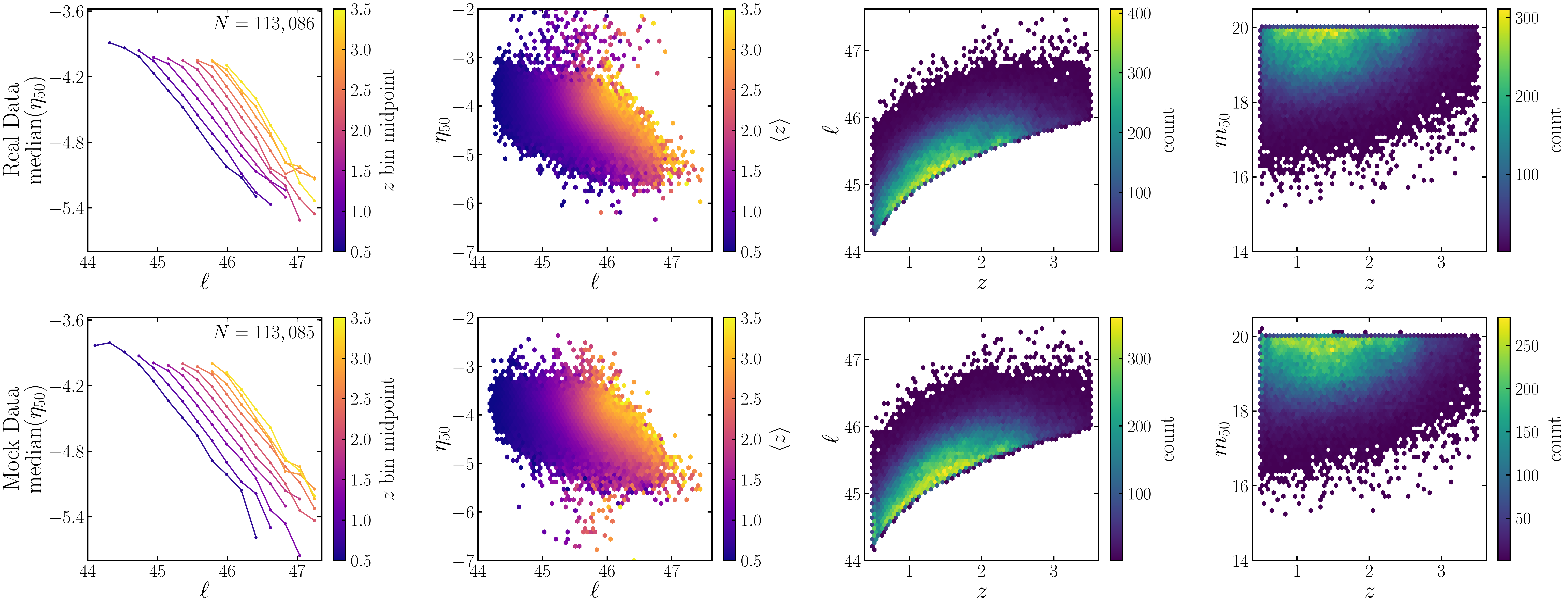}
    \caption{
    Stage--2 run-matching diagnostics for the representative Prior~5 closure sample. Rows compare the selected real Gaia templates, top, with the selected run-matched mock, bottom. Columns show the binned median \(\eta\) versus \(\ell\), density coloured by redshift, the redshift--\(\ell\) distribution, and the \(m_i\)--redshift distribution. Matching is performed in standardized \((z,\ell,m_i)\) space and does not use \(D_i\) or \(\eta_i\).
    }
\label{fig:stage2_mock_fullbox_selected_diagnostics}
\end{figure}

Figure~\ref{fig:stage2_mock_fullbox_selected_diagnostics} summarizes the run-matching diagnostics for the representative Prior~5 closure sample. The comparison uses the definitions above: real templates use \(\ell_j^{\rm real}\) inferred from posterior-median \(m_j^{\rm real}\), while mocks use the injected \(\ell_i\). The agreement indicates that the mock catalogue samples the same selected redshift, luminosity-proxy, apparent-magnitude, and cadence/noise-template domain as the real-data sample after quality cuts.

\subsubsection{Short-lag cadence support and Stage--1 calibration in the closure sample}
\label{sec:sf_drw_cadence_results}

Before testing global parameter recovery, we check that the retained \(I_i=1\) closure sample has both adequate short-lag cadence support for \(\eta_i\) and near-nominal Stage--1 calibration for the propagated summaries. For each selected  light curve, let
\(N_{31.6}\in\mathbb{N}\) be the number of rest-frame epoch pairs with
\(1<\Delta t_{\rm rf}/{\rm d}<31.6\). This is a cadence diagnostic, not an additional likelihood input or catalogue
selection cut. The lower limit avoids near-zero-lag pairs most sensitive to noise and cadence
microstructure, while the upper limit remains short compared with the central
mock damping timescales.

\begin{figure}[t]
    \centering
    \includegraphics[width=1\linewidth]{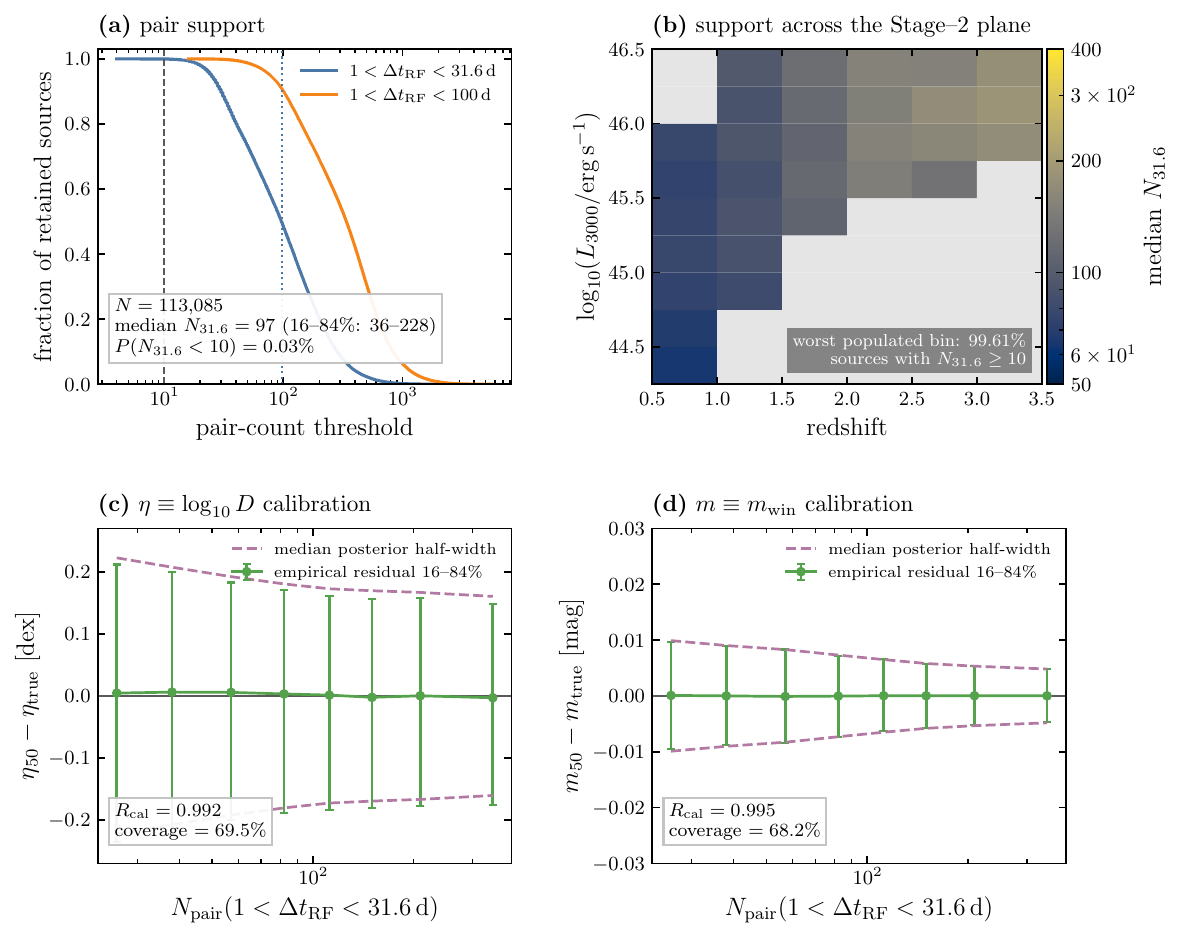}
    \caption{
    Short-lag cadence support and Stage--1 calibration for the representative Prior~5 selected closure sample, corresponding to the run-matched mock shown in the second row of figure~\ref{fig:stage2_mock_fullbox_selected_diagnostics}. Panel (a) shows the survival functions of rest-frame pair counts in the \(1<\Delta t_{\rm rf}<31.6\,{\rm d}\) and \(1<\Delta t_{\rm rf}<100\,{\rm d}\) lag ranges. Panel (b) maps the median \(N_{31.6}\) across the selected Stage--2 redshift--luminosity plane. Panels (c) and (d) show closure-mock residuals in \(\eta_i\) and \(m_i\), binned by \(N_{31.6}\), with median Stage--1 posterior half-widths overlaid for comparison.
    }
\label{fig:shortlag_pair_support_calibration}
\end{figure}

Figure~\ref{fig:shortlag_pair_support_calibration} summarizes this diagnostic for the representative Prior~5 selected closure sample, corresponding to the run-matched mock in figure~\ref{fig:stage2_mock_fullbox_selected_diagnostics}. Panels~(a) and (b) show broad short-lag pair support across the retained Stage--2 redshift--luminosity plane, with very few sources falling below \(N_{31.6}=10\). Panels~(c) and (d) show that the residual widths in \(\eta_i\) and \(m_i\) closely track the Stage--1 posterior half-widths, with near-nominal coverage. Repeating the same audit across the available Stage--1 reference priors from table~\ref{tab:s1_mock_priors} gives near-nominal coverage in all cases: \(68.9\)--\(69.6\%\) for \(\eta\) and \(68.0\)--\(68.7\%\) for \(m\), with \(R_{\rm cal}\simeq0.98\)--\(1.01\).

Additional checks versus redshift, rest-frame baseline, apparent magnitude, epoch count, photometric uncertainty, and median signal-to-noise are near nominal overall, with only a small residual signal-to-noise trend that remains within the Stage--1 posterior scale. Together with the grid-based object-level recovery in section~\ref{sec:stage1_recovery_results}, these diagnostics show that the
retained \(I_i=1\) closure sample provides calibrated Stage--1 proposals for \((m_i,\eta_i)\) and sufficient short-lag information for the Stage--2 test.

\subsubsection{Global-parameter recovery}
\label{sec:full_closure_stage2}

For each Stage--1 reference prior, the unchanged Stage--2 code receives only the posterior draws for objects passing the \(I_i=1\) closure cuts defined above. No truth-level local quantities, \((m_i^{\rm true},\eta_i^{\rm true},\tau_i,\sigma_i,m_{0,i}^{\rm gen})\), are supplied to the global sampler; they enter only through the simulated light curves and the resulting object-level posterior draws. As the main end-to-end cosmological closure, we use the posterior-draw analysis with \(H_0\) and \(\alpha_D\) fixed to their injected values. Fixing \(H_0\) removes the absolute-calibration degeneracy, while fixing \(\alpha_D\) isolates distance--redshift recovery from the single-band wavelength--cosmology degeneracy discussed in section~\ref{subsec:model_degeneracies}. The sampled CPL shape parameters have broad independent top-hat priors,
\(\Omega_{\rm m}\sim\mathcal U(0,1)\),
\(w_0\sim\mathcal U(-5,1)\), and
\(w_a\sim\mathcal U(-27,7)\). For the free variability--luminosity
parameters we use deliberately broad weakly informative priors,
\(
\eta_{D,0}\sim\mathcal U(-20,40),
d_D\sim\mathcal N(0,2^2),
\ln\sigma_{\rm int}\sim\mathcal N(0,1^2),
\)
with \(\sigma_{\rm int}=\exp(\ln\sigma_{\rm int})\). These priors comfortably enclose the literature-motivated and injected values.

\begin{figure}[t]
    \centering
    \includegraphics[width=1\linewidth]{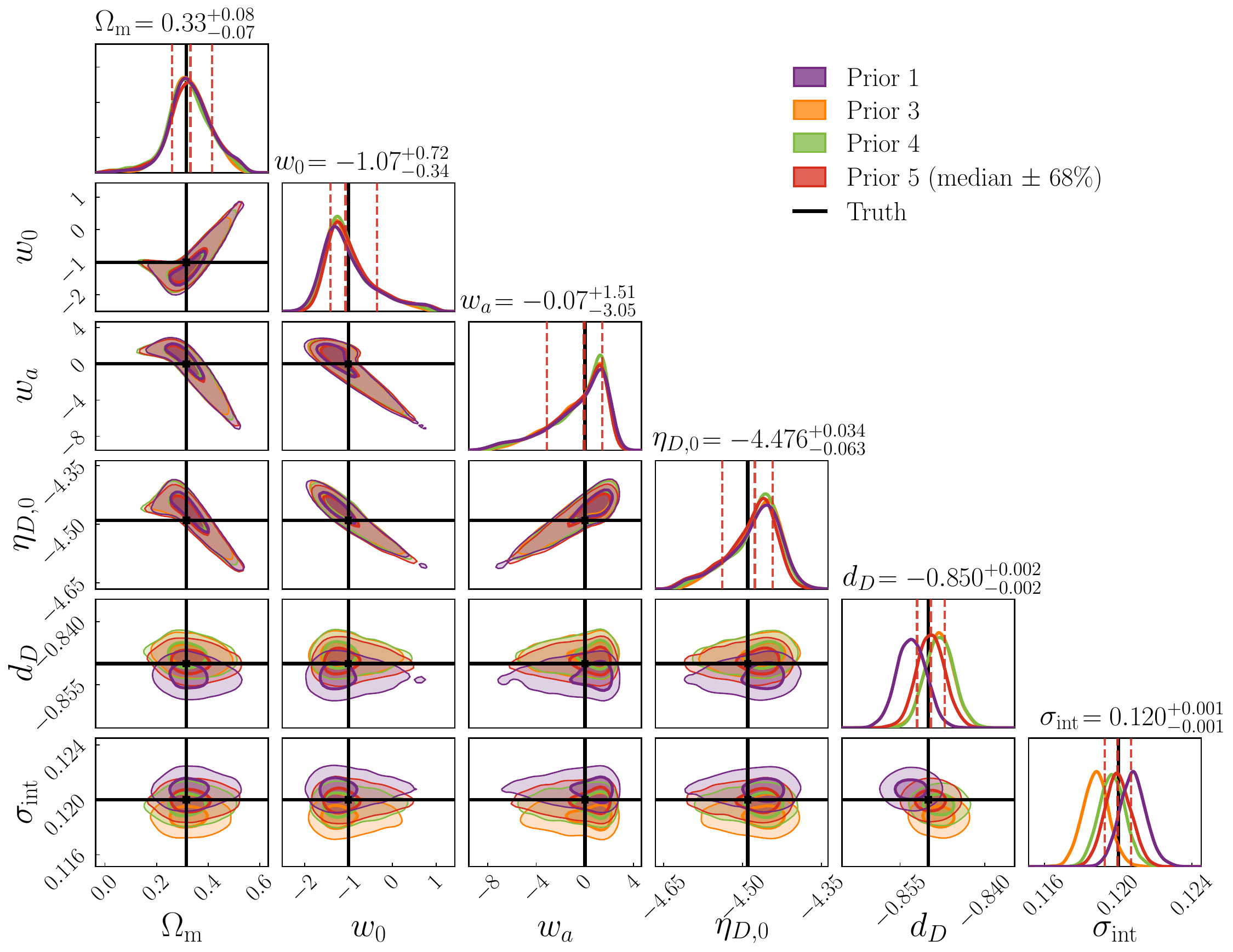}
    \caption{
    End-to-end cosmological closure test for the mock sample shown in figure~\ref{fig:stage2_mock_fullbox_selected_diagnostics}, with \(H_0\) and \(\alpha_D\) fixed to the injected values. Contours show the Stage-2 posterior for \((\Omega_m,w_0,w_a)\), the variability--luminosity parameters \((\eta_{D,0},d_D)\), and \(\sigma_{\rm int}\). Dashed lines mark posterior medians and central \(68\%\) intervals; black lines mark the injected truth. The injected cosmology is recovered within the broad single-band \(w_0\)--\(w_a\) degeneracy.
    }
    \label{fig:fullclosure_stage2_fixH0alpha_corner}
\end{figure}

Figure~\ref{fig:fullclosure_stage2_fixH0alpha_corner} shows the resulting Stage-2 posteriors using 250 thinned Stage--1 draws per object for the closure sample shown in figure~\ref{fig:stage2_mock_fullbox_selected_diagnostics}. Prior~5 is summarized by medians and central 68\% intervals; the other broad-\(\tau\)-support reference priors are overplotted for comparison. The free injected cosmological parameters are recovered within the broad degeneracies expected for a single-band AGN-only analysis, while the variability--luminosity relation and intrinsic scatter are recovered simultaneously. The posterior mass for all parameters remains well inside the broad prior ranges and does not accumulate at the imposed boundaries. Across the broad-\(\tau\)-support Stage--1 reference priors, the cosmological parameters and variability normalization are stable, with shifts well below the posterior uncertainties. The most visible proposal dependence appears in the variability slope and intrinsic scatter, shown in the last two columns of figure~\ref{fig:fullclosure_stage2_fixH0alpha_corner}. Because the closure catalogue contains \(N\simeq 1.1\times10^5\) AGN and the data are generated from the fitted conditional model, the formal statistical uncertainties on these population-level parameters are very small; small absolute changes in the Stage--1 proposal can therefore appear visually amplified. These shifts do not move the posteriors toward the prior boundaries or alter recovery of the injected closure relation.

We repeated the end-to-end closure tests with alternative mock seeds and representative literature-motivated \(\tau_i\) prescriptions for eq.~\eqref{eq:closure_tau_distribution}.  These checks gave qualitatively consistent recovery, indicating that the closure results are not driven by a particular mock seed or by the adopted \(\tau_i\) prescription. At this catalogue size, Stage--2 remains computationally tractable: with 250 retained draws per object, representative NUTS runs require 3.4--5.2 s per transition, corresponding to an effective \(\simeq10\)--16 ms per likelihood-plus-gradient evaluation.

As a higher-scatter stress test, we conducted additional closure experiments with up to \(\sigma_{\rm int,true}=0.30\) dex, keeping \(H_0\) and \(\alpha_D\) fixed as
in the fiducial runs. The injected relation and distance--redshift parameters remain statistically consistent with the recovered posterior, while the
credible intervals broaden by roughly a factor of \(1.5\)--\(2\), as expected from the reduced distance information per object.

Although the fiducial closure in figure~\ref{fig:fullclosure_stage2_fixH0alpha_corner} uses 250 retained Stage--1 draws per object, we verified the importance-sampling requirements of 
section~\ref{subsec:stage2inference} using \(R=200-2000\) retained Stage--1
posterior draws per object.
For representative runs with different Stage--1 priors, we monitored the
effective fraction of contributing draws, the largest single-draw contribution
to each normalized importance average, and the per-object log-normalization.
The effective fraction is typically \(\simeq0.5\), with 5th percentiles still
of order \(0.1\), and the largest single-draw contribution remains only a few
percent even at the 95th percentile. These diagnostics show that the
importance averages are not dominated by rare Stage--1 posterior draws and
that the Stage--1 proposal has adequate overlap with the Stage--2 population
model.

This closure test validates the complete Stage-1--Stage-2 code path for the conditional model: the light-curve fits recover the local summaries, the importance-reweighting likelihood propagates their uncertainty, and the catalogue-level inference recovers the injected variability--luminosity relation, scatter, and distance--redshift shape within the expected degeneracies.

\section{Discussion}
\label{sec:Discussion}

\subsection{Absolute luminosity calibration}
\label{subsec:abslumcalibration}

As discussed in section~\ref{subsec:model_degeneracies}, the Gaia-only conditional analysis fixes \(H_0\) because the absolute distance scale is degenerate with the variability--luminosity intercept. A natural possibility is to cross-calibrate the AGN relation against SN Ia in the overlapping redshift range. This would provide a
well-characterized relative distance ladder, but it would not make the AGN absolute
calibration fully independent: the resulting zero point would inherit the
absolute calibration, covariance structure, and possible catalogue-dependent
systematics of the adopted SN~Ia sample. Existing AGN Hubble-diagram analyses make this trade-off explicit: X-ray--UV quasar work calibrates the quasar intercept by matching or jointly fitting to SN~Ia in the common redshift range, while recent AGN-variability work uses a joint AGN+SN likelihood with Cepheid-calibrated Pantheon+ distances and their covariance
\citep{Risaliti_2015,Risaliti_2019,Lusso_2020,Lusso_2025,Dutra_2025,Brout_2022,Scolnic_2022}.
Thus SN-based calibration supplies the missing absolute scale, but imports the SN zero point, covariance/bias-correction model, and catalogue-specific systematics.

A more independent route is to calibrate the variability--luminosity intercept with low-redshift AGNs that have external, redshift-independent distances. This would tie the high-redshift AGN relation to nearby objects governed by the same accretion and variability physics. The main obstacle, however, is host-galaxy contamination. In unresolved or partially resolved nearby AGN photometry, stellar host light adds a substantial, approximately non-variable flux component. Thus, unmodelled host light both brightens the apparent source and dilutes the nuclear variability. Since the host fraction varies with luminosity, redshift, morphology, aperture, and bandpass, the resulting bias need not be a pure zero-point shift: it can also tilt the inferred relation. A robust low-redshift calibration should therefore include the host component
directly in the zero-point inference, using information from high-resolution
imaging and aperture-matched photometry
\citep[e.g.][]{Bentz_2006,Bentz_2009,Bentz_2013}, spectral host--nucleus
decomposition \citep[e.g.][]{Ren_2024}, or multi-band variability
\citep[e.g.][]{Gianniotis_2022}. Developing such a
low-redshift calibration sample is therefore an important next step beyond the present
Gaia-only implementation.

\subsection{Chromatic calibration, SED assumptions, and extensions}
\label{subsec:chromaticcalibrationandSED}

The chromatic coefficient in the Gaia-only analysis should be interpreted as an
effective \(G\)-band calibration, rather than as a universal physical chromatic
slope. As summarized in section~\ref{subsec:agn_variability_correlations}, the
literature does not imply a single value of \(\alpha_D\): converting the DRW wavelength scalings for
\({\rm SF}_{\infty}\) and \(\tau\) to the short-lag coordinate
\(D\) gives a long-timescale DRW-based expectation
\(\alpha_D\simeq1.1\), fixed-timescale variance measurements imply effective
\(D\)-like chromatic coefficients of \(\alpha_D\simeq1.3\)--\(1.6\)
on \(75\)--\(300\) d rest-frame timescales, and
\(\alpha_D\simeq2.5\)--\(2.6\) on \(\sim30\) d timescales. We therefore use \(\alpha_D\simeq2.5\) as the fiducial Gaia-\(G\) effective coefficient in the closure tests, consistent with the shortest-timescale variance measurements and with the preliminary Gaia DR3-only calibration fits. Closure tests with \(\alpha_D\simeq1.5\) provide a literature-motivated sensitivity check and also show consistent recovery of the injected relation.

This effective coefficient also absorbs single-band modelling effects that are
not part of the intrinsic chromatic variability law. In a single observed band, rest-frame
wavelength is strongly coupled to redshift, and the selected Gaia-like analysis sample has a strong redshift--luminosity correlation, as illustrated by the run-matching diagnostics in figure~\ref{fig:stage2_mock_fullbox_selected_diagnostics}. Residual errors in the
assumed SED, bandpass mapping, selection function, or finite-baseline modelling
can therefore project onto both the chromatic term and the cosmological distance
term. This is especially relevant because the baseline implementation presented here uses a
fixed, luminosity-independent quasar SED template, whereas real quasars show
diversity and variability in continuum slopes, emission-line equivalent widths,
broad emission lines, the Balmer continuum, Fe~{\sc ii} pseudo-continuum emission, and host-galaxy contributions \citep{Wilhite_2005, Shen_2011, Temple_2021}.
Thus, the fitted \(\alpha_D\) should be read as a single-band nuisance
calibration for Gaia rather than as a direct measurement of the physical chromatic variability law.

The modular bandpass treatment of
appendix~\ref{app:bandpass_quantities} allows \(a_b\), \(K_b(z)\), and \(\lambda_{\rm rf}^{\rm eff}(z)\) to be recomputed for alternative SED templates or source-specific spectra. In the closure tests above, the SED is fixed by construction, so this uncertainty is not part of the mock recovery problem. For real-data cosmology, however, SED choice becomes a calibration axis: non-contemporaneous spectra, such as SDSS spectra paired with Gaia light curves, should not be treated as exact Gaia-epoch SED measurements because continuum slopes, line strengths, and normalizations vary with source state. We therefore use the adapted SED from \citep{Temple_2021} as the baseline template, while treating source-specific spectra and alternative templates as calibration tests for real-data applications.

The most direct way to improve the chromatic calibration is to add wavelength
information. Existing ZTF light curves already provide a practical route to
multi-band optical calibration for large AGN samples
\citep{Bellm_2019,Graham_2019}, while repeat or multi-epoch spectroscopy from
DESI and related surveys can constrain continuum slopes, emission-line
contamination, Fe and Balmer pseudo-continuum emission, and host-galaxy light
\citep[e.g.][]{Aghamousa_2016, Adame_2024}. Such extensions require cross-calibration of instruments, bandpasses, cadences,
and observing windows. A dedicated Gaia/ZTF/DESI calibration analysis will be presented separately. Rubin/LSST will also provide
the natural long-term extension, delivering multi-band light curves for much
larger AGN samples over a broad range of observed and rest-frame wavelengths
\citep{Ivezic_2019}. Because the present framework is based on short-lag
variability rather than requiring a fully converged long-baseline DRW
measurement, it can be applied before the full Rubin/LSST time baseline is
available. A dedicated analysis of cadence, seasonal gaps, and selection effects
for this application, building on AGN light-curve cadence studies such as
\citep{Kovacevic_2021}, is left to future work.

A related sample-definition point is that the Gaia DR3 implementation is calibrated within the \(E_i=1\) epoch-photometry parent sample defined in
section~\ref{subsec:real_data_quality_cuts}. This is the appropriate domain for the present proof-of-concept because the standardized observables require
time-domain measurements. Within this parent sample, the relevant conditional-likelihood requirement is internal: after conditioning on \((m_i,z_i,c_i)\), the retained-analysis cuts should control data quality and
posterior support rather than select directly on the fitted variability coordinate. Future releases with broader epoch-data publication should expand
this parent sample and reduce the DR3-specific dependence on variability-pipeline epoch photometry, while leaving the same scalable conditional framework applicable.

\section{Conclusions}
\label{sec:conclusions}

We have developed a scalable hierarchical Bayesian framework for using Type~1
AGN optical variability as a population-level luminosity-distance probe, applicable to the
moderate-baseline regime of current time-domain surveys. The method uses independent light-curve fits to propagate posterior draws of
two finite-baseline summaries, the window-averaged apparent magnitude
\(m_{\rm win}\) and the short-lag variability coordinate
\(\eta=\log_{10}D\), into a conditional catalogue-level
variability--luminosity relation. Separating the per-object light-curve
likelihood evaluations from the global population inference makes the approach
suitable for survey-scale AGN samples.

Gaia DR3-like simulations show that the finite-window
brightness \(m_{\rm win}\) and the short-lag rate \(D\) are the robustly recoverable light-curve summaries in the moderate-baseline regime, whereas the
separate long-timescale DRW quantities \(m_0\), \(\sigma\), and \(\tau\) remain
prior- and baseline-sensitive when the turnover is not sampled. The short-lag pair-count and Stage--1 calibration diagnostics further show
that the retained Gaia-like closure samples contain enough day-to-month
rest-frame lag information for these summaries to be propagated to Stage--2. The end-to-end closure tests then show that these local summaries carry enough
information for the Stage--2 likelihood to recover the injected
variability--luminosity relation, intrinsic scatter, and distance--redshift
shape under the assumed conditional model, after applying the same
analysis-domain and light-curve-quality cuts intended for real-data
applications.

The present Gaia \(G\)-band implementation therefore establishes the statistical and computational core of an AGN-variability distance framework applicable to large-scale survey data. The remaining steps are calibration-driven: setting the absolute \(H_0\) scale
requires an external zero-point anchor, and the single-band chromatic
coefficient must be calibrated with multi-wavelength information. A dedicated Gaia/ZTF/DESI calibration analysis is left to future work. With larger epoch-photometry samples, improved host and SED modelling,
multi-band light curves from ZTF and Rubin/LSST, and DESI-like spectroscopic
information, the same framework can move from Gaia-only closure validation to
an independently calibrated high-redshift probe of the relative
distance--redshift relation.

\acknowledgments
JLM acknowledges support from the ``la Caixa'' Foundation
(ID 100010434; fellowship code LCF\slash BQ\slash EU24\slash 12060091). VB is grateful for the support from the Leverhulme Trust Research Project Grant RPG-2021-205 ‘The Faint Universe Made Visible with Machine Learning’. MC is grateful for support from the Isaac Newton Trust and the Schmidt Sciences AI2050 Early Career Fellowship. We thank Matthew J. Temple for providing the SDSS-photometry \(L_{3000}\)
estimation code used to compute the auxiliary luminosities for sample characterization and diagnostics. Computations were performed on the Swirles high-performance computing cluster at
DAMTP, University of Cambridge, using NVIDIA A100 GPUs.

\appendix

\section{SED and bandpass convention}
\label{app:bandpass_quantities}

This appendix defines the SED/passband convention used in the baseline analysis to map the finite-window magnitude draws to the luminosity coordinate. We describe the fixed reference SED and luminosity scaling; the band-dependent \(K\)-correction \(K_b:\mathbb{R}_{>0}\rightarrow\mathbb{R}\) and
constant \(a_b\in\mathbb{R}\); and the effective rest-frame wavelength function \(\lambda_{{\rm rf},b}^{\rm eff}:\mathbb{R}_{>0}\rightarrow
\mathbb{R}_{>0}\), which is used as the chromatic predictor in the population model.

We use a fixed-shape quasar SED template constructed with \textsc{qsosed} \citep{Temple_2021}. It includes the quasar continuum, hot-dust component, and emission-line components, while host-galaxy light, intrinsic reddening, and IGM absorption are disabled. The template is extended to \(500\,\text{\AA}\) using the composite from \citep{Shen_2020}. It is generated at \(z_{\rm ref}=1\) with continuum anchor \(\ell_{\rm ref}=45\); and after emission lines are added, the reference template has \(\log_{10}[\lambda L_\lambda(3000\,\text{\AA})/{\rm erg\,s^{-1}}]=45.0969\). For any given fixed SED shape, changing the luminosity coordinate \(\ell\in\mathbb{R}\) only rescales the template:
\begin{equation}
L_\lambda(\lambda,\ell)
=
L_\lambda^{\rm ref}(\lambda)
10^{\ell-\ell_{\rm ref}},
\label{eq:app_sed_scaling}
\end{equation}
where \(\lambda\in\mathbb{R}_{>0}\), and
\(L_\lambda^{\rm ref}:\mathbb{R}_{>0}\rightarrow\mathbb{R}_{\ge0}\)
is the reference luminosity density in
\({\rm erg\,s^{-1}\,\text{\AA}^{-1}}\). We note the simplicity of this SED convention, which we only use for our closed mock analyses for computational efficiency.

Synthetic photometry and same-band \(K\)-corrections follow the photon-counting bandpass convention of \citep{Hogg_2002_kcorr}. The emitted and observed wavelengths satisfy
\(\lambda_{\rm rf},\lambda_{\rm obs}\in\mathbb{R}_{>0}\), with
\(\lambda_{\rm obs}=(1+z)\lambda_{\rm rf}\). Let
\(S_b:\mathbb{R}_{>0}\rightarrow\mathbb{R}_{\ge0}\) be the throughput curve of band \(b\); for the Gaia analysis, \(S_G(\lambda)\) is the published Gaia DR3 \(G\)-band passband from \citep{Riello_2021,Gaia_DR3_passbands}. For any rest-frame luminosity density \(L_\lambda\), with wavelength in the same units as the passband table, we define
\begin{equation}
\left\langle L_\lambda\right\rangle_b
=
\frac{
\int L_\lambda(\lambda_{\mathrm{rf}})\,\lambda_{\mathrm{rf}}S_b(\lambda_{\mathrm{rf}})
\,d\lambda_{\mathrm{rf}}
}{
\int \lambda_{\mathrm{rf}}S_b(\lambda_{\mathrm{rf}})\,d\lambda_{\mathrm{rf}}
},
\qquad
\left\langle L_\lambda\right\rangle_{b,z}
=
\frac{
\int L_\lambda(\lambda_{\mathrm{rf}})\,\lambda_{\mathrm{rf}}
S_b[(1+z)\lambda_{\mathrm{rf}}]\,d\lambda_{\mathrm{rf}}
}{
\int \lambda_{\mathrm{rf}}S_b[(1+z)\lambda_{\mathrm{rf}}]\,d\lambda_{\mathrm{rf}}
}.
\label{eq:app_photon_counting_and_redshifted_band_average}
\end{equation}
Here \(\langle L_\lambda\rangle_b\) is the rest-frame band average entering
\(M_b\), while \(\langle L_\lambda\rangle_{b,z}\) is the corresponding
observed-band average written in the rest-frame variable
\(\lambda_{\rm rf}\). Using the sign convention of
Eq.~\eqref{eq:general_magnitude_relation}, the same-band \(K\)-correction is
\begin{equation}
K_b(z)
=
-2.5\log_{10}
\left[
\frac{
\left\langle L_\lambda\right\rangle_{b,z}
}{
(1+z)\left\langle L_\lambda\right\rangle_b
}
\right].
\label{eq:app_kcorr_same_band}
\end{equation}
For Gaia \(G\), we use the VEGAMAG relation
\(G=-2.5\log_{10}(F_G)-Z_G\), with \(F_G\) in
\({\rm W\,m^{-2}\,nm^{-1}}\) and \(Z_G=26.48986\) from \citep{Gaia_EDR3_photcal}.  Placing the reference template at \(10\,{\rm pc}\) and
using Eq.~\eqref{eq:app_sed_scaling} gives 
\begin{equation}
M_G(\ell)= a_G-2.5\,\ell , \qquad
a_G
=
-2.5\log_{10}
\left[
\frac{
\left\langle L_\lambda^{\rm ref}\right\rangle_G
}{
4\pi(10\,{\rm pc})^2\,100
}
\right]
-
Z_G
+
2.5\,\ell_{\rm ref}.
\label{eq:app_aG}
\end{equation}
The division by \(100\) converts
\({\rm erg\,s^{-1}\,cm^{-2}\,\text{\AA}^{-1}}\) to
\({\rm W\,m^{-2}\,nm^{-1}}\). The analogous constant \(a_b\) for another band is obtained by replacing the
Gaia \(G\)-band passband and zero point with those of the given band.

For the chromatic term used in the population model, a single instrumental pivot wavelength is not sufficient: a broad observed band samples a redshift-dependent range of emitted wavelengths, with weights set by both the quasar SED and the bandpass. We therefore define the corresponding SED- and throughput-weighted effective rest-frame wavelength in band \(b\) as
\begin{equation}
\lambda_{{\rm rf},b}^{\rm eff}(z)
=
\frac{
\int
\lambda_{\rm rf}
L_\lambda^{\rm ref}(\lambda_{\rm rf})
S_b[(1+z)\lambda_{\rm rf}]
\,d\lambda_{\rm rf}
}{
\int
L_\lambda^{\rm ref}(\lambda_{\rm rf})
S_b[(1+z)\lambda_{\rm rf}]
\,d\lambda_{\rm rf}
}.
\label{eq:app_lambda_rf_eff}
\end{equation}
For the baseline Gaia analysis, \(b=G\). In the single-band main-text notation, the band label is suppressed and this quantity is written simply
as \(\lambda_{\rm rf}^{\rm eff}(z)\). 

\section{Notation summary}
\label{app:notation_summary}

Table~\ref{tab:notation_summary} summarizes recurring notation used in the hierarchical model. 

\begin{table}[t]
\centering
\tiny
\setlength{\tabcolsep}{2pt}
\renewcommand{\arraystretch}{0.86}

\begin{tabularx}{\textwidth}{@{}>{\raggedright\arraybackslash}p{0.19\textwidth}
                                >{\raggedright\arraybackslash}p{0.27\textwidth}
                                >{\raggedright\arraybackslash}X@{}}
\toprule
Symbol & Space/type & Meaning/use \\
\midrule

\multicolumn{3}{@{}l}{\textit{Indices and observed quantities}}\\[-0.2ex]

\(N,n_i,R_i\) &
\(\mathbb{N}\) &
Numbers of objects, retained epochs, and posterior draws. \\

\(i,j,r\) &
\(i\in\{1,\ldots,N\}\), \(j\in\{1,\ldots,n_i\}\),
\(r\in\{1,\ldots,R_i\}\) &
Object, epoch, and Stage--1 draw indices. \\

\(b,\mathcal B_i\) &
\(b\) band label; \(\mathcal B_i\) SED/passband information &
Bandpass quantities; baseline \(b=G\). \\

\(y_i=(\mathbf t_i^{\rm obs},\mathbf m_i^{\rm obs},\mathbf s_i)\) &
\(\mathcal Y_i=\mathbb{R}^{n_i}\times\mathbb{R}^{n_i}
\times\mathbb{R}_{>0}^{n_i}\) &
Observed single-band light curve used in Stage--1. \\

\(z_i\) &
\(\mathbb{R}_{>0}\) &
Spectroscopic redshift, treated as known. \\

\(F_i,E_i,c_i,I_i\) &
\(F_i,E_i,I_i\in\{0,1\};\; c_i\in\mathcal C_i\) &
Full-forward fitted-sample inclusion, parent-sample indicator, observed
sample-definition variables, and retained-analysis indicator. \\

\addlinespace[0.4ex]
\multicolumn{3}{@{}l}{\textit{Light-curve and variability quantities}}\\[-0.2ex]

\(f_i,\mathbf f_i\) &
\(f_i:\mathbb{R}\to\mathbb{R}\), \(\mathbf f_i\in\mathbb{R}^{n_i}\) &
Latent rest-frame magnitude process and epoch vector. \\

\(\mathbf C_i,\mathbf K_i,\bm{\Sigma}_i\) &
\(\mathbb{R}^{n_i\times n_i}\), with \(\mathbf C_i\in\mathcal S_{++}^{n_i}\) &
Observed, DRW, and noise covariance matrices. \\

\(m_{0,i},m_i\equiv m_{\rm win, i},\eta_i\) &
\(\mathbb{R}\) &
Stationary mean, finite-window magnitude, and log short-lag coordinate. \\

\(\tau_i,\sigma_i,\sigma_{{\rm add},i}\) &
\(\mathbb{R}_{>0}\) &
DRW damping time, standard deviation, and extra white-noise scale. \\

\(D_i\) &
\(\mathbb{R}_{>0}\) &
Short-lag variability rate, \(D_i=\sigma_i^2/\tau_i\). \\

\(\bm{\nu}_i\) &
\(\mathbb{R}^{3}\) &
Local DRW nuisance vector; see Eq.~\eqref{eq:local_nuisance_definition}. \\

\addlinespace[0.4ex]
\multicolumn{3}{@{}l}{\textit{Luminosity, bandpass, and cosmology}}\\[-0.2ex]

\(\ell_i,\ell_{\rm piv}\) &
\(\mathbb{R}\) &
Luminosity proxy and luminosity pivot. \\

\(\lambda_{\rm piv},\lambda_{\rm rf}^{\rm eff}(z_i)\) &
\(\mathbb{R}_{>0}\) &
Wavelength pivot and chosen-band effective rest-frame wavelength. \\

\(K_b,a_b\) &
\(K_b:\mathbb{R}_{>0}\to\mathbb{R}\), \(a_b\in\mathbb{R}\) &
Same-band \(K\)-correction and band constant in \(M_b=a_b-2.5\ell\). \\

\(\mu_{\rm DM}\) &
\(\mathbb{R}_{>0}\times\Theta_{\rm cos}\to\mathbb{R}\) &
Distance modulus under the trial cosmology. \\

\(\theta_{\rm cos}\) &
\(\Theta_{\rm cos}\) &
Cosmological parameter vector. \\

\addlinespace[0.4ex]
\multicolumn{3}{@{}l}{\textit{Variability--luminosity relation and population quantities}}\\[-0.2ex]

\(\eta_{D,0},d_D,\alpha_D\) &
\(\mathbb{R}\) &
Intercept, luminosity slope, and wavelength slope. \\

\(\sigma_{\rm int}\) &
\(\mathbb{R}_{>0}\) &
Intrinsic scatter in \(\eta_i\). \\

\(\theta_{\rm VL}\) &
\(\Theta_{\rm VL}\equiv\mathbb{R}^{3}\) &
Variability--luminosity parameter vector
\((\eta_{D,0},d_D,\alpha_D)\). \\

\(\bm{\theta}\) &
\(\Theta_{\rm cos}\times\Theta_{\rm VL}\times\mathbb{R}_{>0}\) &
Global parameter vector
\((\theta_{\rm cos},\theta_{\rm VL},\sigma_{\rm int})\). \\

\(\mu_{\eta,i}\) &
\(\Theta_{\rm cos}\times\Theta_{\rm VL}\to\mathbb{R}\) &
Conditional population mean of \(\eta_i\). \\

\(p_{\bm{\theta}}(\eta_i\mid m_i,z_i)\) &
Density on \(\mathbb{R}\) with respect to \(d\eta_i\) &
Conditional population relation used in Stage--2. \\

\(\bm{\psi}_{\Phi},\bm{\psi}_{\nu},\bm{\psi}_{S}\) &
\(\Psi_{\Phi}\times\Psi_{\nu}\times\Psi_S\) &
Luminosity-function, nuisance-population, and selection parameters. \\

\(\mathcal S^{\rm full}_{\bm{\psi}_S},\mathcal A_i\) &
\(\mathcal S^{\rm full}_{\bm{\psi}_S}:\cdots\to[0,1]\),
\(\mathcal A_i\in\mathbb{R}_{>0}\) &
Full-forward selection probability and selected-sample normalization. \\

\addlinespace[0.4ex]
\multicolumn{3}{@{}l}{\textit{Inference quantities}}\\[-0.2ex]

\(p_{\rm GP}\) &
Density on \(\mathcal Y_i\times\mathbb{R}\) &
Joint GP density for \((y_i,m_i)\). \\

\(p_{\rm LC}^{(0)},p_{\rm LC}^{\rm win}\) &
Densities on \(\mathcal Y_i\) &
Stationary-mean and finite-window light-curve likelihoods. \\

\(p_1,p_1^\eta,p_{1,\nu},\pi_\nu\) &
Densities/measures on \(\mathbb{R}\times\mathbb{R}^3\) or marginals &
Stage--1 reference prior, interim prior, nuisance marginal, and reference measure. \\

\(p_{\rm S1}\) &
Density on \(\mathbb{R}\times\mathbb{R}^3\) &
Stage--1 posterior for \((\eta_i,\bm{\nu}_i)\). \\

\(\omega_i,q_i\) &
\(\omega_i\in\Omega_i=\mathbb{R}\times\mathbb{R}\times\mathbb{R}^3\),
\(q_i:\Omega_i\to\mathbb{R}_{\ge0}\) &
Augmented local variable and Stage--1 proposal density. \\

\(r_{ir},\widetilde{\mathcal L}_i\) &
\(\mathbb{R}_{\ge0}\) &
Importance ratio and object likelihood estimate. \\

\(\mathcal L_{\rm full},\mathcal L^{\rm exact}_{\rm cond,\pi_\nu}\) &
\(\mathbb{R}_{\ge0}\) &
Full survey-generative likelihood and implemented conditional target. \\

\bottomrule
\end{tabularx}
\caption{Summary of the main recurring notation used in the hierarchical AGN
variability model. The space/type column gives the mathematical type where it
is useful for distinguishing scalars, vectors, matrices, functions, densities,
and sequences.}
\label{tab:notation_summary}
\end{table}

\section{DRW mock light-curve generation}
\label{app:DRWmockLCgeneration}

This appendix specifies the DRW light-curve generation used in the Stage--1 recovery tests and in the end-to-end closure tests. Each mock object is assigned a real Gaia cadence/noise template: the epochs and epoch-level uncertainties are retained, while the measured magnitudes are discarded. Observer-frame Gaia epochs are converted to rest-frame times using the redshift assigned to the mock object: the real template redshift for the Stage--1 recovery tests, and the injected mock redshift for the end-to-end closure tests. The relevant distinction for mock generation is how the latent DRW is anchored before the common photometric-noise step is applied.

For the stationary-mean-anchored mocks in section~\ref{sec:stage1_recovery_results}, the latent Gaia-epoch magnitudes are generated using the exact Ornstein--Uhlenbeck transition \citep{Kelly_2009,Zu_2013}. For ordered epochs \(t_{ij}>t_{i,j-1}\), \(\rho_{ij}=\exp[-(t_{ij}-t_{i,j-1})/\tau_i]\), and
\begin{equation}
f_{ij}\mid f_{i,j-1},m_{0,i},\sigma_i,\tau_i
\sim
\mathcal N
\left[
m_{0,i}
+
\rho_{ij}
\left(
f_{i,j-1}-m_{0,i}
\right),
\;
\sigma_i^2
\left(
1-\rho_{ij}^2
\right)
\right].
\label{eq:app_ou_transition}
\end{equation}
The first latent epoch is drawn from the stationary distribution, \(f_{i1}\sim\mathcal N(m_{0,i},\sigma_i^2)\). For each injected grid
point we set \(m_{0,i}^{\rm true}=G_{{\rm med},i}\) and \(D_i^{\rm true}=(\sigma_i^{\rm true})^2/\tau_i^{\rm true}\), generate the
latent light curve with Eq.~\eqref{eq:app_ou_transition}, and use the
corresponding latent finite-window average \(m_i^{\rm true}\equiv m_{{\rm win},i}^{\rm true}\) as the main
magnitude recovery target.

The full-closure mocks described in section~\ref{sec:full_closure_results} are
finite-window anchored: \(m_i^{\rm true}\equiv m_{{\rm win},i}^{\rm true}\) is fixed by the injected
\((z_i,\ell_i)\), cosmology, and SED/passband information defined in appendix~\ref{app:bandpass_quantities}. After the cadence/noise template fixes the matched rest-frame epochs and window, we draw the nuisance \(\tau_i\), set \(\sigma_i^2=D_i^{\rm true}\tau_i\), and draw
\begin{equation}
m_{0,i}^{\rm gen}
\sim
\mathcal N(m_{i}^{\rm true},v_{{\rm win},i}).
\label{eq:app_m0_gen_from_mwin}
\end{equation}
This allows the stationary mean to differ from the imposed finite-window brightness by the amount expected for a DRW over the matched finite baseline. The latent Gaia-epoch vector is then drawn from the noise-free DRW Gaussian
distribution conditioned on the finite-window average being
\(m_{i}^{\rm true}\):
\begin{equation}
\mathbf f_i
\mid
m_{i}=m_{i}^{\rm true},
m_{0,i}^{\rm gen},D_i^{\rm true},\tau_i
\sim
\mathcal N
\left(
m_{0,i}^{\rm gen}\mathbf 1
+
\frac{\mathbf c_{{\rm win},i}}{v_{{\rm win},i}}
\left[
m_{{\rm win},i}^{\rm true}-m_{0,i}^{\rm gen}
\right],
\;
\mathbf K_i
-
\frac{
\mathbf c_{{\rm win},i}\mathbf c_{{\rm win},i}^{\top}
}{
v_{{\rm win},i}
}
\right).
\label{eq:app_closure_conditional_latent}
\end{equation}
Here \(\mathbf f_i\) is the latent Gaia-epoch magnitude vector, \(\mathbf K_i\) is the noise-free DRW covariance matrix at the matched
rest-frame epochs, and \(\mathbf c_{{\rm win},i}\) and \(v_{{\rm win},i}\) are
the finite-window covariance vector and variance defined in section~\ref{subsec:population_relation}, evaluated at the mock parameters and
matched rest-frame window.

For both anchoring modes, once the latent Gaia-epoch magnitudes \(f_{ij}\) are
generated, photometric noise is added as
\begin{equation}
m_{ij}^{\rm obs}=f_{ij}+\epsilon_{ij}^{\rm phot},
\qquad
\epsilon_{ij}^{\rm phot}\sim\mathcal N(0,s_{ij}^2),
\label{eq:app_mock_phot_noise}
\end{equation}
where \(s_{ij}\) is the Gaia uncertainty from the matched template. No extra
white-noise jitter is injected, although \(\sigma_{{\rm add},i}\) is still fitted in
Stage-1 to replicate the real data procedure.

Thus the Stage-1 recovery tests anchor the stationary mean \(m_0\), whereas the
full-closure tests anchor the finite-window magnitude \(m\), the brightness coordinate entering \(p_{\bm{\theta}}(\eta_i\mid m_i,z_i)\).

\section{Matching mock objects to real Gaia cadence templates}
\label{app:mockdatamatchrealdata}

For the Stage-1 recovery mocks, the 800 Gaia cadence/noise templates are drawn from the \(G_{\rm med}\)--\(N_G\) strata listed in table~\ref{tab:mock_template_strata}.
\begin{table}[h]
\centering
\scriptsize
\setlength{\tabcolsep}{3pt}
\begin{tabular}{lcccc}
\hline
Template magnitude bin
&
\(10\le N_G<29\)
&
\(29\le N_G<40\)
&
\(40\le N_G<52\)
&
\(N_G\ge52\)
\\
\hline
\(G_{\rm med}<19.0\)          & \(8487/50\)  & \(9876/50\)  & \(11476/50\) & \(14892/50\) \\
\(19.0\le G_{\rm med}<19.8\)  & \(11975/50\) & \(13871/50\) & \(17183/50\) & \(21554/50\) \\
\(19.8\le G_{\rm med}<20.3\)  & \(13752/50\) & \(13490/50\) & \(17762/50\) & \(15064/50\) \\
\(G_{\rm med}\ge20.3\)        & \(19737/50\) & \(14262/50\) & \(8810/50\)  & \(3782/50\)  \\
\hline
\end{tabular}
\caption{Stage--1 mock-template stratification. Entries give available/selected templates; 50 templates are selected per \(G_{\rm med}\)--\(N_G\) bin.}
\label{tab:mock_template_strata}
\end{table}

For the end-to-end closure tests from section~\ref{sec:full_closure_results}, matching is used only to assign real Gaia cadence/noise templates to mock objects. Each candidate mock carries its injected \((z_i,\ell_i,m_i^{\rm true})\), where \(m_i^{\rm true}\) is computed from the injected cosmology and SED/passband convention. Each real Gaia template is represented, for matching only, by
\((z_j^{\rm real},\ell_j^{\rm real},m_j^{\rm real})\), where
\(m_j^{\rm real}\) is the posterior-median finite-window magnitude from the real-data Stage--1 fit and
\(\ell_j^{\rm real}\) is the posterior median of the luminosity-proxy draws obtained from \(m_j^{(r)}\) under the reference cosmology and fixed SED/passband convention.
We do not match on \(\eta_i\) or \(D_i\), and the matching summaries are used only to assign templates; the Stage--2 likelihood uses the full paired posterior draws of \((m_i,\eta_i)\).

Matching is performed without replacement in the standardized weighted feature space
\begin{equation}
\mathbf{x}
=
\left(
\sqrt{p_z}\,\frac{z-\tilde{z}}{s_z},\;
\sqrt{p_\ell}\,\frac{\ell-\tilde{\ell}}{s_\ell},\;
\sqrt{p_m}\,\frac{m-\tilde{m}}{s_m}
\right),
\label{eq:stage2_mock_matching_features}
\end{equation}
where tildes and scales are the median and half the 16th--84th percentile range
of the real selected sample. We use \((p_z,p_\ell,p_m)=(1,1,2.5)\), giving extra weight to apparent magnitude because Gaia epoch-level uncertainties are brightness dependent. A KD-tree (\(k\)-dimensional tree) nearest-neighbour index is built from the mock features; real templates are processed from largest to smallest initial nearest-neighbour distance, and each template is assigned to the nearest unused mock candidate.

\bibliographystyle{JCAP}
\bibliography{agn_cosmoprobe}

\providecommand{\href}[2]{#2}\begingroup\raggedright\begin{thebibliography}{100}

\bibitem{Planck_2020}
{Planck Collaboration}, N.~{Aghanim}, Y.~{Akrami}, M.~{Ashdown}, J.~{Aumont}, C.~{Baccigalupi}, M.~{Ballardini}, A.~J. {Banday}, R.~B. {Barreiro}, N.~{Bartolo}, and et~al., {\it {Planck 2018 results. VI. Cosmological parameters}},  {\em \aap} {\bf 641} (Sept., 2020) A6, [\href{http://arxiv.org/abs/1807.06209}{{\tt arXiv:1807.06209}}].

\bibitem{Riess_2022}
A.~G. {Riess}, W.~{Yuan}, L.~M. {Macri}, D.~{Scolnic}, D.~{Brout}, S.~{Casertano}, D.~O. {Jones}, Y.~{Murakami}, G.~S. {Anand}, L.~{Breuval}, and et~al., {\it {A Comprehensive Measurement of the Local Value of the Hubble Constant with 1 km s$^{-1}$ Mpc$^{-1}$ Uncertainty from the Hubble Space Telescope and the SH0ES Team}},  {\em \apjl} {\bf 934} (July, 2022) L7, [\href{http://arxiv.org/abs/2112.04510}{{\tt arXiv:2112.04510}}].

\bibitem{Abdul_Karim_2025}
M.~{Abdul Karim}, J.~{Aguilar}, S.~{Ahlen}, S.~{Alam}, L.~{Allen}, C.~{Allende Prieto}, O.~{Alves}, A.~{Anand}, U.~{Andrade}, E.~{Armengaud}, and et~al., {\it {DESI DR2 results. II. Measurements of baryon acoustic oscillations and cosmological constraints}},  {\em \prd} {\bf 112} (Oct., 2025) 083515, [\href{http://arxiv.org/abs/2503.14738}{{\tt arXiv:2503.14738}}].

\bibitem{Lodha_2025}
K.~{Lodha}, R.~{Calderon}, W.~L. {Matthewson}, A.~{Shafieloo}, M.~{Ishak}, J.~{Pan}, C.~{Garcia-Quintero}, D.~{Huterer}, G.~{Valogiannis}, L.~A. {Ure{\~n}a-L{\'o}pez}, and et~al., {\it {Extended dark energy analysis using DESI DR2 BAO measurements}},  {\em \prd} {\bf 112} (Oct., 2025) 083511, [\href{http://arxiv.org/abs/2503.14743}{{\tt arXiv:2503.14743}}].

\bibitem{Cortes_2024a}
M.~{Cort{\^e}s} and A.~R. {Liddle}, {\it {Interpreting DESI's evidence for evolving dark energy}},  {\em \jcap} {\bf 2024} (Dec., 2024) 007, [\href{http://arxiv.org/abs/2404.08056}{{\tt arXiv:2404.08056}}].

\bibitem{Cortes_2024b}
M.~{Cort{\^e}s} and A.~R. {Liddle}, {\it {On data set tensions and signatures of new cosmological physics}},  {\em \mnras} {\bf 531} (June, 2024) L52--L56, [\href{http://arxiv.org/abs/2309.03286}{{\tt arXiv:2309.03286}}].

\bibitem{JiangPedrotti_2024}
J.-Q. {Jiang}, D.~{Pedrotti}, S.~S. {da Costa}, and S.~{Vagnozzi}, {\it {Nonparametric late-time expansion history reconstruction and implications for the Hubble tension in light of recent DESI and type Ia supernovae data}},  {\em \prd} {\bf 110} (Dec., 2024) 123519, [\href{http://arxiv.org/abs/2408.02365}{{\tt arXiv:2408.02365}}].

\bibitem{Efstathiou_2025}
G.~{Efstathiou}, {\it {Baryon acoustic oscillations from a different angle}},  {\em \mnras} {\bf 540} (July, 2025) 2844--2852, [\href{http://arxiv.org/abs/2505.02658}{{\tt arXiv:2505.02658}}].

\bibitem{Efstathiou_2025_SN}
G.~{Efstathiou}, {\it {Evolving dark energy or supernovae systematics?}},  {\em \mnras} {\bf 538} (Apr., 2025) 875--882, [\href{http://arxiv.org/abs/2408.07175}{{\tt arXiv:2408.07175}}].

\bibitem{Ong_2026}
D.~D.~Y. {Ong}, D.~{Yallup}, and W.~{Handley}, {\it {The Bayesian view of DESI DR2 with unimpeded: Evidence and tension in a combined analysis with CMB and supernovae across cosmological models}},  {\em arXiv e-prints} (Mar., 2026) arXiv:2603.05472, [\href{http://arxiv.org/abs/2603.05472}{{\tt arXiv:2603.05472}}].

\bibitem{Brout_2022}
D.~{Brout}, D.~{Scolnic}, B.~{Popovic}, A.~G. {Riess}, A.~{Carr}, J.~{Zuntz}, R.~{Kessler}, T.~M. {Davis}, S.~{Hinton}, D.~{Jones}, and et~al., {\it {The Pantheon+ Analysis: Cosmological Constraints}},  {\em \apj} {\bf 938} (Oct., 2022) 110, [\href{http://arxiv.org/abs/2202.04077}{{\tt arXiv:2202.04077}}].

\bibitem{Scolnic_2022}
D.~{Scolnic}, D.~{Brout}, A.~{Carr}, A.~G. {Riess}, T.~M. {Davis}, A.~{Dwomoh}, D.~O. {Jones}, N.~{Ali}, P.~{Charvu}, R.~{Chen}, and et~al., {\it {The Pantheon+ Analysis: The Full Data Set and Light-curve Release}},  {\em \apj} {\bf 938} (Oct., 2022) 113, [\href{http://arxiv.org/abs/2112.03863}{{\tt arXiv:2112.03863}}].

\bibitem{Lyke_2020}
B.~W. {Lyke}, A.~N. {Higley}, J.~N. {McLane}, D.~P. {Schurhammer}, A.~D. {Myers}, A.~J. {Ross}, K.~{Dawson}, S.~{Chabanier}, P.~{Martini}, N.~G. {Busca}, and et~al., {\it {The Sloan Digital Sky Survey Quasar Catalog: Sixteenth Data Release}},  {\em \apjs} {\bf 250} (Sept., 2020) 8, [\href{http://arxiv.org/abs/2007.09001}{{\tt arXiv:2007.09001}}].

\bibitem{WuShen_2022}
Q.~{Wu} and Y.~{Shen}, {\it {A Catalog of Quasar Properties from Sloan Digital Sky Survey Data Release 16}},  {\em \apjs} {\bf 263} (Dec., 2022) 42, [\href{http://arxiv.org/abs/2209.03987}{{\tt arXiv:2209.03987}}].

\bibitem{Gaia_DR3}
{Gaia Collaboration}, A.~{Vallenari}, A.~G.~A. {Brown}, T.~{Prusti}, J.~H.~J. {de Bruijne}, F.~{Arenou}, C.~{Babusiaux}, M.~{Biermann}, O.~L. {Creevey}, C.~{Ducourant}, and et~al., {\it {Gaia Data Release 3. Summary of the content and survey properties}},  {\em \aap} {\bf 674} (June, 2023) A1, [\href{http://arxiv.org/abs/2208.00211}{{\tt arXiv:2208.00211}}].

\bibitem{eyer_2023gaia}
L.~{Eyer}, M.~{Audard}, B.~{Holl}, L.~{Rimoldini}, M.~I. {Carnerero}, G.~{Clementini}, J.~{De Ridder}, E.~{Distefano}, D.~W. {Evans}, P.~{Gavras}, and et~al., {\it {Gaia Data Release 3. Summary of the variability processing and analysis}},  {\em \aap} {\bf 674} (June, 2023) A13, [\href{http://arxiv.org/abs/2206.06416}{{\tt arXiv:2206.06416}}].

\bibitem{Risaliti_2015}
G.~{Risaliti} and E.~{Lusso}, {\it {A Hubble Diagram for Quasars}},  {\em \apj} {\bf 815} (Dec., 2015) 33, [\href{http://arxiv.org/abs/1505.07118}{{\tt arXiv:1505.07118}}].

\bibitem{Risaliti_2019}
G.~{Risaliti} and E.~{Lusso}, {\it {Cosmological Constraints from the Hubble Diagram of Quasars at High Redshifts}},  {\em Nature Astronomy} {\bf 3} (Jan., 2019) 272--277, [\href{http://arxiv.org/abs/1811.02590}{{\tt arXiv:1811.02590}}].

\bibitem{Lusso_2020}
E.~{Lusso}, G.~{Risaliti}, E.~{Nardini}, G.~{Bargiacchi}, M.~{Benetti}, S.~{Bisogni}, S.~{Capozziello}, F.~{Civano}, L.~{Eggleston}, M.~{Elvis}, and et~al., {\it {Quasars as standard candles. III. Validation of a new sample for cosmological studies}},  {\em \aap} {\bf 642} (Oct., 2020) A150, [\href{http://arxiv.org/abs/2008.08586}{{\tt arXiv:2008.08586}}].

\bibitem{Lusso_2025}
E.~{Lusso}, G.~{Risaliti}, and E.~{Nardini}, {\it {Are quasars reliable standard candles?}},  {\em \aap} {\bf 697} (May, 2025) A108, [\href{http://arxiv.org/abs/2504.02040}{{\tt arXiv:2504.02040}}].

\bibitem{Lusso_2017}
E.~{Lusso} and G.~{Risaliti}, {\it {Quasars as standard candles. I. The physical relation between disc and coronal emission}},  {\em \aap} {\bf 602} (June, 2017) A79, [\href{http://arxiv.org/abs/1703.05299}{{\tt arXiv:1703.05299}}].

\bibitem{Huang_2024RLQ}
L.~{Huang}, Z.~Y. {Tu}, N.~{Chang}, F.~F. {Song}, F.~{He}, and X.~Y. {Fu}, {\it {Cosmological constraints from type-I radio-loud quasars}},  {\em \prd} {\bf 109} (Feb., 2024) 043529, [\href{http://arxiv.org/abs/2307.07592}{{\tt arXiv:2307.07592}}].

\bibitem{Huang_2023CIV}
L.~{Huang}, H.~{Wang}, Z.~{Gao}, X.~{Zeng}, and Z.~{Chang}, {\it {A measure of cosmological distance using the C IV Baldwin effect in quasars}},  {\em \aap} {\bf 674} (June, 2023) A163, [\href{http://arxiv.org/abs/2207.09456}{{\tt arXiv:2207.09456}}].

\bibitem{Hook_1994}
I.~M. {Hook}, R.~G. {McMahon}, B.~J. {Boyle}, and M.~J. {Irwin}, {\it {The variability of optically selected quasars.}},  {\em \mnras} {\bf 268} (May, 1994) 305--320.

\bibitem{VandenBerk_2004}
D.~E. {Vanden Berk}, B.~C. {Wilhite}, R.~G. {Kron}, S.~F. {Anderson}, R.~J. {Brunner}, P.~B. {Hall}, {\v{Z}}.~{Ivezi{\'c}}, G.~T. {Richards}, D.~P. {Schneider}, D.~G. {York}, and et~al., {\it {The Ensemble Photometric Variability of \raisebox{-0.5ex}\textasciitilde25,000 Quasars in the Sloan Digital Sky Survey}},  {\em \apj} {\bf 601} (Feb., 2004) 692--714, [\href{http://arxiv.org/abs/astro-ph/0310336}{{\tt astro-ph/0310336}}].

\bibitem{Kelly_2009}
B.~C. Kelly, J.~Bechtold, and A.~Siemiginowska, {\it Are the variations in quasar optical flux driven by thermal fluctuations?},  {\em The Astrophysical Journal} {\bf 698} (2009), no.~1 895--910, [\href{http://arxiv.org/abs/0903.5315}{{\tt arXiv:0903.5315}}].

\bibitem{MacLeod_2010}
C.~L. {MacLeod}, {\v{Z}}.~{Ivezi{\'c}}, C.~S. {Kochanek}, S.~{Koz{\l}owski}, B.~{Kelly}, E.~{Bullock}, A.~{Kimball}, B.~{Sesar}, D.~{Westman}, K.~{Brooks}, and et~al., {\it {Modeling the Time Variability of SDSS Stripe 82 Quasars as a Damped Random Walk}},  {\em \apj} {\bf 721} (Oct., 2010) 1014--1033, [\href{http://arxiv.org/abs/1004.0276}{{\tt arXiv:1004.0276}}].

\bibitem{MacLeod_2012}
C.~L. {MacLeod}, {\v{Z}}.~{Ivezi{\'c}}, B.~{Sesar}, W.~{de Vries}, C.~S. {Kochanek}, B.~C. {Kelly}, A.~C. {Becker}, R.~H. {Lupton}, P.~B. {Hall}, G.~T. {Richards}, and et~al., {\it {A Description of Quasar Variability Measured Using Repeated SDSS and POSS Imaging}},  {\em \apj} {\bf 753} (July, 2012) 106, [\href{http://arxiv.org/abs/1112.0679}{{\tt arXiv:1112.0679}}].

\bibitem{Caplar_2017}
N.~{Caplar}, S.~J. {Lilly}, and B.~{Trakhtenbrot}, {\it {Optical Variability of AGNs in the PTF/iPTF Survey}},  {\em \apj} {\bf 834} (Jan., 2017) 111, [\href{http://arxiv.org/abs/1611.03082}{{\tt arXiv:1611.03082}}].

\bibitem{Suberlak_2021}
K.~L. {Suberlak}, {\v{Z}}.~{Ivezi{\'c}}, and C.~{MacLeod}, {\it {Improving Damped Random Walk Parameters for SDSS Stripe 82 Quasars with Pan-STARRS1}},  {\em \apj} {\bf 907} (Feb., 2021) 96, [\href{http://arxiv.org/abs/2012.12907}{{\tt arXiv:2012.12907}}].

\bibitem{Arevalo_2023}
P.~{Ar{\'e}valo}, P.~{Lira}, P.~{S{\'a}nchez-S{\'a}ez}, P.~{Patel}, E.~{L{\'o}pez-Navas}, E.~{Churazov}, and L.~{Hern{\'a}ndez-Garc{\'i}a}, {\it {Optical variability in quasars: scalings with black hole mass and Eddington ratio depend on the observed time-scales}},  {\em Monthly Notices of the Royal Astronomical Society} {\bf 526} (2023), no.~4 6078--6087, [\href{http://arxiv.org/abs/2304.14228}{{\tt arXiv:2304.14228}}].

\bibitem{Dutra_2025}
I.~Dutra, C.~J. Burke, P.~Natarajan, and W.~Yu, {\it Evidence for evolving dark energy from a new cosmic probe},  {\em arXiv e-prints} (Dec., 2025) arXiv:2512.07931, [\href{http://arxiv.org/abs/2512.07931}{{\tt arXiv:2512.07931}}].

\bibitem{Yu_2025}
W.~{Yu}, G.~T. {Richards}, J.~J. {Ruan}, M.~S. {Vogeley}, F.~E. {Bauer}, and M.~J. {Graham}, {\it {Examining Active Galactic Nucleus UV/Optical Variability beyond the Simple Damped Random Walk. II. Insights from 22 yr Observations of SDSS, PS1, and ZTF}},  {\em The Astrophysical Journal} {\bf 992} (2025), no.~1 130, [\href{http://arxiv.org/abs/2508.12076}{{\tt arXiv:2508.12076}}].

\bibitem{Kozlowski_2017}
S.~{Koz{\l}owski}, {\it {Limitations on the recovery of the true AGN variability parameters using damped random walk modeling}},  {\em \aap} {\bf 597} (Jan., 2017) A128, [\href{http://arxiv.org/abs/1611.08248}{{\tt arXiv:1611.08248}}].

\bibitem{Hu_2024}
X.-F. Hu, Z.-Y. Cai, and J.-X. Wang, {\it How can the optical variation properties of active galactic nuclei be unbiasedly measured?},  {\em The Astrophysical Journal} {\bf 961} (2024), no.~1 5, [\href{http://arxiv.org/abs/2310.16223}{{\tt arXiv:2310.16223}}].

\bibitem{deVries_2005}
W.~H. {de Vries}, R.~H. {Becker}, R.~L. {White}, and C.~{Loomis}, {\it {Structure Function Analysis of Long-Term Quasar Variability}},  {\em \aj} {\bf 129} (Feb., 2005) 615--629, [\href{http://arxiv.org/abs/astro-ph/0411348}{{\tt astro-ph/0411348}}].

\bibitem{Schmidt_2010}
K.~B. {Schmidt}, P.~J. {Marshall}, H.-W. {Rix}, S.~{Jester}, J.~F. {Hennawi}, and G.~{Dobler}, {\it {Selecting Quasars by Their Intrinsic Variability}},  {\em \apj} {\bf 714} (May, 2010) 1194--1208, [\href{http://arxiv.org/abs/1002.2642}{{\tt arXiv:1002.2642}}].

\bibitem{Kozlowski_2016}
S.~{Koz{\l}owski}, {\it {Revisiting Stochastic Variability of AGNs with Structure Functions}},  {\em \apj} {\bf 826} (Aug., 2016) 118, [\href{http://arxiv.org/abs/1604.05858}{{\tt arXiv:1604.05858}}].

\bibitem{Wilhite_2008}
B.~C. {Wilhite}, R.~J. {Brunner}, C.~J. {Grier}, D.~P. {Schneider}, and D.~E. {Vanden Berk}, {\it {On the variability of quasars: a link between the Eddington ratio and optical variability?}},  {\em Monthly Notices of the Royal Astronomical Society} {\bf 383} (2008), no.~3 1232--1240, [\href{http://arxiv.org/abs/0711.4844}{{\tt arXiv:0711.4844}}].

\bibitem{Morganson_2014}
E.~{Morganson}, W.~S. {Burgett}, K.~C. {Chambers}, P.~J. {Green}, N.~{Kaiser}, E.~A. {Magnier}, P.~J. {Marshall}, J.~S. {Morgan}, P.~A. {Price}, H.-W. {Rix}, and et~al., {\it {Measuring Quasar Variability with Pan-STARRS1 and SDSS}},  {\em \apj} {\bf 784} (Apr., 2014) 92, [\href{http://arxiv.org/abs/1407.0716}{{\tt arXiv:1407.0716}}].

\bibitem{Li_2018}
Z.~{Li}, I.~D. {McGreer}, X.-B. {Wu}, X.~{Fan}, and Q.~{Yang}, {\it {The Ensemble Photometric Variability of Over {}10$^{5}$ Quasars in the Dark Energy Camera Legacy Survey and the Sloan Digital Sky Survey}},  {\em \apj} {\bf 861} (July, 2018) 6, [\href{http://arxiv.org/abs/1805.07747}{{\tt arXiv:1805.07747}}].

\bibitem{Kovacevic_2021}
A.~B. {Kova{\v{c}}evi{\'c}}, D.~{Ili{\'c}}, L.~{\v{C}}. {Popovi{\'c}}, V.~{Radovi{\'c}}, I.~{Jankov}, I.~{Yoon}, N.~{Caplar}, I.~{{\v{C}}vorovi{\'c}-Hajdinjak}, and S.~{Simi{\'c}}, {\it {On possible proxies of AGN light-curves cadence selection in future time domain surveys}},  {\em \mnras} {\bf 505} (Aug., 2021) 5012--5028, [\href{http://arxiv.org/abs/2105.14889}{{\tt arXiv:2105.14889}}].

\bibitem{Stone_2023}
Z.~{Stone}, Y.~{Shen}, C.~J. {Burke}, Y.-C. {Chen}, Q.~{Yang}, X.~{Liu}, R.~A. {Gruendl}, M.~{Adam{\'o}w}, F.~{Andrade-Oliveira}, J.~{Annis}, and et~al., {\it {Correction to: Optical variability of quasars with 20-year photometric light curves}},  {\em \mnras} {\bf 521} (May, 2023) 836--839.

\bibitem{Kelly_2014}
B.~C. {Kelly}, A.~C. {Becker}, M.~{Sobolewska}, A.~{Siemiginowska}, and P.~{Uttley}, {\it {Flexible and Scalable Methods for Quantifying Stochastic Variability in the Era of Massive Time-domain Astronomical Data Sets}},  {\em The Astrophysical Journal} {\bf 788} (2014), no.~1 33, [\href{http://arxiv.org/abs/1402.5978}{{\tt arXiv:1402.5978}}].

\bibitem{Zu_2013}
Y.~{Zu}, C.~S. {Kochanek}, S.~{Koz{\l}owski}, and A.~{Udalski}, {\it {Is Quasar Optical Variability a Damped Random Walk?}},  {\em \apj} {\bf 765} (Mar., 2013) 106, [\href{http://arxiv.org/abs/1202.3783}{{\tt arXiv:1202.3783}}].

\bibitem{Kozlowski_2010}
S.~{Koz{\l}owski}, C.~S. {Kochanek}, A.~{Udalski}, {\L}.~{Wyrzykowski}, I.~{Soszy{\'n}ski}, M.~K. {Szyma{\'n}ski}, M.~{Kubiak}, G.~{Pietrzy{\'n}ski}, O.~{Szewczyk}, K.~{Ulaczyk}, and et~al., {\it {Quantifying Quasar Variability as Part of a General Approach to Classifying Continuously Varying Sources}},  {\em \apj} {\bf 708} (Jan., 2010) 927--945, [\href{http://arxiv.org/abs/0909.1326}{{\tt arXiv:0909.1326}}].

\bibitem{Mushotzky_2011}
R.~F. Mushotzky, R.~Edelson, W.~Baumgartner, and P.~Gandhi, {\it Kepler observations of rapid optical variability in active galactic nuclei},  {\em The Astrophysical Journal Letters} {\bf 743} (2011), no.~1 L12, [\href{http://arxiv.org/abs/1111.0672}{{\tt arXiv:1111.0672}}].

\bibitem{Kasliwal_2015}
V.~P. {Kasliwal}, M.~S. {Vogeley}, and G.~T. {Richards}, {\it {Are the variability properties of the Kepler AGN light curves consistent with a damped random walk?}},  {\em Monthly Notices of the Royal Astronomical Society} {\bf 451} (2015), no.~4 4328--4345, [\href{http://arxiv.org/abs/1505.00360}{{\tt arXiv:1505.00360}}].

\bibitem{Zhu_2016}
F.-F. {Zhu}, J.-X. {Wang}, Z.-Y. {Cai}, and Y.-H. {Sun}, {\it {The Timescale-dependent Color Variability of Quasars Viewed with GALEX}},  {\em The Astrophysical Journal} {\bf 832} (2016), no.~1 75, [\href{http://arxiv.org/abs/1609.07136}{{\tt arXiv:1609.07136}}].

\bibitem{Yu_2022}
W.~{Yu}, G.~T. {Richards}, M.~S. {Vogeley}, J.~{Moreno}, and M.~J. {Graham}, {\it {Examining AGN UV/Optical Variability beyond the Simple Damped Random Walk}},  {\em The Astrophysical Journal} {\bf 936} (2022), no.~2 132, [\href{http://arxiv.org/abs/2201.08943}{{\tt arXiv:2201.08943}}].

\bibitem{Kozlowski_2021}
S.~{Koz{\l}owski}, {\it A survey length for agn variability studies},  {\em Acta Astronomica} {\bf 71} (2021), no.~2 103--112, [\href{http://arxiv.org/abs/2109.03896}{{\tt arXiv:2109.03896}}].

\bibitem{ShakuraSunyaev_1973}
N.~I. {Shakura} and R.~A. {Sunyaev}, {\it {Black holes in binary systems. Observational appearance.}},  {\em \aap} {\bf 24} (Jan., 1973) 337--355.

\bibitem{Krolik_1991}
J.~H. {Krolik}, K.~{Horne}, T.~R. {Kallman}, M.~A. {Malkan}, R.~A. {Edelson}, and G.~A. {Kriss}, {\it {Ultraviolet Variability of NGC 5548: Dynamics of the Continuum Production Region and Geometry of the Broad-Line Region}},  {\em \apj} {\bf 371} (Apr., 1991) 541.

\bibitem{Cackett_2007}
E.~M. {Cackett}, K.~{Horne}, and H.~{Winkler}, {\it {Testing thermal reprocessing in active galactic nuclei accretion discs}},  {\em \mnras} {\bf 380} (Sept., 2007) 669--682, [\href{http://arxiv.org/abs/0706.1464}{{\tt arXiv:0706.1464}}].

\bibitem{Cackett_2021}
E.~M. {Cackett}, M.~C. {Bentz}, and E.~{Kara}, {\it {Reverberation mapping of active galactic nuclei: from X-ray corona to dusty torus}},  {\em iScience} {\bf 24} (June, 2021) 102557, [\href{http://arxiv.org/abs/2105.06926}{{\tt arXiv:2105.06926}}].

\bibitem{KubotaDone_2018}
A.~{Kubota} and C.~{Done}, {\it {A physical model of the broad-band continuum of AGN and its implications for the UV/X relation and optical variability}},  {\em \mnras} {\bf 480} (Oct., 2018) 1247--1262, [\href{http://arxiv.org/abs/1804.00171}{{\tt arXiv:1804.00171}}].

\bibitem{Hagen_2024}
S.~Hagen, C.~Done, and R.~Edelson, {\it What drives the variability in {AGN}? explaining the {UV--X-ray} disconnect through propagating fluctuations},  {\em Monthly Notices of the Royal Astronomical Society} {\bf 530} (2024), no.~4 4850--4867, [\href{http://arxiv.org/abs/2401.03452}{{\tt arXiv:2401.03452}}].

\bibitem{SanchezSaez_2018}
P.~{S{\'a}nchez-S{\'a}ez}, P.~{Lira}, J.~{Mej{\'\i}a-Restrepo}, L.~C. {Ho}, P.~{Ar{\'e}valo}, M.~{Kim}, R.~{Cartier}, and P.~{Coppi}, {\it {The QUEST-La Silla AGN Variability Survey: Connection between AGN Variability and Black Hole Physical Properties}},  {\em \apj} {\bf 864} (Sept., 2018) 87, [\href{http://arxiv.org/abs/1808.00967}{{\tt arXiv:1808.00967}}].

\bibitem{Patel_2025}
P.~{Patel}, P.~{Lira}, P.~{Ar{\'e}valo}, M.~{Sun}, S.~{Bernal}, and M.~L. {Mart{\'\i}nez-Aldama}, {\it {Probing the rest-frame wavelength dependence of quasar variability: Insights from the Zwicky Transient Facility Survey}},  {\em \aap} {\bf 695} (Mar., 2025) A162, [\href{http://arxiv.org/abs/2409.14999}{{\tt arXiv:2409.14999}}].

\bibitem{Goncalves_2025}
H.~B. Gon{\c{c}}alves, S.~Panda, T.~Storchi-Bergmann, E.~M. Cackett, and M.~Eracleous, {\it Exploring quasar variability with ztf at $0 < z < 3$: A universal relation with eddington ratio},  \href{http://arxiv.org/abs/2505.09779}{{\tt arXiv:2505.09779}}.

\bibitem{Burke_2021}
C.~J. Burke, Y.~Shen, O.~Blaes, C.~F. Gammie, K.~Horne, Y.-F. Jiang, X.~Liu, I.~M. McHardy, C.~W. Morgan, S.~Scaringi, and Q.~Yang, {\it A characteristic optical variability time scale in astrophysical accretion disks},  {\em Science} {\bf 373} (2021), no.~6556 789--792, [\href{http://arxiv.org/abs/2108.05389}{{\tt arXiv:2108.05389}}].

\bibitem{Vestergaard_2006}
M.~{Vestergaard} and B.~M. {Peterson}, {\it {Determining Central Black Hole Masses in Distant Active Galaxies and Quasars. II. Improved Optical and UV Scaling Relationships}},  {\em \apj} {\bf 641} (Apr., 2006) 689--709, [\href{http://arxiv.org/abs/astro-ph/0601303}{{\tt astro-ph/0601303}}].

\bibitem{ShenKelly_2010}
Y.~{Shen} and B.~C. {Kelly}, {\it {The Impact of the Uncertainty in Single-epoch Virial Black Hole Mass Estimates on the Observed Evolution of the Black Hole-bulge Scaling Relations}},  {\em \apj} {\bf 713} (Apr., 2010) 41--45, [\href{http://arxiv.org/abs/0911.5208}{{\tt arXiv:0911.5208}}].

\bibitem{Shen_2013}
Y.~{Shen}, {\it {The mass of quasars}},  {\em Bulletin of the Astronomical Society of India} {\bf 41} (Mar., 2013) 61--115, [\href{http://arxiv.org/abs/1302.2643}{{\tt arXiv:1302.2643}}].

\bibitem{Runnoe_2012}
J.~C. {Runnoe}, M.~S. {Brotherton}, and Z.~{Shang}, {\it {Updating quasar bolometric luminosity corrections}},  {\em \mnras} {\bf 422} (May, 2012) 478--493, [\href{http://arxiv.org/abs/1201.5155}{{\tt arXiv:1201.5155}}].

\bibitem{Krawczyk_2013}
C.~M. {Krawczyk}, G.~T. {Richards}, S.~S. {Mehta}, M.~S. {Vogeley}, S.~C. {Gallagher}, K.~M. {Leighly}, N.~P. {Ross}, and D.~P. {Schneider}, {\it {Mean Spectral Energy Distributions and Bolometric Corrections for Luminous Quasars}},  {\em \apjs} {\bf 206} (May, 2013) 4, [\href{http://arxiv.org/abs/1304.5573}{{\tt arXiv:1304.5573}}].

\bibitem{Aird_2015}
J.~{Aird}, A.~L. {Coil}, A.~{Georgakakis}, K.~{Nandra}, G.~{Barro}, and P.~G. {P{\'e}rez-Gonz{\'a}lez}, {\it The evolution of the x-ray luminosity functions of unabsorbed and absorbed agns out to z~$\sim$~5},  {\em Monthly Notices of the Royal Astronomical Society} {\bf 451} (2015), no.~2 1892--1927, [\href{http://arxiv.org/abs/1503.01120}{{\tt arXiv:1503.01120}}].

\bibitem{Kulkarni_2019}
G.~{Kulkarni}, G.~{Worseck}, and J.~F. {Hennawi}, {\it Evolution of the agn uv luminosity function from redshift 7.5},  {\em Monthly Notices of the Royal Astronomical Society} {\bf 488} (2019), no.~1 1035--1065, [\href{http://arxiv.org/abs/1807.09774}{{\tt arXiv:1807.09774}}].

\bibitem{Shen_2020}
X.~{Shen}, P.~F. {Hopkins}, C.-A. {Faucher-Gigu{\`e}re}, D.~M. {Alexander}, G.~T. {Richards}, N.~P. {Ross}, and R.~C. {Hickox}, {\it The bolometric quasar luminosity function at z = 0--7},  {\em Monthly Notices of the Royal Astronomical Society} {\bf 495} (2020), no.~3 3252--3275, [\href{http://arxiv.org/abs/2001.02696}{{\tt arXiv:2001.02696}}].

\bibitem{Kelly_2007}
B.~C. Kelly, {\it Some aspects of measurement error in linear regression of astronomical data},  {\em The Astrophysical Journal} {\bf 665} (2007), no.~2 1489--1506, [\href{http://arxiv.org/abs/0705.2774}{{\tt arXiv:0705.2774}}].

\bibitem{March_2018}
M.~C. {March}, R.~C. {Wolf}, m.~{Sako}, C.~{D'Andrea}, and D.~{Brout}, {\it {A Bayesian approach to truncated data sets: An application to Malmquist bias in Supernova Cosmology}},  {\em arXiv e-prints} (Apr., 2018) arXiv:1804.02474, [\href{http://arxiv.org/abs/1804.02474}{{\tt arXiv:1804.02474}}].

\bibitem{Mantz_2019}
A.~B. Mantz, {\it Coping with selection effects: A primer on regression with truncated data},  {\em Monthly Notices of the Royal Astronomical Society} {\bf 485} (2019), no.~4 4863--4872, [\href{http://arxiv.org/abs/1901.10522}{{\tt arXiv:1901.10522}}].

\bibitem{Mandel_2019}
I.~Mandel, W.~M. Farr, and J.~R. Gair, {\it Extracting distribution parameters from multiple uncertain observations with selection biases},  {\em Monthly Notices of the Royal Astronomical Society} {\bf 486} (2019), no.~1 1086--1093, [\href{http://arxiv.org/abs/1809.02063}{{\tt arXiv:1809.02063}}].

\bibitem{Freedman_2019}
W.~L. {Freedman}, B.~F. {Madore}, D.~{Hatt}, T.~J. {Hoyt}, I.~S. {Jang}, R.~L. {Beaton}, C.~R. {Burns}, M.~G. {Lee}, A.~J. {Monson}, J.~R. {Neeley}, and et~al., {\it {The Carnegie-Chicago Hubble Program. VIII. An Independent Determination of the Hubble Constant Based on the Tip of the Red Giant Branch}},  {\em \apj} {\bf 882} (Sept., 2019) 34, [\href{http://arxiv.org/abs/1907.05922}{{\tt arXiv:1907.05922}}].

\bibitem{Foreman-Mackey_2017}
D.~{Foreman-Mackey}, E.~{Agol}, S.~{Ambikasaran}, and R.~{Angus}, {\it {Fast and Scalable Gaussian Process Modeling with Applications to Astronomical Time Series}},  {\em \aj} {\bf 154} (Dec., 2017) 220, [\href{http://arxiv.org/abs/1703.09710}{{\tt arXiv:1703.09710}}].

\bibitem{Goodman_2010}
J.~{Goodman} and J.~{Weare}, {\it {Ensemble samplers with affine invariance}},  {\em Communications in Applied Mathematics and Computational Science} {\bf 5} (Jan., 2010) 65--80.

\bibitem{Foreman-Mackey_2013}
D.~{Foreman-Mackey}, D.~W. {Hogg}, D.~{Lang}, and J.~{Goodman}, {\it {emcee: The MCMC Hammer}},  {\em \pasp} {\bf 125} (Mar., 2013) 306, [\href{http://arxiv.org/abs/1202.3665}{{\tt arXiv:1202.3665}}].

\bibitem{Bradbury_2021}
J.~{Bradbury}, R.~{Frostig}, P.~{Hawkins}, M.~J. {Johnson}, C.~{Leary}, D.~{Maclaurin}, G.~{Necula}, A.~{Paszke}, J.~{VanderPlas}, S.~{Wanderman-Milne}, and et~al., ``{JAX: Autograd and XLA}.'' Astrophysics Source Code Library, record ascl:2111.002, Nov., 2021.

\bibitem{Phan_2019}
D.~{Phan}, N.~{Pradhan}, and M.~{Jankowiak}, {\it {Composable Effects for Flexible and Accelerated Probabilistic Programming in NumPyro}},  {\em arXiv e-prints} (Dec., 2019) arXiv:1912.11554, [\href{http://arxiv.org/abs/1912.11554}{{\tt arXiv:1912.11554}}].

\bibitem{HoffmanGelman_2011}
M.~D. {Hoffman} and A.~{Gelman}, {\it {The No-U-Turn Sampler: Adaptively Setting Path Lengths in Hamiltonian Monte Carlo}},  {\em arXiv e-prints} (Nov., 2011) arXiv:1111.4246, [\href{http://arxiv.org/abs/1111.4246}{{\tt arXiv:1111.4246}}].

\bibitem{GelmanRubin_1992}
A.~{Gelman} and D.~B. {Rubin}, {\it {Inference from Iterative Simulation Using Multiple Sequences}},  {\em Statistical Science} {\bf 7} (Jan., 1992) 457--472.

\bibitem{Vehtari_2021}
A.~{Vehtari}, A.~{Gelman}, D.~{Simpson}, B.~{Carpenter}, and P.-C. {B{\"u}rkner}, {\it {Rank-normalization, folding, and localization: An improved R-hat for assessing convergence of MCMC (with Discussion)}},  {\em Bayesian Analysis} {\bf 16} (June, 2021) 667--718, [\href{http://arxiv.org/abs/1903.08008}{{\tt arXiv:1903.08008}}].

\bibitem{Temple_2021}
M.~J. {Temple}, P.~C. {Hewett}, and M.~{Banerji}, {\it {Modelling type 1 quasar colours in the era of Rubin and Euclid}},  {\em \mnras} {\bf 508} (Nov., 2021) 737--754, [\href{http://arxiv.org/abs/2109.04472}{{\tt arXiv:2109.04472}}].

\bibitem{rimoldini_2023gaia}
L.~{Rimoldini}, B.~{Holl}, P.~{Gavras}, M.~{Audard}, J.~{De Ridder}, N.~{Mowlavi}, K.~{Nienartowicz}, G.~{Jevardat de Fombelle}, I.~{Lecoeur-Ta{\"\i}bi}, L.~{Karbevska}, and et~al., {\it {Gaia Data Release 3. All-sky classification of 12.4 million variable sources into 25 classes}},  {\em \aap} {\bf 674} (June, 2023) A14, [\href{http://arxiv.org/abs/2211.17238}{{\tt arXiv:2211.17238}}].

\bibitem{Wang_2026}
H.~{Wang}, X.-B. {Wu}, N.~{Yao}, B.~{Lyu}, Y.~{Pang}, Y.~{Fu}, R.~{Zhu}, and Q.~{Yang}, {\it {Systematic Analysis of Changing-Look Active Galactic Nucleus Variability Using ZTF Light Curves}},  {\em \apj} {\bf 997} (Jan., 2026) 100, [\href{http://arxiv.org/abs/2511.10217}{{\tt arXiv:2511.10217}}].

\bibitem{Stone_2022}
Z.~{Stone}, Y.~{Shen}, C.~J. {Burke}, Y.-C. {Chen}, Q.~{Yang}, X.~{Liu}, R.~A. {Gruendl}, M.~{Adam{\'o}w}, F.~{Andrade-Oliveira}, J.~{Annis}, and et~al., {\it {Optical variability of quasars with 20-yr photometric light curves}},  {\em \mnras} {\bf 514} (July, 2022) 164--184, [\href{http://arxiv.org/abs/2201.02762}{{\tt arXiv:2201.02762}}].

\bibitem{Yu_2026_EztaoX}
W.~{Yu}, J.~J. {Ruan}, C.~J. {Burke}, R.~J. {Assef}, T.~T. {Ananna}, F.~E. {Bauer}, D.~{De Cicco}, K.~{Horne}, L.~{Hern{\'a}ndez-Garc{\'\i}a}, D.~{Ili{\'c}}, and et~al., {\it {Scalable and Robust Multiband Modeling of AGN Light Curves in Rubin-LSST}},  {\em \apj} {\bf 998} (Feb., 2026) 144, [\href{http://arxiv.org/abs/2511.21479}{{\tt arXiv:2511.21479}}].

\bibitem{Brewer_2026}
B.~J. {Brewer}, G.~F. {Lewis}, X.~{Yu}, and Y.~{Li}, {\it {The Information Content of Quasar Variability Light Curves: How Well Can we Infer Stochastic Model Parameters?}},  {\em arXiv e-prints} (May, 2026) arXiv:2606.01496, [\href{http://arxiv.org/abs/2606.01496}{{\tt arXiv:2606.01496}}].

\bibitem{Chevallier_2001}
M.~{Chevallier} and D.~{Polarski}, {\it {Accelerating Universes with Scaling Dark Matter}},  {\em International Journal of Modern Physics D} {\bf 10} (Jan., 2001) 213--223, [\href{http://arxiv.org/abs/gr-qc/0009008}{{\tt gr-qc/0009008}}].

\bibitem{Linder_2003}
E.~V. {Linder}, {\it {Exploring the Expansion History of the Universe}},  {\em \prl} {\bf 90} (Mar., 2003) 091301, [\href{http://arxiv.org/abs/astro-ph/0208512}{{\tt astro-ph/0208512}}].

\bibitem{Madau_1995}
P.~{Madau}, {\it {Radiative Transfer in a Clumpy Universe: The Colors of High-Redshift Galaxies}},  {\em \apj} {\bf 441} (Mar., 1995) 18.

\bibitem{Inoue_2014}
A.~K. {Inoue}, I.~{Shimizu}, I.~{Iwata}, and M.~{Tanaka}, {\it {An updated analytic model for attenuation by the intergalactic medium}},  {\em \mnras} {\bf 442} (Aug., 2014) 1805--1820, [\href{http://arxiv.org/abs/1402.0677}{{\tt arXiv:1402.0677}}].

\bibitem{Bentz_2006}
M.~C. {Bentz}, B.~M. {Peterson}, R.~W. {Pogge}, M.~{Vestergaard}, and C.~A. {Onken}, {\it {The Radius-Luminosity Relationship for Active Galactic Nuclei: The Effect of Host-Galaxy Starlight on Luminosity Measurements}},  {\em \apj} {\bf 644} (June, 2006) 133--142, [\href{http://arxiv.org/abs/astro-ph/0602412}{{\tt astro-ph/0602412}}].

\bibitem{Bentz_2009}
M.~C. {Bentz}, B.~M. {Peterson}, H.~{Netzer}, R.~W. {Pogge}, and M.~{Vestergaard}, {\it {The Radius-Luminosity Relationship for Active Galactic Nuclei: The Effect of Host-Galaxy Starlight on Luminosity Measurements. II. The Full Sample of Reverberation-Mapped AGNs}},  {\em \apj} {\bf 697} (May, 2009) 160--181, [\href{http://arxiv.org/abs/0812.2283}{{\tt arXiv:0812.2283}}].

\bibitem{Bentz_2013}
M.~C. {Bentz}, K.~D. {Denney}, C.~J. {Grier}, A.~J. {Barth}, B.~M. {Peterson}, M.~{Vestergaard}, V.~N. {Bennert}, G.~{Canalizo}, G.~{De Rosa}, A.~V. {Filippenko}, and et~al., {\it {The Low-luminosity End of the Radius-Luminosity Relationship for Active Galactic Nuclei}},  {\em \apj} {\bf 767} (Apr., 2013) 149, [\href{http://arxiv.org/abs/1303.1742}{{\tt arXiv:1303.1742}}].

\bibitem{Ren_2024}
W.~{Ren}, H.~{Guo}, Y.~{Shen}, J.~D. {Silverman}, C.~J. {Burke}, S.~{Wang}, and J.~{Wang}, {\it {Prior-informed Active Galactic Nucleus Host Spectral Decomposition Using PyQSOFit}},  {\em \apj} {\bf 974} (Oct., 2024) 153, [\href{http://arxiv.org/abs/2406.17598}{{\tt arXiv:2406.17598}}].

\bibitem{Gianniotis_2022}
N.~{Gianniotis}, F.~{Pozo Nu{\~n}ez}, and K.~L. {Polsterer}, {\it {Disentangling the optical AGN and host-galaxy luminosity with a probabilistic flux variation gradient}},  {\em \aap} {\bf 657} (Jan., 2022) A126, [\href{http://arxiv.org/abs/2109.03619}{{\tt arXiv:2109.03619}}].

\bibitem{Wilhite_2005}
B.~C. {Wilhite}, D.~E. {Vanden Berk}, R.~G. {Kron}, D.~P. {Schneider}, N.~{Pereyra}, R.~J. {Brunner}, G.~T. {Richards}, and J.~V. {Brinkmann}, {\it {Spectral Variability of Quasars in the Sloan Digital Sky Survey. I. Wavelength Dependence}},  {\em \apj} {\bf 633} (Nov., 2005) 638--648, [\href{http://arxiv.org/abs/astro-ph/0504309}{{\tt astro-ph/0504309}}].

\bibitem{Shen_2011}
Y.~{Shen}, G.~T. {Richards}, M.~A. {Strauss}, P.~B. {Hall}, D.~P. {Schneider}, S.~{Snedden}, D.~{Bizyaev}, H.~{Brewington}, V.~{Malanushenko}, E.~{Malanushenko}, D.~{Oravetz}, K.~{Pan}, and A.~{Simmons}, {\it {A Catalog of Quasar Properties from Sloan Digital Sky Survey Data Release 7}},  {\em The Astrophysical Journal Supplement Series} {\bf 194} (2011), no.~2 45, [\href{http://arxiv.org/abs/1006.5178}{{\tt arXiv:1006.5178}}].

\bibitem{Bellm_2019}
E.~C. {Bellm}, S.~R. {Kulkarni}, M.~J. {Graham}, R.~{Dekany}, R.~M. {Smith}, R.~{Riddle}, F.~J. {Masci}, G.~{Helou}, T.~A. {Prince}, S.~M. {Adams}, and et~al., {\it {The Zwicky Transient Facility: System Overview, Performance, and First Results}},  {\em \pasp} {\bf 131} (Jan., 2019) 018002, [\href{http://arxiv.org/abs/1902.01932}{{\tt arXiv:1902.01932}}].

\bibitem{Graham_2019}
M.~J. {Graham}, S.~R. {Kulkarni}, E.~C. {Bellm}, S.~M. {Adams}, C.~{Barbarino}, N.~{Blagorodnova}, D.~{Bodewits}, B.~{Bolin}, P.~R. {Brady}, S.~B. {Cenko}, and et~al., {\it {The Zwicky Transient Facility: Science Objectives}},  {\em \pasp} {\bf 131} (July, 2019) 078001, [\href{http://arxiv.org/abs/1902.01945}{{\tt arXiv:1902.01945}}].

\bibitem{Aghamousa_2016}
{DESI Collaboration}, A.~{Aghamousa}, J.~{Aguilar}, S.~{Ahlen}, S.~{Alam}, L.~E. {Allen}, C.~{Allende Prieto}, J.~{Annis}, S.~{Bailey}, C.~{Balland}, and et~al., {\it {The DESI Experiment Part I: Science,Targeting, and Survey Design}},  {\em arXiv e-prints} (Oct., 2016) arXiv:1611.00036, [\href{http://arxiv.org/abs/1611.00036}{{\tt arXiv:1611.00036}}].

\bibitem{Adame_2024}
{DESI Collaboration}, A.~G. {Adame}, J.~{Aguilar}, S.~{Ahlen}, S.~{Alam}, G.~{Aldering}, D.~M. {Alexander}, R.~{Alfarsy}, C.~{Allende Prieto}, M.~{Alvarez}, and et~al., {\it {The Early Data Release of the Dark Energy Spectroscopic Instrument}},  {\em \aj} {\bf 168} (Aug., 2024) 58, [\href{http://arxiv.org/abs/2306.06308}{{\tt arXiv:2306.06308}}].

\bibitem{Ivezic_2019}
{\v Z}.~{Ivezi{\'c}}, S.~M. {Kahn}, J.~A. {Tyson}, B.~{Abel}, E.~{Acosta}, R.~{Allsman}, D.~{Alonso}, Y.~{AlSayyad}, S.~F. {Anderson}, J.~{Andrew}, et~al., {\it {LSST}: From science drivers to reference design and anticipated data products},  {\em The Astrophysical Journal} {\bf 873} (2019), no.~2 111, [\href{http://arxiv.org/abs/0805.2366}{{\tt arXiv:0805.2366}}].

\bibitem{Hogg_2002_kcorr}
D.~W. {Hogg}, I.~K. {Baldry}, M.~R. {Blanton}, and D.~J. {Eisenstein}, {\it {The K correction}},  {\em arXiv e-prints} (Oct., 2002) astro--ph/0210394, [\href{http://arxiv.org/abs/astro-ph/0210394}{{\tt astro-ph/0210394}}].

\bibitem{Riello_2021}
M.~{Riello}, F.~{De Angeli}, D.~W. {Evans}, P.~{Montegriffo}, J.~M. {Carrasco}, G.~{Busso}, L.~{Palaversa}, P.~W. {Burgess}, C.~{Diener}, M.~{Davidson}, and et~al., {\it {Gaia Early Data Release 3. Photometric content and validation}},  {\em \aap} {\bf 649} (May, 2021) A3, [\href{http://arxiv.org/abs/2012.01916}{{\tt arXiv:2012.01916}}].

\bibitem{Gaia_DR3_passbands}
{European Space Agency}, ``{Gaia DR3} passbands.'' {ESA Gaia Cosmos}, 2022.
\newblock Accessed 2025 Feb 14.

\bibitem{Gaia_EDR3_photcal}
{European Space Agency}, ``{Gaia Early Data Release 3 Documentation: Photometric calibration}.'' {Gaia Archive documentation}, 2021.
\newblock Section 5.4.1, photometric zero points and conversion factors; Accessed 2025 Feb 14.

\end{thebibliography}\endgroup

\end{document}